\documentclass[12pt]{iopart}

\usepackage{citesort}

\usepackage{iopams} 
\usepackage{amstext}

\usepackage{graphicx}
\usepackage{color}
\usepackage{bm}
\usepackage{enumerate}
\usepackage{array}
\usepackage[perpage]{footmisc} 
\usepackage{titletoc} 

\newcommand{\bk}{\mathbf{k}}
\newcommand{\abs}[1]{\ensuremath{| #1 |}}
\newcommand{\norm}[1]{\ensuremath{|| #1 ||}}
\newcommand{\vect}[1]{\mathbf{#1}}
\newcommand{\ket}[1]{\ensuremath{|#1\rangle}}

\newcommand{\avg}[1]{\ensuremath{\langle #1\rangle}}

\renewcommand{\contentsname}{} 

\begin{document}

\title{Topology by dissipation}

\author{C-E Bardyn$^1$, M A Baranov$^{2, 3, 4}$, C V Kraus$^{2, 3}$, E Rico$^{2, 3}$, A \.Imamo\u glu$^1$, P Zoller$^{2, 3}$ and S Diehl$^{2, 3}$}

\address{$^1$ Institute for Quantum Electronics, ETH Zurich, 8093 Zurich, Switzerland}
\address{$^2$ Institute for Quantum Optics and Quantum Information of the Austrian Academy of Sciences, A-6020 Innsbruck, Austria}
\address{$^3$ Institute for Theoretical Physics, University of Innsbruck, A-6020 Innsbruck, Austria}
\address{$^4$ NRC ``Kurchatov Institute'', Kurchatov Square 1, 123182 Moscow, Russia}

\eads{\mailto{bardync@phys.ethz.ch}, \mailto{sebastian.diehl@uibk.ac.at}}

\begin{abstract}

Topological states of fermionic matter can be induced by means of a suitably engineered dissipative dynamics. Dissipation then does not occur as a perturbation, but rather as the main resource for many-body dynamics, providing a targeted cooling into a topological phase starting from an arbitrary initial state. We explore the concept of topological order in this setting, developing and applying a general theoretical framework based on the system density matrix which replaces the wave function appropriate for the discussion of Hamiltonian ground-state physics. We identify key analogies and differences to the more conventional Hamiltonian scenario. Differences mainly arise from the fact that the properties of the spectrum and of the state of the system are not as tightly related as in a Hamiltonian context. We provide a symmetry-based topological classification of bulk steady states and identify the classes that are achievable by means of quasi-local dissipative processes driving into superfluid paired states. We also explore the fate of the bulk-edge correspondence in the dissipative setting, and demonstrate the emergence of Majorana edge modes. We illustrate our findings in one- and two-dimensional models that are experimentally realistic in the context of cold atoms.

\end{abstract}

\pacs{To be defined.}

\submitto{\NJP}

\maketitle
\tableofcontents 
\newpage


\section{Introduction}
\label{sec:introduction}

Symmetries, their spontaneous breaking, and related order parameters were considered for a long time as the paradigm for understanding ordered states of matter. A paradigm shift was initiated in the late 1980s when the Landau-Ginzburg broken-symmetry theory of ordered phases---widely thought to be exhaustive---proved unable to characterize a new kind of phases with no local order parameter: \emph{topological} phases, or phases with \emph{topological order}~\cite{Witten89,Wen90}. Instead of being distinguished by symmetries, topological phases are characterized by distinct values of a non-local, topological order parameter, and phases transitions occur whenever the topology changes, signaled by discontinuities in this topological invariant. The existence of topological order may be conditioned on the existence of symmetries. However, as long as topological order is present, the underlying system generally exhibits topological features, i.e., features that are robust against arbitrary (symmetry-preserving) quasi-local perturbations. 

Spectral gap and ground-state degeneracy are typical topological properties which have been theoretically shown to be robust for wide classes of Hamiltonians~\cite{Bravyi10,Bravyi111,Bravyi112,Michalakis11}. Whereas the spectral gap is a property of the bulk---as topological order itself---the ground-state degeneracy  generally depends on the boundary conditions imposed at the edges of the system and on the existence of topological defects in the bulk (e.g., vortices in a superfluid). Most importantly, the degeneracy can be traced to the existence of zero-energy modes localized at the edges or bound to topological defects, which are robust topological features as well. These objects can exhibit exotic behavior under spatial exchange (or ``braiding'') such as non-Abelian statistics~\cite{Moore91,Ivanov01,Nayak08}, which opens up exciting possibilities for practical applications such as topological quantum memories and topological quantum computation~\cite{Freedman02,Kitaev03,Nayak08}.

The search for topological phases exhibiting quasiparticles with non-Abelian statistics has brought $p$-wave paired superfluids and superconductors to the forefront of theoretical and experimental condensed-matter research~\cite{Read00,Ivanov01,Kitaev03,Kitaev06,Nayak08}. Such systems have first been studied in two dimensions (2D), where they have been predicted to support topological phases with gapless chiral edges modes and quasiparticles known as \emph{Majorana zero modes}, giving rise to Ising-type non-Abelian exchange statistics~\cite{Volovik88,Read00,Ivanov01}. Following a seminal paper by Kitaev~\cite{Kitaev01}, the focus has moved more recently to networks of one-dimensional (1D) systems, which were shown to allow for similar topological features (non-Abelian statistics, in particular) as genuine 2D systems~\cite{Alicea11,Halperin12}. Recent proposals for solid-state~\cite{PhysRevLett.105.177002,PhysRevLett.105.077001} and cold atom~\cite{PhysRevLett.106.220402} systems have made it possible for Majorana zero modes to enter the experimental stage, with promising first results in solid-state devices~\cite{Mourik12,Rokhinson12,Das12,Williams12} and the perspective of increased future experimental efforts.

In recent years, the quest for topological states was extended to non-equilibrium systems, going beyond the Hamiltonian ground-state scenario. A first step in this direction was taken with periodically driven Hamiltonian systems~\cite{Lindner11,PhysRevB.82.235114,Rudner13}, in which the time coordinate plays the role of an extra dimension, allowing for the realization of topological invariants with no equilibrium ground-state counterpart. In this work, we focus on a different paradigm in which Hamiltonian unitary dynamics is replaced by specifically designed dissipative dynamics described by a quantum master equation. Such a scenario was originally proposed as a means for quantum state preparation and quantum computation~\cite{Diehl08,Verstraete09} and relies on the proper engineering of a coupling of the system to a suitable reservoir. In the context of cold atoms, such reservoir engineering may be seen as a natural extension of the more conventional Hamiltonian engineering, with similar advantages as compared to solid-state systems such as precise microscopic control and tunability. In previous works, we have shown how this concept can be utilized to ``cool'' or drive ensembles of atomic fermions into topologically ordered states in one~\cite{Diehl11} and two~\cite{Bardyn12} dimensions in a targeted way, starting from an arbitrary initial state described a density matrix. The analysis of the many-body properties of the phases and phase transitions arising in these examples has revealed similarities but also differences between the physics of topological ground states of Hamiltonians and topological steady states resulting from a purely dissipative evolution.

In this work we put the results obtained in our two previous case studies into a broader theoretical perspective. We provide a framework for investigating non-equilibrium topological states that can be reached by means of engineered dissipation, developing a formalism and physical understanding that can also be used in situations where dissipation occurs as a perturbation. The natural object to study is the density matrix of the system, which does not necessarily correspond to a pure state described by a wave function alone. In the present article we focus on quadratic master equations with the aim of classifying topological states described by density matrices in analogy to the Hamiltonian ground-state scenario. All information contained in the density matrix is then equivalently encoded in the \emph{covariance matrix} gathering all static single-particle correlations. By identifying and exploiting the analogy between this object and a quadratic Hamiltonian in a ``first-quantized'' representation, we demonstrate how to classify topological phases in a non-equilibrium context where mixed states are allowed. Our analysis focuses on both bulk and edge properties.

As compared to a Hamiltonian ground-state scenario, key differences arise from the fact that the dynamics---or the spectral properties of the system---and the properties of its ``ground'' (steady) state---or the static correlation properties---are not as tightly related as in the Hamiltonian context. As far as the bulk is concerned, this crucial difference manifests itself in the fact that two independent spectral properties must be present to guarantee that the system is in a stable topological state: The first quantity that we identify is the \emph{dissipative gap}, which corresponds to the slowest damping rate associated with modes belonging to the bulk of the system and is a direct counterpart of the excitation gap of a Hamiltonian spectrum. The second is the \emph{purity gap}, which describes the purity of the mode belonging to the bulk which is most strongly mixed. Clearly, a purity gap is always present in the Hamiltonian context, since Hamiltonian ground states are by definition pure states. In our context, however, we argue that the system can undergo a topological phase transition if \emph{either} (or both) of these two different gaps vanishes in a particular parameter regime.

The purity of the state plays a key role not only in the bulk, but also for the edge physics. In the Hamiltonian context, bulk-edge correspondence theorems describe a tight relation between the number of edge zero modes (i.e., modes that are decoupled from the Hamiltonian dynamics and thus do not evolve) found at the interface between two topologically distinct phases and the value of the topological invariant associated with each of the phases~\cite{Hatsugai93,Kitaev06,Ryu10,Gurarie11,Essin11}. We formulate a dissipative variant of such bulk-edge correspondence: Topological order ensures the existence, at the interface, of a fermionic subspace that is isolated from the bulk (with a dimension determined by the value of the topological invariant on both sides of the interface). However, in contrast to the Hamiltonian case, topology does not guarantee the decoupling of this subspace from the dynamics. As a result, the modes corresponding to this subspace can be either be \emph{zero-damping} modes---i.e., modes that are decoupled from the dynamics similarly as in the Hamiltonian setting---or emerge as \emph{zero-purity} modes---i.e., modes that are in a completely mixed state; in which no information can be stored. In the context of engineered dissipation, the simultaneous appearance of both zero-damping and zero-purity modes may give rise to intriguing physical effects, as we discuss in this work.

The fact that actual physical implementations of model Hamiltonians need often be properly modelled as open systems due to particle losses or dephasing, e.g., has been recognized in a number of theoretical works focusing on the stability of the edge modes~\cite{PhysRevB.84.205109,PhysRevB.85.165124,PhysRevB.85.121405,PhysRevB.85.174533} or on the very definition of topological order in such circumstances~\cite{Rivas13}. We emphasize that our approach is fundamentally different here, since dissipation does not occur as a perturbation but is rather harnessed as the main resource to generate the dynamics.

Our paper is organized as follows. In section~\ref{sec:dissipativeFramework}, we discuss the dissipative framework of interest: We introduce the concept of ``dark states'' in a many-body context, and explain the main ideas behind the physical implementation of a dissipative counterpart of Kitaev's quantum wire, thereby illustrating how to engineer more general dissipative evolutions giving rise to superfluid paired states. We also provide both a second- and a first-quantized formulation of quadratic dissipative dynamics, and discuss the key properties that are necessary to understand the bulk and edge physics of Gaussian states in terms of either the corresponding density matrix or the associated covariance matrix. In section~\ref{sec:topPropertiesBulk}, we construct a symmetry-based topological classification of driven-dissipative systems using the covariance matrix, and identify relevant topological invariants in one and two dimensions. In section~\ref{sec:physicalConstraints}, we then apply this framework to identify the classes D and BDI of Altland and Zirnbauer as the symmetry classes that can be dissipatively targeted under physical constraints related to ``typical'' implementation schemes. As is well known, the edge modes of systems belonging to these two classes are Majorana fermions, which explains the potential of dissipatively induced superfluids to exhibit such modes. We also show that, in two dimensions, the quasi-locality of the dissipative operations acting on the system density matrix alone implies a vanishing Chern number. In section~\ref{sec:edgePhysics}, we discuss the fate of the bulk-edge correspondence in the dissipative setting. We also show how to construct dissipative Majorana modes explicitly in a translation-invariant setting, and examine the robustness of such modes in the presence of typical imperfections. Section~\ref{sec:nonAbelianStatistics} is devoted to the discussion of the physical role of the dissipative gap for adiabatic manipulations---in particular, for the braiding of dissipative Majorana modes---showing that dissipative Majorana modes exhibit non-Abelian exchange statistics just as their Hamiltonian counterparts. The remainder of the paper deals with illustrative examples of our general framework and results: In section~\ref{sec:1D}, we analyze a ``zigzag'' dissipative quantum wire exhibiting topological phase transitions of the three possible types allowed by the closure of the dissipative and/or purity gaps. In section~\ref{sec:2D}, we illustrate in a two-dimensional model a dissipative mechanism that makes it possible to obtain unpaired Majorana modes in a topological phase characterized by an even-valued integer topological invariant.

\section{Dissipative framework}
\label{sec:dissipativeFramework}

\subsection{Quantum master equations for many-body systems}
\label{sec:masterEquations}

The quantum master equation describing the time evolution of the reduced system density matrix $\rho$ is given by 
\begin{equation}
\label{opendyn} 
\dot{\rho} = - \rmi [ H , \rho ] + \sum_i \left( \ell_i \rho   \ell_i^\dagger - \frac{1}{2} \{   \ell_i^\dagger \ell_i , \rho  \}\right).
\end{equation}
The commutator term familiar from the Heisenberg equation describes the coherent dynamics generated by a system Hamiltonian $H$. The second part, often referred to as \emph{Liouville operator} or \emph{Liouvillian}, describes the dissipative dynamics resulting from the interaction of the system with an environment, or ``bath''. In particular, the set of \emph{Lindblad operators} (or ``quantum jump'' operators) $\ell_i$ describe the coupling to that bath. The Liouville operator $\mathcal L$ has a characteristic Lindblad form: The anticommutator term describes dissipation and must be accompanied by fluctuations in order to conserve the ``norm'' $\mathrm{Tr}(\rho)$ of the system density matrix. The corresponding term, where the Lindblad operators act from both sides onto the density matrix, is called ``recycling'' or ``quantum jump'' term. Note that we have absorbed the rates $\kappa_i$ associated with each dissipative process into the definition of the Lindblad operators, making them carry dimension of square root of energy. The rates are non-negative, so that the density matrix evolution is completely positive, i.e., the eigenvalues of $\rho$ remain positive under the combined dynamics generated by $H$ and $\mathcal L$~\cite{PhysRevA.64.062302}.

The quantum master equation provides an accurate description of a system-bath setting with a strong separation of scales. More precisely, there must be a fast energy scale in the bath (as compared to the system-bath coupling) that justifies to integrate out the bath in second-order time-dependent perturbation theory. If, in addition, the bath has a broad bandwidth, the combined Born-Markov and rotating-wave approximations are appropriate, resulting in equation~\eref{opendyn}. Such a situation is generically realized in quantum optical few-level systems, e.g., for a laser-driven atom undergoing spontaneous emission. On the other hand, typical condensed matter systems do not display the necessary scale separations to justify a microscopic description in terms of a master equation. In systems with engineered dissipation~\cite{Mueller} as investigated in this paper, we are interested in scenarios that share features from both quantum optical systems---in that they are coupled to Markovian quantum baths---and condensed matter systems---in that they dispose of a continuum of spatial degrees of freedom on a lattice. Using the manipulation tools of quantum optics, the validity of a Markovian master equation can be ensured, giving rise to a well-defined microscopic dissipative many-body dynamics. A similar level of microscopic control as obtained in Hamiltonian engineering in a cold atom context can be expected for this ``Liouvillian engineering", which therefore is a natural extension of the former to a more general non-equilibrium situation. In this context, dissipation does not occur as a perturbation, but rather as the dominant resource of the many-body dynamics. In particular, here we will consider the case where the Hamiltonian is absent $H = 0$. Such a scenario can be useful from a practical point of view---for cooling systems into desired states---but also gives rise to interesting new many-body physics.

\subsection{Dark states}
\label{sec:darkStates}

In the long-time limit, a quantum system governed by equation~\eref{opendyn} approaches a \emph{stationary} or \emph{steady state} $\rho_{\text{f}} = \rho(t\to \infty )$ which generically corresponds to a mixed state. An interesting situation appears if, instead, the many-body density matrix is driven towards a pure stationary state, $\rho_{\text{f}} = |\psi\rangle _D \langle \psi |_D$~\cite{Diehl08a,Verstraete09}. In quantum optics, such pure states $|\psi\rangle _D$ that are obtained as a result of a dissipative evolution are called \emph{dark states}. Mathematically, such dark states are zero modes of the Liouville operator. More precisely, a dark state is a zero mode shared by all Lindblad operators,
\begin{eqnarray}\label{eq:darkState}
\ell_{i} |\psi\rangle_D = 0 \text{ for all } i.
\end{eqnarray}
The dark-state solution is the unique solution of the Liouville dynamics if (i) there exists exactly one normalized dark state $ |\psi\rangle_D$, and (ii) there are no stationary solutions other than this dark state~\cite{Diehl08b,Verstraete09}. In the specific case of interacting Liouville operators (higher than quadratic in the creation and annihilation operators)~\cite{Diehl08b,Verstraete09} or non-interacting Liouville operators (quadratic in the creation and annihilation operators), the fact that these conditions are satisfied can be proved rigorously. If present, the dynamics described by equation~\eref{opendyn} for $H = 0$ corresponds to a directed motion---in the Hilbert space of the system---into the ``sink'' provided by the dark state, which is reached independently of the initial density matrix. In recent years, a number of theoretical~\cite{Diehl08,weimer-nphys-6-382} and experimental~\cite{barreiro-nature-470-486,Krauter} studies have focused on how to construct Liouville operators such that, in the long-time limit, a quantum system reaches a well-defined, pure many-body steady state or exhibits novel phase transitions resulting from the competition between coherent and dissipative dynamics~\cite{Diehl10a,Eisert,Hoening12,PhysRevA.87.012108}. In particular, in the context of atomic fermions, it has been shown how to engineer number-conserving dissipative dynamics that drives the system into a pure BCS-type paired state in the absence of conservative forces~\cite{Diehl10b,Yi}. The dissipative pairing mechanism forms a basis for the targeted cooling into states with non-trivial topological properties far from thermodynamic equilibrium.

\begin{figure}[t]
    \begin{center}
        \includegraphics[width=\columnwidth]{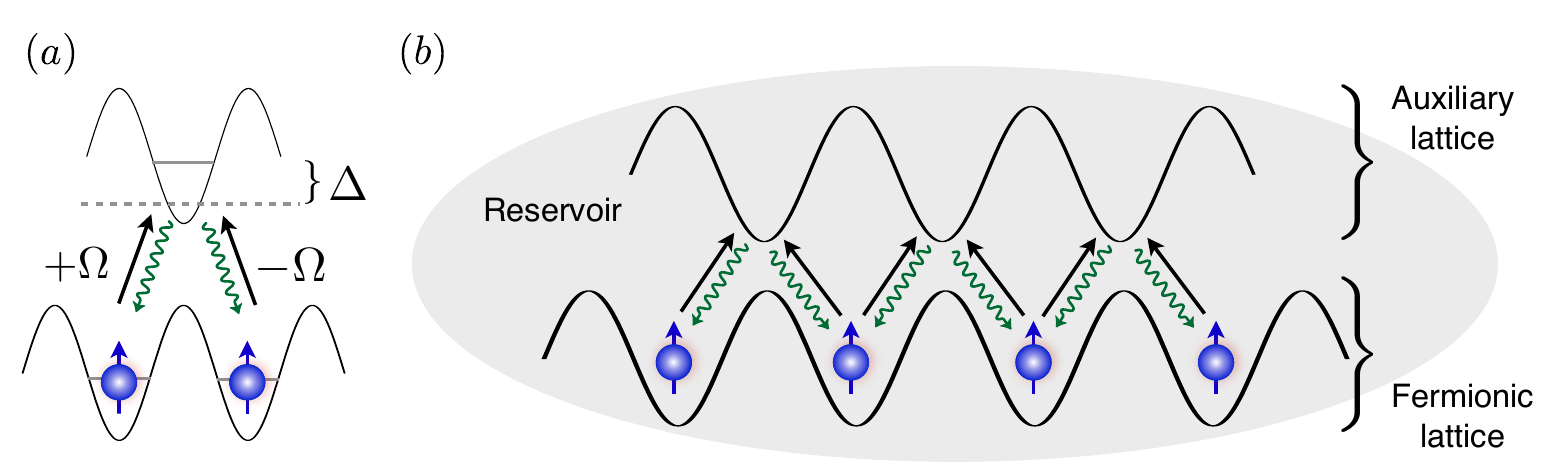}
        \caption{(a) Unit cell for dissipation engineering. The lower potential wells correspond to the physical sites, whereas the upper site in between is an auxiliary site. Atoms in the lower sites are coherently coupled to the auxiliary site with opposite Rabi frequency $\pm\Omega$. Decay back to the lower sites occurs via spontaneous emission, where energy is released into a surrounding reservoir (see text). If the coupling to the upper level is sufficiently far detuned $(\Delta \gg \Omega)$, the latter can be integrated out, so that an effective dynamics for the lower sites is obtained. (b) In an optical lattice setup, this unit cell is repeated in a translation-invariant fashion multiple times in a natural way. The quantum wire corresponds to the lower sites---as anticipated above---which are populated by spinless (or spin-polarized) fermions in the cases discussed in this work. Rabi frequencies with alternating signs are realized by a commensurability condition on lattice and drive lasers (see text). Dissipation results from spontaneous phonon emission into a BEC reservoir (light grey).}
        \label{fig:dissipationEngineering}
    \end{center}
\end{figure}

\subsection{Physical realization}
\label{sec:physicalRealization}

Here we briefly sketch the implementation idea common to the dissipative models studied in this paper. The basic setting consists of an atomic ensemble confined in an optical lattice (the system), which is driven coherently and immersed into a larger reservoir consisting of a different atomic species and playing the role of the dissipative bath. In the case of interest here, the constituents of the system are fermions. In cold atomic gases, the associated spin is realized in terms of hyperfine states, and thus both the cases of spinless and spinful fermions can be meaningfully considered. The bath constituents are chosen as bosonic atoms, so that the conservation of fermion parity in the system is guaranteed. 

The working of the driven-dissipative mechanism is best illustrated by the unit cell $\Lambda$-configuration displayed in figure~\ref{fig:dissipationEngineering}. The complete driven-dissipative process consists of two steps: The first step is a coherent excitation from the system (lower sites) to the auxiliary site in between. In the example of figure~\ref{fig:dissipationEngineering}, we quasi-locally excite fermions on the system sites $i$ and $i+1$ (with annihilation and creation operators $a_i$ and $a^\dagger_i$ corresponding to each site) into an antisymmetric superposition $\sim a_i - a_{i+1}$, which can be controlled by a commensurability condition of the driving laser to the standing wave laser generating the optical superlattice (see reference~\cite{Mueller} for details): If the driving laser has twice the wavelength of the lattice laser, there is phase shift of $\pi$ in the effective Rabi frequency from one site to the next, leading to  a relative minus sign. The auxiliary level is unstable if coupled to the reservoir. In this case, spontaneous phonon emission into the surrounding bath can occur, thereby giving rise to the second, dissipative step. The atoms are ``brought back'' to the lower sites in a quasi-local way $\sim a_i^\dag + a_{i+1}^\dag$; since this process is isotropic, there is now a relative plus sign. If this driven-dissipative process is generated using a drive laser that is far detuned ($\Omega/\Delta \ll1$) from the auxiliary site resonance frequency, the auxiliary site can be integrated out and we obtain a Lindblad operator of the form
\begin{eqnarray}\label{eq:ell}
\ell_i = C_i^\dag A_i \text{ for all } i  ,  \quad C_i^\dag = a_i^\dag + a_{i+1}^\dag, \quad A_i = a_i - a_{i+1} . 
\end{eqnarray}
A few remarks are in order: (i) For a driving laser superimposed over the extent of the whole optical superlattice, we obtain \emph{translation-invariant} Lindblad operators as depicted in figure~\ref{fig:dissipationEngineering}(b), up to system boundaries which are not shown. (ii) The \emph{quasi-locality} of the operators is controlled by the Wannier function overlap between the onsite wave functions involved in the combined excitation and de-excitation processes. (iii) The Lindblad operators that can be realized in this setting have a generic form $\ell_i = C_i^\dag A_i$, where $C_i^\dag (A_i)$ is a linear translation-invariant superposition of creation (annihilation) operators with generic properties: The excitation part ($A_i$) can be controlled to high accuracy---involving, in particular, the control of the relative phases in the superposition---since it is generated by a coherent laser beam. In two dimensions, for example, this allows to imprint angular momentum by choosing relative laser phases in different primitive directions of the lattice. On the other hand, the de-excitation or decay part ($C_i^\dag$) results from spontaneous emission and is therefore unavoidably isotropic (or $s$-wave symmetric). (iv) The \emph{system particle number is conserved} in our dissipative framework. This is reflected in the property $[\ell_i , \hat N] =0$ for all Lindblad operators, where $\hat N = \sum_i \hat n_i$ the total particle number operator, $\hat n_i= a_i^\dag a_i$. This exact microscopic property of the system, which implies an exact \emph{conservation of parity}, is of importance to the discussion of the possible imperfections that may occur in the dissipative setting after performing approximations. Physically, this property originates from the fact that typical interactions in cold atomic systems are local density-density interactions. In particular, the system and bath constituents will interact via such coupling.  On an even more elementary level, the fact that the bath is bosonic provides a further reason for fermion parity conservation. This aspect is crucially different from solid-state implementations which are not perfectly closed systems: There the environment is typically fermionic, which facilitates system-bath exchange processes affecting the parity of the system. (v) While no particle number exchange is possible between the system and the reservoir, energy can be exchanged. This enables the outflow of entropy from the system into the (infinite) reservoir, and the targeted cooling into pure many-body states. A crucial prerequisite for the entropy removal from the system is the coherent driving of the system. (vi) The fast energy scale ensuring the validity of the Born-Markov and rotating-wave approximations underlying our construction is provided by the band separation between the system and the auxiliary sites, which is the largest energy scale in the problem. (vii) The creation and annihilation part of the Lindblad operators is respectively symmetric and antisymmetric under the exchange $i\leftrightarrow i+1$. Such property is important for reaching pure stationary states, as will become clear below.

\subsection{From interacting Liouvillians to quadratic master equations}
\label{sec:interactingLiouvillians}

The physical implementation discussed above realizes a number-conserving microscopic dynamics, with the key advantage of conserving fermion parity as a consequence. The dynamics generated by the corresponding bilinear Lindblad operators $\ell_i$ is described by an interacting (quartic) Liouvillian. The Lindblad operators are constructed in such a way that the stationary state is a unique dark state for a given even particle number $2N$, characterized by a BCS pair wave function with fixed particle number $|\text{BCS, N}\rangle$ that satisfies $\ell_i |\text{BCS, N}\rangle = 0$ for all $i$. The Liouville operator ensuring this property thus represents a \emph{parent Liouville operator} for a given fixed number BCS pair wave function (see~\ref{app:meanField} for more details). Starting from the exact knowledge of the fixed-number dark-state wave function, we can switch in the thermodynamic limit from a fixed-number to a fixed-phase ensemble. In particular, the long-time evolution of the interacting master equation can be linearized based on this knowledge. The calculation presented in~\ref{app:meanField} can be summarized as 
\begin{eqnarray}
\ell_ i = C_i^\dag A_i \stackrel{N,t\to \infty}{\longrightarrow} L_i = C_i^\dag + \alpha A_i. 
\end{eqnarray} 
That is, the product of creation and annihilation parts in the quadratic Lindblad operators transforms into a sum, giving rise to linear Lindblad operators. This relies on the property that $C_i^\dag (A_i)$ is symmetric (antisymmetric). It provides a dynamical connection between the fixed-number and fixed-phase settings at the level of the equation of motion.  The long-time dynamics is universal, in the sense that it is independent of the initial state.

We note that $\alpha = r e^{\mathrm i \theta}$ is a complex number in the above equation. Its phase is not fixed by the dynamics, but rather reflects spontaneous symmetry breaking in the dissipative setting of interest. The modulus $r$, on the other hand, is determined by the average particle number in the system (see~\ref{app:meanField}). In particular, for half filling and in the example of equation~\eref{eq:ell}, we find $r=1$; such that for $\theta = 0$, without loss of generality,
\begin{eqnarray}
L_i = (a_i^\dag + a_{i+1}^\dag) + (a_i - a_{i+1}).
\end{eqnarray}
We recognize the quasi-local Bogoliubov quasiparticle operators of Kitaev's Hamiltonian quantum wire \cite{Kitaev01} (at half filling and with equal pairing and hopping amplitudes, up to normalization), emerging here naturally in the long-time limit of a dissipative dynamics. The ground-state condition of the Hamiltonian quantum wire, $L_i |\text{BCS, } \theta\rangle = 0$ for all $i$, is now interpreted as the dark-state condition of the linearized dissipative evolution. The corresponding wave function now has a fixed phase $\theta$ instead of a fixed number. Since the Lindblad operators form a complete Dirac algebra, $\{ L_i , L_j^\dag \} \sim \delta_{ij}, \{ L_i , L_j \}=  \{ L^\dag_i , L_j^\dag \} = 0$ for an infinite system with no boundaries, the uniqueness of the dark-state solution is obvious.

The quadratic dynamics obtained in the long-time limit makes the systems considered in this work amenable to a treatment analogous to the discussion of quadratic Hamiltonians when examining their topological properties. This dynamics was obtained by giving up exact particle number conservation, which is justified in the thermodynamic limit. The absence of exact particle number conservation thus emerges similarly as in the Hamiltonian scenario. There is, on a formal level, however, a crucial difference between dissipative and Hamiltonian settings. While a quadratic number non-conserving BCS Hamiltonian still conserves parity, formally such a property is not present in a quadratic Liouville evolution (for example, single fermions can be ejected into the environment, giving rise to quasiparticle poisoning \cite{PhysRevB.85.121405,PhysRevB.85.174533}). However, remembering that the microscopic dissipative processes do conserve particle number and thus parity exactly, we can rule out parity non-conserving processes as potential imperfections arising in our scenario. The number non-conserving nature of the system is introduced ``within the system only'', and there is no exchange of particles with the reservoir. Physically, the number non-conserving processes describe pairwise creation and annihilation out of or into the mean field provided by the system itself.

\subsection{Gaussian master equations}
\label{sec:gaussianMasterEquations}

Having discussed how quadratic fermionic master equations naturally emerge in the long-time limit of interacting Liouville operators, we now summarize some general properties of such master equations. We do this in both a second- and a first-quantized formulation, working with operators or matrices, respectively, as familiar from the Hamiltonian setting~\cite{Altland97}. For this discussion, it is useful to work in the real (or Majorana) basis of fermionic quadrature component operators. For a system with $N$ sites, $2N$ real fermionic modes are introduced according to
\begin{equation}
c_{2n-1} =\rmi \left( a_n - a_n^{\dagger} \right),  ~ \, ~ c_{2n} = a_n + a_n^{\dagger}, ~ \, ~ \{ c_n , c_m \} = 2 \delta_{n,m}, ~ \, ~ c^{\dagger}_n = c_n. 
\end{equation}
The fact that the master equation is quadratic in the fermion operators implies solutions in terms of Gaussian density operators. In the second-quantized formulation, this can be written as
\begin{eqnarray}\label{eq:densop}
\rho(t) \sim \exp{\left( - \frac{\text{i}}{4} \mathbf c^{T} G(t) \mathbf c \right) },
\end{eqnarray}
where $\mathbf c^T = (c_1, ... , c_{2N})$ is a column vector defined from the $2N$ Majorana operators and $G$ is a real antisymmetric matrix (so that $\mathrm i G$ is Hermitian). Formally, $\rho$ thus has the form of a canonical density matrix $\rho_c \sim \rme^{-\beta H}$ for a quadratic Hamiltonian.

Instead of working in second quantization, we can move to a first-quantized formulation. As in the Hamiltonian scenario, the latter allows us to discuss symmetry and topological classifications in a more direct way. The key object here is the covariance or (equal-time) correlation matrix collecting the second moments, which is the only information contained in a Gaussian density operator. It is defined as
\begin{eqnarray}\label{eq:gammatr}
\Gamma_{nm} =  \mathrm{Tr} (\rho  \, \hat \Gamma_{nm} ), \quad \hat \Gamma_{nm} = \frac{\rmi}{2} [ c_n , c_m ].
\end{eqnarray}
We now look for an equation of motion for this object \cite{Prosen10,Eisert}. A straightforward way to derive such an equation is via the adjoint equation to a given master equation for the density operator \cite{breuer}, describing the evolution of an operator in the Heisenberg picture. For the correlation operator $\hat \Gamma_{nm}  = \frac{\rmi}{2} \sum_{j,k=1}^{2N} c_j \mathcal{G}^{nm}_{jk} c_k$ with real antisymmetric $\mathcal{G}^{nm}_{jk}= \left( \delta_{j,n} \delta_{m,k} - \delta_{j,m} \delta_{n,k}  \right) $, this reads
\begin{equation}\label{eq:hatgamma}
\partial_t \hat{\Gamma}_{nm} =   \sum_i L^{\dagger}_i \hat{\Gamma}_{nm} L_i - \frac{1}{2} \{ L^{\dagger}_i L_i , \hat{\Gamma}_{nm} \} = -\rmi\sum_{j,k=1}^{2N} c_j \left(  \{ M , \mathcal{G}^{nm} \} \right)_{jk} c_k .
\end{equation}
Here we have written the linear quantum jump operators as $ L_i \equiv \sum_{k=1}^{2N} l^{i}_k c_k ,  L^{\dagger}_i \equiv \sum_{k=1}^{2N} l^{i*}_k c_k$ and introduced the matrix
\begin{equation} \label{eq:bathMatrix}
M_{jk} \equiv \sum_{i=1}^{N}  l^{i*}_j l^{i}_k   = \frac{1}{2} \,\big(X_{jk} - \frac{\mathrm{i}}{2} Y_{jk}\big) ,
\end{equation}
where $X = X^T$ and $Y = - Y^T$ are real symmetric and antisymmetric matrices, respectively. Furthermore, $X$ by construction is positive semidefinite. Taking the expectation value of equation (\ref{eq:hatgamma}), we readily find the fluctuation-dissipation equation describing the evolution of the real antisymmetric correlation matrix $\Gamma$,
\begin{equation}
\partial_t  \Gamma =  - \{ X ,  \Gamma  \} +   Y,
\end{equation}
where we have suppressed the matrix indices. Denoting the steady-state correlation matrix as $\tilde{  \Gamma  }$, which satisfies the equation $\{ X , \tilde{  \Gamma  } \} =  Y$, we can give a clear physical meaning to the matrices $X$ and $Y$. Writing $ \Gamma =  \tilde{ \Gamma }+\delta\Gamma$, the approach to the steady state is governed by $\partial_{t} \delta\Gamma=  -\{X ,\delta\Gamma\}$; i.e., $X$ alone governs the damping dynamics towards that steady state. The matrix $Y$ describes fluctuations, which come along with dissipation in a probability preserving  ($\partial_t \mathrm{Tr} (\rho (t)) =0$) quantum mechanical evolution. The steady state $\tilde \Gamma$ depends on both $X$ and $Y$. 

Finally, we remark that the correlation matrix is related to the density operator equation (\ref{eq:densop}) by 
\begin{equation}
\Gamma_{nm} (t)=\frac{\rmi}{2} \text{Tr} \left( \rho (t) [ c_n , c_m ]\right) = \text{i} \tanh{\left[\text{i}\frac{G(t)}{2} \right]}_{nm}.
\end{equation}
We may compare this to a Gaussian Hamiltonian equilibrium setting: Introducing the first-quantized, real and antisymmetric Hamiltonian matrix $h$ in the Majorana basis via $H = \frac{\mathrm{i}}{4}\sum_{i, j} h_{ij} c_i c_j$, we have at an arbitrary temperature $T= 1/\beta$
\begin{equation} \label{eq:gammah}
\Gamma^{(\text{eq})}_{nm}  = \rmi \tanh{\left[\text{i}\frac{\beta h}{2} \right]}_{nm},
\end{equation}
which reduces to $\Gamma^{(\text{eq})} = \rmi \,\mathrm{sgn}(\rmi h)$ at $T=0$. 

\subsection{Purity and purity gap}
\label{sec:purityGap}

The purity of a Gaussian state defined by a particular correlation matrix $\Gamma$ can be revealed by examining the spectrum of the Hermitian positive semidefinite matrix $(\rmi \Gamma)^2$, which we refer to as the \emph{purity spectrum}. Pure Gaussian states are characterized by a ``flat'' purity spectrum with eigenvalues all equal to $1$, whereas mixed Gaussian states exhibit eigenvalues smaller than $1$, each zero eigenvalue indicating the existence of a completely mixed subspace. Intuition regarding the purity spectrum can be gained by constructing a \emph{fictitious} quadratic Hamiltonian $H_\Gamma$ from the Hermitian matrix $\rmi \Gamma$, namely,
\begin{eqnarray} \label{eq:HGamma}
    H_\Gamma = \rmi \sum_{i, j} \Gamma_{ij} c_i c_j,
\end{eqnarray}
where the $c_i$ are the $2N$ Majorana basis operators introduced in the previous section. Since $\Gamma$ is a real antisymmetric matrix, the spectrum of this Hamiltonian consists of real eigenvalues $\pm \epsilon_n$ ($n = 1, 2, \ldots, N$). Most importantly, the positive part of the spectrum of $H_\Gamma$ is the purity spectrum of the Gaussian state represented by the correlation matrix $\Gamma$ (up to a square root). Exploiting this analogy further, we will  identify the eigenvectors of $(\rmi \Gamma)^2$ as ``eigenmodes'' or ``modes'' (of the fictitious Hamiltonian $H_\Gamma$). In particular, we will refer to modes of $H_\Gamma$ associated with zero eigenvalues as \emph{zero-purity modes} and to the spectral gap of the latter as the \emph{purity gap}. Such modes are defined in the \emph{mode space} $\mathcal{M}$ consisting of operators of the form $\gamma = \vect{v}^T \vect{c}$ with $\vect{v} \in \mathbb{R}^{2N}$. Modes corresponding to a unit vector $\vect{v}$ will be referred to as ``Majorana'' modes since they satisfy $\gamma^\dagger = \gamma$ and $\gamma^2 = 1$. Moreover, we will distinguish two types of zero-purity modes: \emph{intrinsic} ones, which are determined by the dissipative dynamics, and \emph{extrinsic} ones, which result from mixed initial conditions (and thus disappear when starting from pure initial states). From a topological perspective, Majorana zero-purity modes that have a topological origin will be referred to as \emph{genuine}, as opposed to \emph{spurious} ones.

The purity of the steady state is determined by the dissipative dynamics and, if the steady state is not unique, by the purity of the initial state (i.e., by the initial conditions). In the case of interest in this work where the dissipative dynamics is quadratic, one can show (we refer to our previous work~\cite{Bardyn12} for an explicit proof) that there exist initial conditions leading to a pure steady state whenever the corresponding Lindblad operators $L_i$ form a set of anticommuting operators, i.e., whenever $\{ L_i, L_j \} = 0$ for all $i, j$~\footnote{Note that Lindblad operators satisfying this condition generate the same (exterior) algebra as fermionic annihilation operators. They need not \emph{be} fermionic annihilation operators, however. The anticommutation relation $\{ L_i, L_j \} = 0$ (for all $i, j$) is a necessary and \emph{sufficient} condition to ensure the existence of a pure state $\ket{\psi}$ such that $L_i \ket{\psi} = 0$ for all $i$, which is all that we need.}. In the matrix representation defined in the previous section, one can then establish a one-to-one correspondence between the matrices $X$ and $Y$ encoding the dynamics. Intuitively, this can be understood by examining the steady-state equation $\{ X, \Gamma \} = Y$ (where $\Gamma$ now denotes the steady-state correlation matrix): if the steady state is pure, the spectrum of the associated correlation matrix $\Gamma$ (i.e., the purity spectrum) essentially contains no information, since all of its eigenvalues are equal to $1$. The information contained in $X$ must therefore be exactly the same as that contained in $Y$, otherwise the steady-state equation would not be satisfied. In other words, $X$ and $Y$ both contain \emph{full} information about the dissipative dynamics when the steady state is pure. In that case, one can construct yet another useful object encoding all information about the dynamics: the so-called \emph{parent Hamiltonian} $H_\text{parent}$ naturally associated with the Hermitian matrix $\rmi Y$, defined as
\begin{eqnarray} \label{eq:Hparent}
    H_\text{parent} \equiv H_Y = \rmi \sum_{i, j} Y_{ij} c_i c_j.
\end{eqnarray}
Clearly, the spectrum of $H_\text{parent}$ is directly related to that of $Y$ and therefore to that of $X$ as well for dissipative dynamics whose steady state is pure. Remembering the definition of the matrix $Y$ in terms of the Lindblad operators, one can rewrite the parent Hamiltonian in the equivalent form
\begin{eqnarray} \label{eq:HparentL}
    H_\text{parent} = \sum_i L^\dagger_i L_i.
\end{eqnarray}
This shows that pure steady states $\ket{\psi}$, which are ``dark states'' satisfying $L_i \ket{\psi} = 0$ for all $i$ (see equation~\eref{eq:darkState}), can equivalently be seen as ground states of $H_\text{parent}$. As opposed to the purely fictitious Hamiltonian $H_\Gamma$ that we have constructed above to quantitatively assess the purity of an arbitrary Gaussian state, independently of any dynamics, the parent Hamiltonian therefore describes features associated with the actual (dissipative) dynamics of the system---as expected from its definition from the matrix $Y$. In fact, we will argue in the next section that the spectrum of $H_\text{parent}$ encodes all information regarding pure steady states. We emphasize, however, that the parent Hamiltonian does not play such a prominent role in the more general case where the steady state of the dissipative dynamics is mixed (even when starting from pure initial states).

As demonstrated in our previous work~\cite{Bardyn12}, the above discussion can be formalized and summarized as the following equivalent statements:
\begin{enumerate}[\hspace{0.5cm}(i)]
    \item The steady state is pure (at least for pure initial states);
    \item $\{ L_i, L_j \} = 0$ for all $i,j$ (i.e., the Lindblad operators form a set of anticommuting operators);
    \item $[X, Y] = 0 \text{ and } X^2 = -\frac{1}{4} Y^2$ (in particular, the spectra of $X$ and $Y$ are directly related);
    \item The dissipative dynamics can be fully described using the parent Hamiltonian $H_\text{parent} = \sum_i L^\dagger_i L_i$.
\end{enumerate}
This last point will be clarified in the next section.

\subsection{Dissipative gap and Majorana zero-damping modes}
\label{sec:dissipativeGap}

In the case where the dissipative dynamics is quadratic, the dynamical approach to the steady state is governed by the associated matrix $X$, as mentioned in the previous section. This matrix, which is by construction real, symmetric and positive semidefinite, can be spectrally decomposed in the form $X = \sum_{j = 1}^{2N} \kappa_j (\vect{v}_j \otimes \vect{v}^T_j)$ with eigenvalues $\kappa_j \geq 0$ and associated eigenvectors $\vect{v}_j \in \mathbb{R}^{2N}$. The eigenvectors of $X$ define ``modes'' in the mode space defined in the previous section. Assuming that they are normalized to unity, each eigenvector $\vect{v}_j$ can be identified with a corresponding Majorana mode $\gamma_j = \vect{v}^T_j \vect{c}$. Physically, the eigenvalues of $X$ then correspond to particular \emph{damping rates} associated with particular Majorana modes. Accordingly, we will refer to the spectrum of $X$ as the \emph{damping spectrum} and to Majorana modes $\gamma_j$ associated with a vanishing damping rate as \emph{Majorana zero-damping modes}. One can show (for an explicit proof, see our previous work~\cite{Bardyn12}) that a Majorana mode $\gamma = \vect{v}^T \vect{c}$ is a zero-damping mode whenever the following equivalent conditions are satisfied:
\begin{enumerate}[\hspace{0.5cm}(i)]
    \item $X \vect{v}$ = 0;
    \item $\{ L_i, \gamma \} = 0$ for all $i$;
    \item $\vect{l}^T_i \vect{v} = 0$ for all $i$ ($\vect{l}_i$ being the vector corresponding to the Lindblad operator $L_i$ in mode space, i.e., $L_i = \vect{l}^T_i \vect{c}$).
\end{enumerate}
We remark, however, that Majorana zero-damping modes satisfying the above conditions need not be spatially localized. The topological nature of the system will play a crucial role in ensuring such localization, as we will demonstrate in section \ref{sec:generalFormMZMs} below. In general, we will distinguish \emph{genuine} Majorana zero-damping modes that have a topological origin from \emph{spurious} ones which a mere artefacts of the dissipative dynamics. A simple example of spurious modes is provided by a lattice site on which no dissipative dynamics takes place. This gives rise to two Majorana zero-damping modes decoupled from the dynamics, which obviously do not have a topological origin.

The damping spectrum describes the \emph{dynamical} separation (i.e., in time) between particular modes in a similar way as the spectrum of a Hamiltonian determines the \emph{energy} separation between specific modes. Pushing the analogy further, one can see that the existence of a \emph{dissipative gap} (or ``damping gap'') in the damping spectrum leads to the \emph{dynamical} isolation of bulk and edge modes (through the quantum Zeno effect~\cite{Beige00}), thereby providing a dissipative counterpart to the gap protection arising in the Hamiltonian context. Majorana zero-damping modes form a so-called \emph{decoherence-free subspace}~\cite{Lidar98} unaffected by the dissipative dynamics and, in the presence of a finite dissipative gap, completely isolated from the rest of the system. The dissipative counterpart of a topological Hamiltonian ground-state degeneracy is then  provided by the existence of a non-local decoherence-free subspace associated with zero-damping Majorana modes.

We finally clarify the role of the parent Hamiltonian defined in the previous section in light of the considerations above. When the steady state of the dissipative dynamics is pure, the spectrum of $X$ (i.e., the damping spectrum) directly maps to the spectrum of $Y$, which in turn trivially maps to that of $H_\text{parent}$. The parent Hamiltonian thus contains all information about the dissipative dynamics in that case. When the steady state is mixed (independently of the initial state), however, the tight relationship between $X$ and $Y$ (or $H_\text{parent}$) breaks down and the parent Hamiltonian becomes insufficient to describe the dynamics. In this more general case, one can show (we refer to our previous work~\cite{Bardyn12}) that a zero mode of $H_{\text{parent}}$ (or, equivalently, of the matrix $Y$) does not necessarily correspond to a zero mode of $X$ (i.e., to a zero-damping mode), although the converse is always true. In other words, zero modes of the parent Hamiltonian need not be Majorana zero-damping modes of the dissipative dynamics. In fact, any zero mode of $H_{\text{parent}}$ which does not coincide with a zero mode of $X$ gives rise, in steady state, to an intrinsic zero-purity mode. This crucial phenomenology will be key to understanding the non-equilibrium topological effects that will be exemplified below.

To conclude this section, we remark that Majorana zero-damping modes do not benefit from the protection mechanism featured by $H_{\text{parent}}$ due to the antisymmetry of the matrix $Y$. While the antisymmetry of $Y$ forces $H_\text{parent}$ to have an \emph{even} number of Majorana zero modes, such that spatially isolated modes cannot be affected by local perturbations, the fact that $X$ is symmetric does not lead to such restriction. Although this is potentially harmful for Majorana zero-damping modes, we will demonstrate in the sections below that this can also lead to intriguing physics with no Hamiltonian ground state counterpart.

\section{Topological properties of the bulk}
\label{sec:topPropertiesBulk}

In this section, we focus on the topological properties of the bulk of driven-dissipative fermionic systems with Gaussian steady states. In particular, we identify the correlation matrix, which fully describes such states, as a fictitious first-quantized Hamiltonian and use the latter to construct a topological classification in complete analogy to the conventional Hamiltonian scenario.

The topological classification of gapped states of non-interacting fermions can be achieved on the basis of symmetry properties of the corresponding Hamiltonian under time-reversal, charge-conjugation (or particle-hole) and chiral (or sublattice) transformations, as was proposed by Schnyder \etal~\cite{Schnyder08} following the classification of random matrices developed by Altland and Zirnbauer~\cite{Altland97}. Ten symmetry classes were proved to be sufficient for an exhaustive classification of topological phases in any spatial dimension, and an alternative approach was later introduced by Kitaev in the powerful framework of topological K-theory~\cite{Kitaev09,Stone11,Abramovici12,Wen12} (see reference~\cite{Freedman11}, e.g., for a self-contained review). We argue below that all classification schemes developed in the Hamiltonian setting can be automatically applied to classify the Gaussian steady states of a dissipative dynamics. We do not provide an exhaustive classification, however, since this can be done straightforwardly based on the references cited above. Instead, we construct an explicit topological classification for two symmetry classes of particular interest for this work, namely, for dissipative systems belonging to the symmetry classes D and BDI.

\subsection{Steady-state symmetries}
\label{sec:steadyStateSymmetries}

We first study how symmetries of the Lindblad operators translate into symmetries of the correlation matrix. To this end, we consider a Gaussian dissipative dynamics with unique steady state, i.e., the corresponding correlation matrix $\Gamma$ is a unique solution of 
\begin{eqnarray}
    \{ X, \Gamma \} = Y,
\end{eqnarray}
with matrices $X$ and $Y$ defined as in section~\ref{sec:gaussianMasterEquations} (see equation~\eref{eq:bathMatrix}, in particular). We assume that the Lindblad operators $L_{i}$ are invariant, up to a phase factor, under some symmetry group $G$:
\begin{eqnarray}
    g^{-1} L_i \, g = \rme^{\rmi \phi_i} L_i,
\end{eqnarray}
where $g \in G$. In order to preserve the linearity of the Lindblad operators in the fermionic operators, $G$ must act linearly on the $2N$ Majorana operators $c_j$ introduced in section~\ref{sec:gaussianMasterEquations} above. These operators form an orthonormal basis of the mode space $\mathcal{M} \cong \mathbb{R}^{2N}$ of operators $A = \vect{a}^T \vect{c}$ with $\vect{a} \in \mathbb{R}^{2N}$ with respect to the inner product $\avg{A, B} = (1/2) \{ A, B \}$~\cite{Fidkowski11}. Clearly, any symmetry $g \in G$ (unitary or antiunitary) must act linearly on $\mathcal{M}$ and transform the Majorana basis defined by the operators $c_j$ into another Majorana basis. In mode space, a symmetry $g \in G$ must therefore be represented by a real orthogonal matrix $S_g \in O(2N)$,
\begin{eqnarray}\label{eq:trafoc}
  g^{-1} \mathbf{c} \, g = S_g \mathbf{c}.
\end{eqnarray}
Note that this formula allows to analyze the symmetry properties of the state also in the more general case where the dynamics is governed by both a Liouvillian and a Hamiltonian, see equation \eref{eq:steadyStateSymmetries} below. 

The Lindblad operators are defined in the \emph{Nambu space} $\mathcal{N} \cong \mathbb{C}^{2N}$ of operators $L = \vect{l}^T \vect{c}$ with $\vect{l} \in \mathbb{C}^{2N}$, which can be viewed as a complexification of $\mathcal{M}$. In Nambu space, the relevant representation of $g \in G$ is given by $S_g$ if $g$ is unitary and $S_g K$ ($ = K S_g$) if $g$ is antiunitary. Here $K$ is the complex conjugation operator defined such that $K \rmi = -\rmi K$. The Lindblad operators $L_i = \vect{l}_i^T\vect{c}$ are therefore invariant (up to a phase factor) under the symmetry $g \in G$ if and only if
\begin{eqnarray}
    \vect{l}_i = \rme^{\rmi \phi_i} S_g \vect{l}_i && \qquad \text{if $g$ is unitary}, \\
    \vect{l}_i = \rme^{\rmi \phi_i} S_g K \vect{l}_i && \qquad \text{if $g$ is antiunitary},
\end{eqnarray}
where $\phi_i \in [0, 2\pi)$ (note that the phase factors $\rme^{\rmi \phi_i}$ do not affect the form of the Liouvillian). Using equation~\eref{eq:bathMatrix}, we then find that the matrices $X$ and $Y$ encoding the dissipative dynamics have the properties
\begin{eqnarray}\label{eq:xytr}
    X = S_g X S^T_g, \qquad Y = \pm S_g Y S^T_g,
\end{eqnarray}
where the positive and negative signs corresponds to the cases where $g$ is unitary or antiunitary, respectively. The steady-state equation can then be written as
\begin{eqnarray}
    \{ X, \pm S^T_g \Gamma S_g \} = Y
\end{eqnarray}
and, since we have assumed that the steady state is unique, we obtain
\begin{eqnarray} \label{eq:steadyStateSymmetries}
    \Gamma = \pm S^T_g \Gamma S_g.
\end{eqnarray}
This shows that the symmetries of the Lindblad operators are naturally reflected in symmetries of the steady-state correlation matrix $\Gamma$~\footnote{Clearly, using equations~\eref{eq:gammatr} and~\eref{eq:trafoc}, the transformation of the correlation matrix is $\Gamma \to \pm S_g^T \Gamma S_g$ for unitary or antiunitary symmetries, respectively, irrespective of the dynamics---which in particular may involve both Hamiltonian and Liouvillian parts. In contrast, here we focus on how the invariance of the Lindblad operators (under some symmetry) translates as that of the correlation matrix.}. In turn, this implies that the matrix $\Gamma$ can be used to construct a symmetry-based topological classification of Gaussian steady states. Of particular importance for this purpose are the two discrete symmetries corresponding to particle-hole (PHS) and time-reversal (TRS) symmetry, respectively. The former corresponds to the ``+'' sign in equation~\eref{eq:steadyStateSymmetries} and is trivially satisfied as $S_g = 1$ in our case, as we argue below, while the latter corresponds to the ``-'' sign and depends on the specific form of the Lindblad operators.

We remark that chiral symmetry (defined as the combination of PHS and TRS~\cite{Ryu10}) is automatically satisfied whenever TRS is present, since the system always has PHS, by construction. In that case, there exists matrices $S_C$ and $S_T$ corresponding to PHS and TRS, respectively, such that equation~\eref{eq:steadyStateSymmetries} is satisfied (with a ``+'' sign for $S_C$ and a ``-'' sign for $S_T$), and one can easily verify that the combination of PHS and TRS (i.e., chiral symmetry) leads to $\{ S_C S^T_T, \Gamma \} = \{ S_T S^T_C, \Gamma \} = 0$.

\subsection{Topological classification and topological invariants}
\label{sec:topClassification}

Let us consider an arbitrary Gaussian steady state $\rho$, fully characterized by its correlation matrix $\Gamma_{ij} = (\rmi/2) \Tr( \rho [c_i, c_j])$. Using that the matrix $\Gamma$ is real and antisymmetric, we construct a corresponding \emph{fictitious} free-fermion Hamiltonian as in section~\ref{sec:purityGap} where the purity spectrum was defined (not to be confused with the parent Hamiltonian introduced in the same section),
\begin{eqnarray} \label{eq:mappingGammaHam}
    H_\Gamma = \rmi \sum_{i, j} \Gamma_{ij} c_i c_j,
\end{eqnarray}
thereby establishing a one-to-one correspondence between the set of Gaussian steady states and the set of free-fermion Hamiltonians. It is clear that the symmetries of the dissipative system---embedded in $\Gamma$---are the same as that of the Hamiltonian system defined by $H_\Gamma$, since $\Gamma$ can be viewed as the ``first-quantized'' Hamiltonian corresponding to the ``second-quantized'' Hamiltonian $H_\Gamma$. The problem of classifying steady states according to topological properties is therefore equivalent to that of classifying Hamiltonian systems of non-interacting fermions, which is the main message of this section. Consequently, both the symmetry-based classification of references~\cite{Altland97,Schnyder08} and the K-theory approach of references~\cite{Kitaev09,Stone11,Abramovici12} can be directly applied in the dissipative framework. Note that the Hamiltonian $H_\Gamma$ always takes a Bogoliubov-de Gennes form when expressed in terms of the original fermionic operators $a_i, a^\dagger_i$, and is therefore automatically particle-hole symmetric.

The topological classification crucially relies on the existence of a bulk spectral gap and is essentially based on the mathematical concept of \emph{homotopy equivalence}, or equivalence under continuous deformations~\footnote{In the framework of K-theory, \emph{stable equivalence} also plays a crucial role in the definition of ``topological equivalence'' (see reference~\cite{Kitaev09}).}. More specifically, two gapped Hamiltonians $H_\Gamma$ and $H_{\Gamma'}$ are considered as ``topologically equivalent'' if they can be continuously deformed into each other \emph{without closing the gap}. The dissipative counterpart of this equivalence is provided by the mapping of equation~\eref{eq:mappingGammaHam}: two Gaussian steady states corresponding to correlation matrices $\Gamma$ and $\Gamma'$ will be considered as topologically equivalent and referred to as belonging to the same topological phase if and only if they can be continuously deformed into each other without closing the bulk \emph{purity} gap~\footnote{Note that the spectrum of $H_\Gamma$ is defined by that of the Hermitian matrix $\rmi \Gamma$, and is therefore in one-to-one correspondence with the spectrum of $(\rmi \Gamma)^2$, i.e., with the purity spectrum.}. The existence of a bulk purity gap is therefore the key ingredient required to define topological order in the dissipative setting.

The mapping defined by equation~\eref{eq:mappingGammaHam} and the results of references~\cite{Schnyder08,Kitaev09,Stone11,Abramovici12} provide us, in principle, with a general topological classification of all possible (purity) gapped Gaussian steady states according to symmetries and to the spatial dimension of the system. We now sketch this construction focusing on the symmetry classes that are most relevant for the dissipative systems considered in this work.

An arbitrary $2N \times 2N$ steady-state correlation matrix $\Gamma$ ($N$ being the number of fermionic modes, or the number of sites for systems of spinless fermions defined on a lattice) can be brought to a block diagonal form
\begin{eqnarray}
    \Gamma = Q \bigoplus_{n = 1}^N \left(
    \begin{array}{cc}
        0 & \epsilon_n \\
        -\epsilon_n & 0
    \end{array}
    \right) Q^{T},
\end{eqnarray}
where $Q$ is an orthogonal matrix and $\pm \epsilon_n$ are the real eigenvalues forming the spectrum of the Hermitian matrix $\rmi \Gamma$. The purity spectrum is defined by the spectrum of the real symmetric matrix $(\rmi \Gamma)^2$, which is doubly degenerate with positive eigenvalues $\epsilon^2_n$. Assuming that it is gapped, such that $\abs{\epsilon_n} > 0$ for all $n$, the matrix $\Gamma$ can be continuously deformed into a \emph{topologically equivalent} matrix $\bar{\Gamma}$ with a ``flat'' purity spectrum
\begin{eqnarray}
    \bar{\Gamma} = Q \bigoplus_{n = 1}^N \left(
    \begin{array}{cc}
        0 & 1 \\
        -1 & 0
    \end{array}
    \right) Q^{T}.
\end{eqnarray}
Since $(\rmi \bar{\Gamma})^2 = 1$, this ``spectrally flattened'' correlation matrix defines a \emph{pure} Gaussian state which is topologically equivalent to the not necessarily pure original steady state of interest. The matrix $\bar{\Gamma}$ allows us to construct an orthogonal \emph{spectral projection operator} $P$ (see, e.g., reference~\cite{Kitaev06}) defined as
\begin{eqnarray} \label{eq:spectralProjOp}
    P = \frac{1}{2} (1 - \rmi \bar{\Gamma}),
\end{eqnarray}
which projects onto the $N$-dimensional subspace associated with eigenvectors of $\rmi \Gamma$ with negative eigenvalues~\footnote{From the point of view of the Hamiltonian $H_\Gamma$ (see equation~\eref{eq:mappingGammaHam}), $P$ projects onto the subspace of eigenstates of $H_\Gamma$ with negative energy.}. This operator plays a crucial role in the topological classification of Gaussian steady states as well as in the construction of associated topological invariants, as we will demonstrate below.

We first consider the case of spinless fermions on a $d$-dimensional lattice with periodic boundary conditions evolving under a translation-invariant dissipative dynamics. It will be convenient to label the local Majorana operators $c_i$ as $c_{i, \lambda}$, where $i$ refers to a particular lattice site at position $\vect{r}_{i}$ and $\lambda = 1, 2$ distinguishes the two local Majorana operators associated with the corresponding fermionic creation and annihilation operators $a_i$ and $a^\dagger_i$, i.e., $c_{i, 1} = a^\dagger_i + a_i$ and $c_{i, 2} = \rmi (a^\dagger_i - a_i)$. In momentum space, the steady-state correlation matrix $\Gamma$ then takes the form of a $2 \times 2$ complex antihermitian matrix $\Gamma(\bk)$ with components ($\lambda, \mu = 1, 2$)
\begin{eqnarray}
    (\Gamma(\bk))_{\lambda \mu} = \frac{1}{N} \sum_{i,j} \rme^{\rmi \bk \cdot (\mathbf{r}_j - \mathbf{r}_i)} \Gamma_{i \lambda, j \mu} = \frac{\rmi}{2} \Tr(\rho [ c_{\bk, \lambda}, c_{-\bk, \mu} ]),
\end{eqnarray}
where $N$ is the total number of lattice sites and $c_{\bk,\lambda}=\frac{1}{\sqrt{N}}\sum_{i}\rme^{-\rmi\bk\cdot\mathbf{r}_{i}}c_{i,\lambda}$. Since the matrix $\Gamma(\bk)$ satisfies the condition $\Gamma(\bk)^* = \Gamma(-\bk)$, the spectrum of the Hermitian matrix $\rmi \Gamma(\bk)$ is given by eigenvalues $\pm \epsilon(\bk)$ with $\epsilon(\bk) = \epsilon(-\bk) > 0$. This allows us to introduce the ``spectrally flattened'' momentum-space correlation matrix $\bar{\Gamma}(\bk)$ (the eigenvalues of $\rmi \bar{\Gamma}(\bk)$ being $\pm 1$) and the associated spectral projection operator
\begin{eqnarray}
    P(\bk) = \frac{1}{2} (1 - \rmi \bar{\Gamma}(\bk))
\end{eqnarray}
projecting onto the subspace $\mathbb{C}^2_{-}(\bk)$ of the $2$-dimensional complex vector space $\mathbb{C}^2$ spanned by the (complex) eigenvectors of $\rmi \bar{\Gamma}(\bk)$ associated with the negative eigenvalue $-1$. The operators $P(\bk)$ form a family of operators labeled by the wavevectors $\bk$ belonging to the Brillouin zone, i.e., to the $d$-dimensional torus $\mathcal{T}^d = [-\pi, \pi]^d$. They define a fiber bundle $\oplus_{\bk \in \mathcal{T}^d} \mathbb{C}^2_{-}(\bk)$ on the Brillouin zone manifold, with fibers $\mathbb{C}^2_{-}(\bk)$ assigned to each point $\bk \in \mathcal{T}^d$. The problem of classifying Gaussian steady states according to topology therefore reduces to that of classifying fiber bundles over a torus $\mathcal{T}^d$ (the Brillouin zone). In the present case, the fibers are vector spaces and the fiber bundle is thus a vector bundle, for which a complete topological classification can be constructed using the approach of reference~\cite{Schnyder08} or K-theory~\cite{Kitaev09,Stone11,Abramovici12}. We present below the results pertaining to the two types of systems that are most relevant to this work (see section~\ref{sec:physicalConstraints}), namely: 2D dissipative systems without TRS (symmetry class D of Altland and Zirnbauer) and 1D dissipative systems with TRS (symmetry class BDI).

The topological classification generally reduces to the identification of all possible homotopy classes of continuous maps $\bk \mapsto P(\bk)$ from the Brillouin zone manifold to the space of spectral projection operators. Since the $2 \times 2$ complex matrix $\rmi \bar{\Gamma}(\bk)$ is Hermitian and unitary, with $(\rmi \bar{\Gamma}(\bk))^2 = 1$, the spectral projection operator can be expressed as
\begin{eqnarray} \label{eq:kspaceProjector}
    P(\bk) = \frac{1}{2} (1 + \vect{n}(\bk) \cdot \boldsymbol{\sigma}),
\end{eqnarray}
where $\vect{n}(\bk)$ is a unit vector on the 2-sphere ($S^2$) and $\boldsymbol{\sigma}$ is a vector of Pauli matrices. In the absence of additional symmetries, the spectral projection operator thus describes a mapping from the Brillouin zone torus to the sphere~\footnote{$P(\bk)$ can alternatively be seen as an element of $U(2) / U(1) \times U(1) \simeq G(1,2)$, $G(m,n)$ being the set of all $n$-dimensional subspaces in $\mathbb{C}^m$ (also known as a complex Grassmann manifold)~\cite{Schnyder08}.}.

In 2D dissipative systems where the steady state belongs to the symmetry class D, the only symmetry that $\bar{\Gamma}(\bk)$ (or $P(\bk)$) has is the PHS automatically embedded in Gaussian steady states. The homotopy classes of maps $\bk \rightarrow \vect{n}(\bk)$ then form a group which can be identified with the homotopy group $\pi_2 (S^2) = \mathbb{Z}$~\footnote{$\pi_2 (S^2)$ formally classifies maps from $S^2$ to $S^2$, but since $\pi_1 (S^2)$ is trivial, it equivalently classifies maps from $\mathcal{T}^2$ to $S^2$.} (which is non-trivial) and one can have non-trivial topological steady states and distinguish them via an integer topological invariant known as the \emph{Chern number}, defined as
\begin{eqnarray} \label{eq:chernNumber}
    \nu_{\text{2D}} & = & \frac{1}{2\pi \rmi} \int_{-\pi}^\pi \int_{-\pi}^\pi \rmd k_x \rmd k_y \Tr \left( P \left( \frac{\partial P}{\partial k_x} \frac{\partial P}{\partial k_y} - \frac{\partial P}{\partial k_y} \frac{\partial P}{\partial k_x} \right) \right) \nonumber \\
    & = & \frac{1}{16\pi} \int_{-\pi}^\pi \int_{-\pi}^\pi \rmd k_x \rmd k_y \Tr \left( \bar{\Gamma} \left( \frac{\partial \bar{\Gamma}}{\partial k_x} \frac{\partial \bar{\Gamma}}{\partial k_y} - \frac{\partial \bar{\Gamma}}{\partial k_y} \frac{\partial \bar{\Gamma}}{\partial k_x} \right) \right) \nonumber \\
    & = & \frac{1}{4\pi} \int_{-\pi}^\pi \int_{-\pi}^\pi \rmd k_x \rmd k_y \left( \vect{n} \cdot \left( \frac{\partial \vect{n}}{\partial k_x} \times \frac{\partial \vect{n}}{\partial k_y} \right) \right),
\end{eqnarray}
where in the last line we have used the fact that $\bar{\Gamma}(\bk)=\rmi(\mathbf{n}(\bk)\cdot\boldsymbol{\sigma})$ (see equation~\eref{eq:kspaceProjector}).

In 1D dissipative systems where PHS is the only symmetry, the Brillouin zone corresponds to a circle $S^1$ and the homotopy classes of maps $k \mapsto \mathbf{n}(k)$ form a homotopy group $\pi_1(S^2)$ which is trivial. This means that the vector $\mathbf{n}(k)$ cannot ``twist'' in a way that cannot be continuously untwisted as we move along the Brillouin zone circle, such that all steady states are necessarily topologically trivial. The situation can change in the presence of additional symmetries, however, since additional constraints on $P(\bk)$ (or, equivalently, on $\bar{\Gamma}(\bk)$ or $\mathbf{n}(\bk)$) can potentially restrict the set of allowed continuous deformations. If the 1D dissipative system has TRS, then it also has chiral symmetry and, hence, belongs to the symmetry class BDI. In that case, as argued below equation~\eref{eq:steadyStateSymmetries}, one can find a unitary matrix $\Sigma$ such that $\Sigma^2 = 1$ and
\begin{eqnarray}
    \{ \Sigma, \bar{\Gamma}(\bk) \} = 0
\end{eqnarray}
for all $\bk$. (In our case, TRS takes the form $c_{i, 1} \rightarrow c_{i, 1}$, $c_{i, 2} \rightarrow -c_{i, 2}$, such that $\Sigma = \sigma_z$.) The matrix $\Sigma$ has eigenvalues $\pm 1$ and can be expressed as $\Sigma = \vect{a} \cdot \boldsymbol{\sigma}$, where $\vect{a}$ is a real unit vector that does not depend on $\bk$. The above condition then takes the form $\vect{a} \cdot \vect{n}(\bk) = 0$ for all $\bk$. In other words, chiral symmetry restricts the vector $\vect{n}(\bk)$ from the sphere to a \emph{circle} in the plane perpendicular to $\vect{a}$. As a result, the relevant homotopy group becomes \emph{non-trivial}, given by $\pi_{1}(S^{1})=\mathbb{Z}$. The corresponding topological invariant can be constructed by choosing a basis where
\begin{eqnarray}
    \Sigma = \sigma_z = \left(
    \begin{array}{cc}
        1 & 0 \\
        0 & -1
    \end{array}
    \right), \qquad \bar{\Gamma}(\bk) = \left(
    \begin{array}{cc}
        0 & \rme^{\rmi \theta(\bk)} \\
        -\rme^{-\rmi \theta(\bk)} & 0
    \end{array}
    \right),
\end{eqnarray}
with $\theta(\bk) \in [0, 2\pi)$ and $\rme^{\rmi \theta(\bk)} = n_y(\bk) + \rmi n_x(\bk)$. The $U(1)$ phase $\theta(\bk)$ therefore fully characterizes $\bar{\Gamma}(\bk)$ and contains all required information to construct a topological invariant distinguishing between different topological classes of steady states. The relevant topological invariant, the \emph{winding number}, takes the form
\begin{eqnarray} \label{eq:windingNumber}
    \nu_{\text{1D}} & = & \frac{2}{\pi} \int_{-\pi}^\pi \rmd k \Tr \left( (P_{-} P P_{+}) \frac{\rmd (P_{+} P P_{-})}{\rmd k} \right) \nonumber \\
    & = & \frac{1}{\pi} \int_{-\pi}^\pi \rmd k \left( \vect{a} \cdot \left( \vect{n} \times \frac{\partial \vect{n}}{\rmd k} \right) \right) \nonumber \\
    & = & \frac{1}{2\pi} \int_{-\pi}^\pi \rmd \theta = \frac{1}{2\pi}(\theta(\pi) - \theta(-\pi)),
\end{eqnarray}
where we have introduced projection operators $P_\pm = (1 \pm \Sigma)/2$ defined so that
\begin{eqnarray}
    P_{+} \bar{\Gamma}(\bk) P_{-} = \left(
    \begin{array}{cc}
        0 & \rme^{\rmi \theta(\bk)} \\
        0 & 0
    \end{array}
    \right), \quad P_{-} \bar{\Gamma}(\bk) P_{+} = \left(
    \begin{array}{cc}
        0 & 0 \\
        -\rme^{-\rmi \theta(\bk)} & 0
    \end{array}
    \right).
\end{eqnarray}
We remark that the above construction of topological classes and topological invariants crucially relies on translational symmetry. However, one expects bulk topological properties to be robust against weak dissipative perturbations \emph{that preserve the purity gap as well as the symmetries of the system}. This can be rigorously demonstrated in the K-theoretic framework of reference~\cite{Kitaev09} and in the more general framework of noncommutative geometry developed in references~\cite{Bellissard94,Kitaev06}, allowing to extend the above construction to disordered (and possibly finite) systems. Although the details are beyond the scope of this work, we emphasize that the existence of \emph{quantized} (integer) topological invariants in such systems crucially relies on the quasi-local nature of the spectral projection operator. Here the quasi-local nature of the corresponding steady-state correlation matrix is ensured in the presence of a \emph{dissipative} gap, which is a spectral property of the model in contrast to the purity gap, which relates to mode occupations. More specifically, the existence of a finite dissipative gap $\Delta$ implies an exponentially fast relaxation to the steady state (on a characteristic time scale $\tau \sim 1/\Delta$) and leads to the exponential decay of all spatial correlations, such that the spectral projection operator (see equation~\eref{eq:spectralProjOp}) satisfies
\begin{eqnarray}
    \abs{P_{i \lambda, j \mu}} \leq c \exp(-\alpha \abs{\vect{r}_i - \vect{r}_j})
\end{eqnarray}
for some constants $c, \alpha > 0$~\footnote{Note that the purity gap of $\Gamma$ ensures that the exponential decay of all spatial correlations remains under the continuous ``spectral flattening'' transformation $\Gamma \to \bar{\Gamma}$.}. Note that the converse is also true~\cite{Hoening12}. The operator $P_{ij}$ defined by $(P_{ij})_{\lambda \mu} = P_{i \lambda, j \mu}$ (see reference~\cite{Kitaev06}) is thus \emph{quasi-diagonal}, i.e., there exists some constants $c', c'' > 0$ and $\alpha' > d$ ($d$ being the spatial dimension of the system) such that
\begin{eqnarray}
    \norm{P_{ij}}_{\mathrm{HS}} \leq c' \abs{\vect{r}_i - \vect{r}_j}^{-\alpha'}, \qquad \norm{P_{ii}} \leq c'',
\end{eqnarray}
where $\norm{.}_\mathrm{HS}$ is the Hilbert-Schmidt norm and $\norm{.}$ the usual operator norm. This ``localization'' condition satisfied by the spectral projection operator in the presence of a finite dissipative gap crucially allows for the definition of the Chern number and winding number topological invariants in disordered finite systems. Such a construction can be found in the appendix C of reference~\cite{Kitaev06}, for example.

\subsection{Phenomenology of non-equilibrium topological phase transitions}
\label{sec:phenomenology}

The above discussion shows that the existence of \emph{two} gaps is necessary to identify the topological properties of the bulk (Gaussian) steady state: (i) the purity gap, which allows to identify a particular topological class (e.g., $\mathbb{Z}$) to which the steady state belongs, and (ii) the dissipative gap, which ensures that the steady-state correlations are quasi-local (as defined above), so that in particular the Chern and winding number topological invariants defined above (see equations~\eref{eq:chernNumber} and~\eref{eq:windingNumber}) remain quantized in the presence of disorder. Since the dissipative and purity spectra are in general independent quantities---which can be seen from the fact that two independent matrices $X$ and $Y$ are required to describe the dissipative dynamics---these two gaps can close independently of each other. However, the closure of the dissipative gap alone gives rise to a critical behavior in steady state. As a consequence, non-equilibrium \emph{topological} phase transitions can occur in three distinct ways:
\begin{enumerate}[\hspace{0.5cm}(i)]
    \item Via the closure of the bulk dissipative gap only (criticality);
    \item Via the closure of the bulk purity gap only (no criticality);
    \item Via the closure of the both the bulk dissipative and purity gaps (criticality).
\end{enumerate}
These three possibilities will be exemplified in explicit models in section~\ref{sec:1D} below. Clearly, the same phenomenology must appear at interfaces between distinct non-equilibrium topological phases. We will investigate such physics in detail in section~\ref{sec:edgePhysics} below.

\section{Physical constraints}
\label{sec:physicalConstraints}

Our discussion so far was very general, with the only assumption that the steady state of the dissipative dynamics is a Gaussian fermionic state. We now discuss restrictions that arise in ``typical'' experimental realizations, as sketched in section~\ref{sec:physicalRealization} above. More specifically, we consider \emph{quasi-local} dissipative processes in one- and two-dimensional lattice systems; assuming that the corresponding Lindblad operators act on a quasi-local subset of sites and have a spatially isotropic (or $s$-wave symmetric) creation part resulting from spontaneous emission processes. We additionally assume that the system is translation-invariant and that the steady state of the dissipative dynamics is unique and pure. Under this purity assumption, the steady state of the system can be viewed as the ground state of the parent Hamiltonian associated with the dissipative dynamics (see equation~\eref{eq:Hparent} and the discussion of section~\ref{sec:purityGap}), and the existence of topological order can be assessed in the exact same way as in the Hamiltonian setting. In particular, as for the topological classification of section~\ref{sec:topClassification} above, one can rely on the general Hamiltonian classification of gapped topological phases of non-interacting fermions (according to symmetry and spatial dimension) developed in references~\cite{Zirnbauer96,Altland97,Schnyder08,Ryu10}.

Although we assume a pure steady state, we remark that the conclusions drawn below will also hold in the more general case where the steady state is not necessarily pure but has a \emph{gapped purity spectrum}. We have argued in section~\ref{sec:topClassification} that an arbitrary state exhibiting a finite purity gap is topologically equivalent to a pure state exhibiting a completely ``flat'' purity spectrum. To identify the possible classes of topological steady states that can be reached when taking account the ``typical'' physical constraints described above, one can therefore focus exclusively on pure states, without loss of generality. We emphasize that in that case the (automatically ``flat'') correlation matrix $\rmi \bar{\Gamma}$ describing the pure steady state precisely corresponds to the parent Hamiltonian in its ``first-quantized'' form, namely, $H_\text{parent} = \rmi \sum_{i, j} \bar{\Gamma}_{ij} c_i c_j = H_{\bar{\Gamma}}$, where $H_{\bar{\Gamma}}$ is the fictitious Hamiltonian that can always be constructed from the (Gaussian) steady state (see equations~\eref{eq:mappingGammaHam} and~\eref{eq:spectralProjOp} and discussion thereof).

Translational symmetry makes it most convenient to express the Lindblad operators in the momentum-space form $L_\bk = u_\bk a_\bk + v_\bk a^\dagger_{-\bk}$. In order for our assumption of a pure and unique steady state to be satisfied, the Lindblad operators must form a complete set of anticommuting operators (i.e., $\{ L_\bk, L_{\bk'} \} = 0$ for all $\bk, \bk'$), which implies that
\begin{eqnarray} \label{eq:oddConstraintFromPurity}
    u_\bk v_\bk = -u_{-\bk} v_{-\bk}.
\end{eqnarray}
Translational symmetry additionally restricts the possible lattices for the system to Bravais lattices. In that case, the set of Lindblad operators is complete whenever there are as many Lindblad operators as lattice sites.

As argued above, the pure steady state can equivalently be described as the ground state of the parent Hamiltonian $H_\mathrm{parent} = \sum_\bk L^\dagger_\bk L_\bk$ (see section~\ref{sec:purityGap}), which takes here the explicit form
\begin{eqnarray}
    H_\mathrm{parent} & = & \sum_\bk (u^*_\bk a^\dagger_\bk + v^*_\bk a_{-\bk})(u_\bk a_\bk + v_\bk a^\dagger_{-\bk}) \nonumber \\
    & = & \sum_\bk (\abs{u_\bk}^2 - \abs{v_{-\bk}}^2) a^\dagger_\bk a_\bk + \sum_\bk \abs{v_\bk}^2 \nonumber \\
    && +  \frac{1}{2} \sum_\bk \left[ (u^*_\bk v_\bk - u^*_{-\bk} v_{-\bk}) a^\dagger_\bk a^\dagger_{-\bk} + h.c. \right].
\end{eqnarray}
Dropping the constant term $\sum_\bk \abs{v_\bk}^2$, it becomes
\begin{eqnarray} \label{eq:parentHamBdGForm}
    H_\mathrm{parent} = \sum_\bk \left[ \xi_\bk a^\dagger_\bk a_\bk + \frac{1}{2} \left( \Delta_\bk a^\dagger_\bk a^\dagger_{-\bk} + \Delta^*_\bk a_{-\bk} a_\bk \right) \right],
\end{eqnarray}
where we have defined
\begin{eqnarray} \label{eq:xikDeltak}
    \xi_\bk & = & \abs{u_\bk}^2 - \abs{v_{-\bk}}^2, \quad  \Delta_\bk  =  u^*_\bk v_\bk - u^*_{-\bk} v_{-\bk}=\Delta_{-\bk} .
\end{eqnarray}
Since the Lindblad operators are generally defined up to a phase, we can assume $v_\bk$ to be real, without loss of generality. Using equation~\eref{eq:oddConstraintFromPurity}, we thus find
\begin{eqnarray} \label{eq:gapFunction}
    \Delta_\bk = 2 u^*_\bk v_\bk.
\end{eqnarray}
Our assumption that the creation part of the Lindblad operators is isotropic implies that $v_\bk = v_{-\bk}$~\footnote{In general, it can occur that $v_\bk = v_{-\bk}$ is only satisfied up to a phase $\rme^{\rmi \phi(\bk)}$ where $\phi_\bk$ is a linear function of $\bk$, since the creation part of the Lindblad operators need not be isotropic with respect to a center of symmetry that corresponds to a lattice site. In that case, however, $u_\bk$ carries the opposite phase---since the annihilation part of the Lindblad operators is engineered with respect to the same origin (see section~\ref{sec:generalFormMZMs})---and our analysis remains the same, without loss of generality.}. Since equation~\eref{eq:oddConstraintFromPurity} must be satisfied, we conclude that the condition $u_\bk = -u_{-\bk}$ must be \emph{imposed}. In practice, this can easily be achieved by tuning the phases of the driving lasers, as argued in our previous work~\cite{Bardyn12}, and will therefore be assumed in the following. We then have
\begin{eqnarray} \label{eq:singlePartSpectrum}
    \xi_\bk = \xi_{-\bk}; \qquad \xi_\bk = \abs{u_\bk}^2 - \abs{v_{\bk}}^2.
\end{eqnarray}

\subsection{Symmetry classes}
\label{sec:symmetryClasses}

Equations~\eref{eq:parentHamBdGForm},~\eref{eq:gapFunction}, and~\eref{eq:singlePartSpectrum} show that the parent Hamiltonian takes the generic form of a Bogoliubov-de Gennes (BdG) Hamiltonian and therefore exhibits, by construction, particle-hole (or charge-conjugation) symmetry (PHS). In the scheme of reference~\cite{Ryu10}, further classification can be achieved by considering time-reversal symmetry (TRS). In the Nambu representation, the parent Hamiltonian can be written as
\begin{eqnarray} \label{eq:parentHamFirstQuantizedForm}
    H_\mathrm{parent} = \frac{1}{2} \sum_\bk \Psi^\dagger_\bk \mathcal{H}_\bk \Psi_\bk
    \qquad \text{with } \mathcal{H}_\bk = \left( \begin{array}{cc}
    \xi_\bk & \Delta_\bk \\
    \Delta^*_\bk & -\xi_{-\bk} \end{array} \right),
\end{eqnarray}
where $\mathcal{H}_\bk$ is the ``first-quantized'' parent Hamiltonian and $\Psi^\dagger_\bk = (a^\dagger_\bk, a_{-\bk})$ the Nambu field operator. Particle-hole and time-reversal symmetries are then present whenever there exists $2 \times 2$ unitary matrices $U_C$ and $U_T$, respectively, such that the following conditions on the Hamiltonian are satisfied:
\begin{eqnarray}
    U^\dagger_C \mathcal{H}^*_{-\bk} U_C & = -\mathcal{H}_\bk, \qquad & \mathrm{(PHS)} \label{eq:PHS} \\
    U^\dagger_T \mathcal{H}^*_{-\bk} U_T & = +\mathcal{H}_\bk. \qquad & \mathrm{(TRS)} \label{eq:TRS}
\end{eqnarray}
The symmetry class of the system is determined by whether or not such matrices can be identified and, if so, on the sign of $U^*_C U_C$ and $U^*_T U_T$ (which can only take values $\pm 1$, as argued in reference~\cite{Ryu10}). In the present case, the parent Hamiltonian describes a spin-polarized superfluid and exhibits---again, by construction---PHS with $U_C = \sigma_x$, such that $U^*_C U_C = +1$. It is clear from the form of $H_\mathrm{parent}$ (see equations~\eref{eq:parentHamBdGForm},~\eref{eq:gapFunction}, and~\eref{eq:singlePartSpectrum}) that the system additionally exhibits TRS if and only if $\Delta^*_{-\bk} = \Delta_\bk$ which, remembering equation~\eref{eq:xikDeltak}, is satisfied provided that $\Delta_\bk$ is real up to a global phase~\footnote{One must keep in mind that the gap function $\Delta_\bk$ is defined up to a global phase, since $H_\mathrm{parent}$ is invariant under the global $U(1)$ gauge transformation $a^\dagger_\bk \to \mathrm{e}^{\mathrm{i} \theta} a^\dagger_\bk$.}. In that case, equation~\eref{eq:TRS} holds for $U_T = 1$ (such that $U^*_T U_T = +1$) and the system exhibits chiral symmetry.

In summary, two \emph{symmetry} classes of steady states can be realized through ``typical'' quantum-reservoir engineering (see section~\ref{sec:physicalRealization}) in driven-dissipative systems of spin-polarized fermions: if the Lindblad operators break TRS, the steady state belongs to the symmetry class D of Altland and Zirnbauer; otherwise it belongs to the symmetry class BDI. According to the \emph{topological} classification of reference~\cite{Ryu10}, one must therefore consider either (i) 2D systems in the symmetry class D (with broken TRS) or (ii) 1D systems in the symmetry class BDI (with TRS) in order to have the (a priori) possibility of reaching topologically non-trivial steady states. In these systems, distinct topological states are characterized by distinct values of an integer-valued ($\mathbb{Z}$) topological invariant: the Chern and winding number, respectively. Most importantly, the zero-energy edge modes appearing in such systems in the Hamiltonian setting are Majorana fermions. This opens up the possibility to generate spatially isolated Majorana zero-\emph{damping} modes in the dissipative setting of interest in this work.

\subsection{Chern number}
\label{sec:chernNumber}

Let us now examine more closely the case of a 2D translation-invariant dissipative system with \emph{broken TRS}, with a pure and unique steady state that accordingly belongs to the symmetry class D of Altland and Zirnbauer and is characterized by the Chern number topological invariant of equation~\eref{eq:chernNumber}. We show below that \emph{quasi-local} Lindblad operators generating such dissipative dynamics do not allow to obtain phases with a non-zero Chern number, in contrast to the Hamiltonian setting. Although this result seems to be a no-go statement for Majorana zero modes in such systems, we will demonstrate in section~\ref{sec:2D} that unpaired Majorana zero-damping modes can exist in dissipative systems with vanishing Chern number provided that the steady state is not pure but mixed.

As we have discussed above, a pure steady state can equivalently be viewed as the ground state of the parent Hamiltonian $H_\mathrm{parent}$ associated with the dissipative dynamics, given by equation~\eref{eq:parentHamFirstQuantizedForm} (see also equations~\eref{eq:parentHamBdGForm},~\eref{eq:gapFunction} and~\eref{eq:singlePartSpectrum}), and its bulk topological properties can be characterized by means of the spectral projector $P_\bk = \frac{1}{2} (1 + \vect{n}_\bk \cdot \boldsymbol{\sigma})$, where $\vect{n}_\bk \cdot \boldsymbol{\sigma} = \mathcal{H}_\bk / N_\bk$ ($N_\bk$ being a normalization factor defined such that $\norm{\vect{n}_\bk} = 1$). The corresponding Chern number is determined by the vector $\vect{n}_\bk$ (see equation~\eref{eq:chernNumber}), which takes the explicit form
\begin{eqnarray}
    \vect{n}_\bk = \frac{1}{N_\bk} \left( \begin{array}{c}
        \phantom{-} \mathrm{Re}(\Delta_\bk) \\
        -\mathrm{Im}(\Delta_\bk) \\
        \xi_\bk
    \end{array} \right)
    = \frac{1}{N_\bk} \left( \begin{array}{c}
        2 v_\bk \mathrm{Re}(u_\bk) \\
        2 v_\bk \mathrm{Im}(u_\bk) \\
        \abs{u_\bk}^2 - \abs{v_{\bk}}^2
    \end{array} \right),
\end{eqnarray}
where $N_\bk = \sqrt{\xi_\bk^2 + \abs{\Delta_\bk}^2} = \sqrt{\abs{u_\bk}^2 + \abs{v_\bk}^2}$~\footnote{We assume that the system is infinite such that the components of $\vect{n}_\bk$ are continuous functions of $\bk$ in the Brillouin zone, in which case the Chern number given by equation~\eref{eq:chernNumber} is a well-defined quantity.}. Note that $N_\bk$ corresponds to the spectrum of $H_\mathrm{parent}$ and thus defines the \emph{dissipative} gap (see section~\ref{sec:dissipativeGap}). In order for the vector $\vect{n}_k$ to be well defined, the functions $u_\bk$ and $v_\bk$ must not vanish simultaneously anywhere in the Brillouin zone. Introducing $\varphi_\bk = \abs{\varphi_{\bk}} \rme^{\rmi \theta_\bk} = v_\bk / u_\bk$ and a ``Fermi surface'' $\mathcal{F_\epsilon} = \{ \bk: \, \abs{\varphi_{\bk}} > 1 / \epsilon \}$ (defined for some arbitrary $\epsilon > 0$) which generally~\footnote{More pathological Fermi surfaces can in principle be encountered, but the Chern number is a mere definition in that case.} takes the form of a finite union of piecewise smooth, simple closed curves $\mathcal{F}_{\epsilon, \lambda}$ (i.e., $\mathcal{F_\epsilon} = \bigcup_\lambda \mathcal{F}_{\epsilon, \lambda}$), the Chern number can be expressed (see reference~\cite{Bardyn12}) as a sum
\begin{eqnarray}
    \nu_{\text{2D}} = \sum_\lambda W_{\epsilon, \lambda}
\end{eqnarray}
of winding numbers defined by
\begin{eqnarray}
    W_{\epsilon, \lambda} & = & \frac{1}{2\pi} \oint_{\mathcal{F}_{\epsilon, \lambda}} \nabla_\bk \theta_\bk \cdot d\bk = \frac{1}{2\pi} \oint_{\mathcal{F}_{\epsilon, \lambda}} (\partial_{k_x} \theta_\bk dk_x + \partial_{k_y} \theta_\bk dk_y).
\end{eqnarray}
Choosing $\epsilon \ll 1$, the above expressions tell us that the value of the Chern number can be inferred from the behavior of $\varphi_\bk = v_\bk / u_\bk$ around the zeroes of $u_\bk$---which do not coincide with zeroes of $v_\bk$ since $N_\bk$ must be non-zero. Since the norm of $\varphi_\bk$ diverges upon approaching such points, it is the phase winding of $\varphi_\bk$ around these points that encodes all information about the Chern number. More specifically, each of the zeroes of $u_\bk$ contributes to the Chern number by an integer value corresponding to the \emph{winding number} of the phase $\theta_\bk$ around the latter (in a conventional, e.g., counter-clockwise direction). Since the function $v_\bk$ can always be chosen as real (see discussion below equation~\eref{eq:gapFunction}), $\theta_\bk$ corresponds to the phase of $u_\bk$ (up to a sign that can be absorbed into the definition of the Chern number by a change of convention $\nu_{\text{2D}} \to -\nu_{\text{2D}}$) and the Chern number simply reduces to the sum of the phase windings of the function $u_\bk$ around its zeroes.

In the Hamiltonian setting, the phase of $u_\bk$ is defined by the phase of the gap function $\Delta_\bk$, whereas the norm of $u_\bk$ is defined by the norm of $\Delta_\bk$ \emph{and} the value of the \emph{independent} quantity $\xi_\bk$ (the single-particle dispersion)~\cite{Read00}, such that the phase and the modulus of $u_\bk$ are essentially independent. In our dissipative setting, however, $u_\bk$ is a \emph{single} complex function defined on the Brillouin zone and, therefore, the phase and the norm of $u_\bk$ are closely related. This crucially restricts the possible values of the Chern number: the function $u_\bk$ defines a smooth vector field $(\mathrm{Re}(u_\bk), \mathrm{Im}(u_\bk))$ on a torus and, according to the Poincar\'{e}-Hopf theorem, the sum of its phase windings around its zeroes must vanish. We thus conclude that the only value of the Chern number that can be achieved in the above dissipative setting is zero.

Note that the vanishing of the Chern number is a direct consequence of the smoothness of the vector field corresponding to $u_\bk$, which in turn is due to the \emph{quasi-local nature of the Lindblad operators}. The function $u_\bk$ defines a smooth ($C^\infty$) vector field if and only if the norm of the coefficients $u_{j-i} = u(\vect{r}_j - \vect{R}_i)$ encoding the annihilation part of the Lindblad operators (see equation~\eref{eq:translationInvariantLindbladOp}) decays faster than any power of $1/\abs{\vect{r}_j - \vect{R}_i}$. A power-law decay of the Lindblad operators would therefore be necessary to be able to engineer phases characterized by a non-zero Chern number in our dissipative setting.

\section{Dissipative edge physics}
\label{sec:edgePhysics}

Thus far we have concentrated on \emph{bulk} properties of purely dissipative systems described by a quadratic Lindblad master equation. It is clear, however, that the existence of non-trivial topological bulk properties must somehow be reflected in the physics occurring at physical edges or in topological defects---which we both refer to as ``edges'' in the following. In the Hamiltonian setting, this intimate connection has been rigorously demonstrated and formalized---at least, for non-interacting systems---as a \emph{bulk-edge correspondence} (or bulk-boundary correspondence) relating the topological nature of the bulk to the number of gapless modes occurring at an edge~\cite{Hatsugai93,Kitaev06,Ryu10,Gurarie11,Essin11}. Of particular interest here is the bulk-edge correspondence pertaining to (i) 1D systems characterized by a winding number topological invariant $\nu_{\text{1D}}$ (see equation~\eref{eq:windingNumber}) and (ii) 2D systems characterized by a Chern number topological invariant $\nu_{\text{2D}}$ (see equation~\eref{eq:chernNumber}), which tells us that a number $m = \abs{\nu^{(1)} - \nu^{(2)}}$ of gapless edge modes must be present at the interface between two topological phases characterized by integer topological invariants (winding or Chern numbers, as appropriate) $\nu^{(1)}$ and $\nu^{(2)}$, respectively. In general, such modes are \emph{robust} (e.g., against disorder) \emph{as long as the Hamiltonian system remains in the same topological class}. Since the purity gap can close in the dissipative setting of interest in this work, it is highly unclear whether similar universal signatures of bulk topological properties can be observed. We thus proceed, in the next sections, to investigate the edge physics that may arise in systems described by a quadratic dissipative dynamics.

\subsection{Dissipative bulk-edge correspondence}
\label{sec:bulkEdge}

Let us consider a generic dissipative dynamics characterized by a finite dissipative gap and described by Lindblad operators $L_i$ or, equivalently, by matrices $X$ and $Y$ as defined in section~\ref{sec:gaussianMasterEquations}. It is clear that the bulk-edge correspondence pertaining to Hamiltonian systems must also be satisfied when the steady state of the dissipative dynamics is pure, since such a state can equivalently be regarded as a ground state of the parent Hamiltonian $H_\text{parent} = \sum_i L^\dagger_i L_i = \rmi \sum_{i, j} Y_{ij} c_i c_j$ (see section~\ref{sec:purityGap}). In that case, the spectrum of $H_\text{parent}$ defines the damping spectrum, and at least $m = \abs{\nu^{(1)} - \nu^{(2)}}$ Majorana \emph{zero-damping} modes must be present at the interface between two non-equilibrium topological phases characterized by Chern or winding number topological invariants $\nu^{(1)}$ and $\nu^{(2)}$. In the more general situation where the steady state exhibits a finite purity gap but is not pure, the Hamiltonian bulk-edge correspondence obviously still applies to the parent Hamiltonian, such that $H_\text{parent}$ still exhibits a minimum of $m = \abs{\nu^{(1)} - \nu^{(2)}}$ Majorana zero modes at the interface, but none of these modes is guaranteed to be Majorana \emph{zero-damping} modes of the dissipative dynamics. As argued in section~\ref{sec:dissipativeGap}, Majorana zero mode of $H_\text{parent}$ must either correspond to Majorana zero-damping modes or give rise to intrinsic Majorana \emph{zero-purity} modes in steady state. The dissipative counterpart of the Hamiltonian bulk-edge correspondence can therefore be stated as follows: \emph{As long as the topological nature of the steady state remains the same in the bulk---which is true if the purity gap remains finite and if the symmetries of the Lindblad operators are preserved---a minimum total number of $m = \abs{\nu^{(1)} - \nu^{(2)}} \equiv \Delta \nu $ Majorana zero-damping and intrinsic zero-purity modes must be present at the interface between two non-equilibrium topological phases characterized by topological invariants $\nu^{(1)}$ and $\nu^{(2)}$, respectively}. With obvious notations, this result can be summarized in the form
\begin{eqnarray} \label{eq:dissipativeBulkEdgeCorrespondence}
    (m_\text{damping})_\textit{edge} + (m_\text{purity})_\textit{edge} \geq (\Delta \nu)_\textit{bulk}.
\end{eqnarray}
We remark that such a dissipative bulk-edge correspondence is only guaranteed to hold in the presence of a finite dissipative gap, since the existence of such a gap is necessary to ensure the exact quantization of the topological invariants, as discussed in section~\ref{sec:topClassification}. Most importantly, the inequality appearing in equation~\eref{eq:dissipativeBulkEdgeCorrespondence} originates from the fact that ``spurious'' Majorana zero modes may arise ``accidentally'', as opposed to Majorana zero modes which we refer to as ``genuine'' that have topological origins. A strict equality is therefore only obtained when restricting to \emph{genuine} Majorana zero-damping and zero-purity modes.

Despite its important predictive power, the above result does not provide specific information regarding the exact number of genuine Majorana zero-damping modes that appears at an edge---as we will argue below, such knowledge must come from additional considerations unrelated to the topological nature of the system. One may therefore wonder whether this number can be a robust quantity and, if so, under what conditions. This will be the focus of the next sections: we will first investigate systems whose edge dissipative dynamics emerges in a very natural way by extending the bulk dissipative dynamics as close as possible to a physical edge. Since the topological nature of generic Gaussian states is only well-defined in the presence of a finite purity gap and since all Gaussian states featuring such a gap can  be continuously deformed to pure states (see section~\ref{sec:topClassification}), we will focus on dissipative dynamics driving the system into a pure steady state. In that case, the purity of the system obviously forces the number $(m_\text{purity})_\textit{edge}$ of Majorana zero-purity modes to vanish in equation~\eref{eq:dissipativeBulkEdgeCorrespondence}, and one can try to construct Majorana zero-damping modes explicitly. Having examined this simple situation where the number $(m_\text{damping})_\textit{edge}$ of genuine Majorana zero-damping modes is solely determined by the topological nature of the bulk, we will then extend our discussion to the more general case where the dissipative dynamics can affect the purity of the system while maintaining a finite purity gap in the bulk, and conclude by clarifying the role of topology in the dissipative setting central to this work.

\subsection{General form of Majorana zero-damping modes for translation-invariant dissipative processes}
\label{sec:generalFormMZMs}

We have introduced above a dissipative bulk-edge correspondence which crucially relies on the existence of \emph{two} gaps: the dissipative and purity gaps. Both these gaps must be finite in the bulk in order for equation~\eref{eq:dissipativeBulkEdgeCorrespondence} to hold and---as argued in section~\ref{sec:topClassification}---for the bulk to have a well-defined topological nature with associated topological invariants that are quantized. Any of these two gaps can in principle close at an edge, which gives rise to the variety of possibilities allowed by equation~\eref{eq:dissipativeBulkEdgeCorrespondence}. In the following section, however, we focus on the case where the purity gap remains open at the edge, thus forbidding the appearance of intrinsic Majorana zero-purity modes. In that case, the topological properties of the steady state are equivalent to that of a pure state---as argued in section~\ref{sec:topClassification}---and the bulk-edge correspondence describes the behavior of a \emph{single} gap at the edge (namely, the dissipative gap) exactly as in the Hamiltonian setting. From a topological point of view, all steady-state properties are the same as if the steady state were completely pure. We therefore restrict ourselves, without loss of generality, to pure (Gaussian) steady states. In addition we assume translation invariance in the following in order to make the explicit construction of Majorana zero-damping modes analytically tractable.

\begin{figure}[t]
    \begin{center}
        \includegraphics[width=\columnwidth]{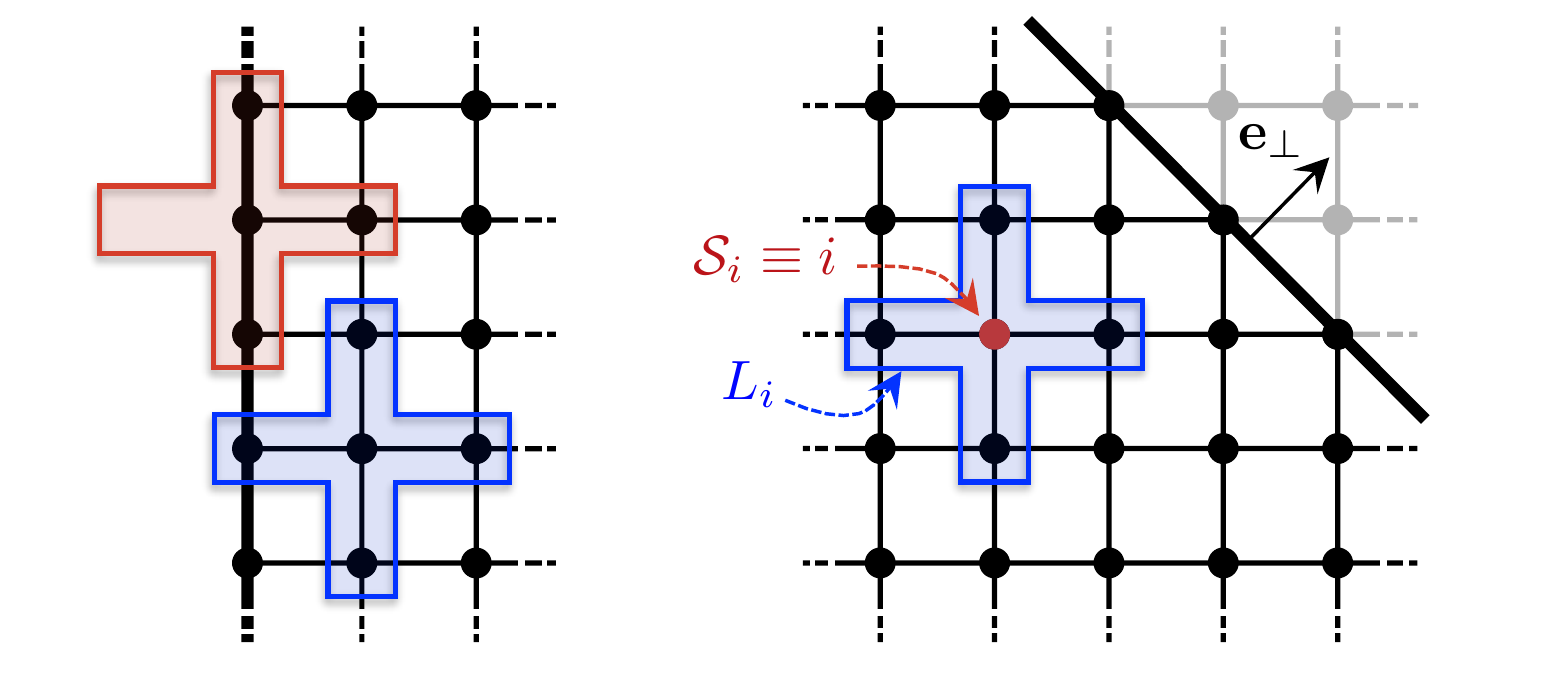}
        \caption{Left: Translation-invariant dissipative dynamics terminated at the edge in the case of ``cross-shaped'' Lindblad operators acting on a subset of $5$ sites (as in the explicit example of section~\ref{sec:2D}). Lindblad operators acting within the system (in blue) are taken into account into the dissipative dynamics, whereas Lindblad operators requiring truncation at the edge (in red) are not considered. Right: Lindblad operator $L_i$ associated with a lattice site $i$ that coincides with its center of symmetry $\mathcal{S}_i$, and example of an edge whose orientation is according to a vector $\vect{e}_\perp$ (see main text).}
        \label{fig:edgePhysics}
    \end{center}
\end{figure}

We consider a $d$-dimensional lattice system with a $(d-1)$-dimensional physical edge and assume that the system evolves under a translation-invariant dissipative dynamics consisting of a periodic repetition, on the lattice, of a single quasi-local dissipative process (or Lindblad operator) everywhere in the bulk and as close as possible to the edge (see figure~\ref{fig:edgePhysics}), so that every lattice site $i$ of the bulk becomes associated with a Lindblad operator $L_i$ whose form is independent of $i$. In order to ensure that the steady state is pure we further assume that the Lindblad operators form a set of anticommuting operators. For simplicity (and in order to incorporate the typical physical constraints discussed in section~\ref{sec:physicalConstraints}), we moreover restrict ourselves to Lindblad operators $L_i$ that possess a center of symmetry, which we denote as $\mathcal{S}_i$ (the index $i$ used to distinguish Lindblad operators then corresponds, by convention, to the index of the lattice site located closest to $\mathcal{S}_i$~\footnote{If there exists many such points, an arbitrary convention can be chosen.}). We define a position vector $\vect{R}_i$ corresponding to $\mathcal{S}_i$ and similarly define the position of the lattice sites $j$ as $\vect{r}_j$. In the translation-invariant setting defined above, the Lindblad operators then take the generic form
\begin{eqnarray} \label{eq:translationInvariantLindbladOp}
    L_i = \sum_{j \in \mathcal{I}(i)} u_{j-i} \, a_j + v_{j-i} \, a^\dagger_j,
\end{eqnarray}
where $\mathcal{I}(i)$ denotes the subset of sites $j$ onto which $L_i$ acts in a non-trivial way (see figure~\ref{fig:edgePhysics}), and $u_{j-i}$, $v_{j-i}$, and $a_j$ are shorthand notations for $u(\vect{r}_j - \vect{R}_i)$, $v(\vect{r}_j - \vect{R}_i)$, and $a(\vect{r}_j)$, respectively~\footnote{Note that the form of the coefficients $u_{j-i}$ and $v_{j-i}$ ensures that $L_i$ has the same structure independently of $i$, as required by translational symmetry.}. As argued in section~\ref{sec:dissipativeGap} above, a necessary and sufficient condition for a Majorana mode $\gamma$ to correspond to a Majorana zero-damping mode of the dissipative dynamics is $\{ L_i, \gamma \} = 0$ (for all $i$). Using this condition as well as the translation-invariant form of the Lindblad operators, one can show (see~\ref{app:explicit}) that any Majorana zero-damping mode $\gamma$ must take the generic form
\begin{eqnarray} \label{eq:generalSolutionZeroMode}
    \gamma = \mathcal{N} \rme^{\rmi \phi_i / 2} \sum_{\{ m_1, m_2, \atop \phantom{\{} \ldots, m_d \}} (\beta_{\vect{e}_1})^{n_1} (\beta_{\vect{e}_2})^{n_2} \ldots (\beta_{\vect{e}_d})^{n_d} a(\vect{r}_i + \sum_{n = 1}^d m_n \vect{e}_n) + h.c. \; ,
\end{eqnarray}
where $\mathcal{N} > 0$ is a normalization factor, $\phi_i \in [0, 2\pi)$ a phase, $\{ \vect{e}_n \}_{n = 1, 2, \ldots, d}$ a set of primitive vectors associated with the $d$-dimensional Bravais lattice on which the system is defined, and $\{ m_1, m_2, \ldots, m_d \}$ a set of integers defined such that the vectors $\vect{r}_i + \sum_{n = 1}^d m_n \vect{e}_n$ span the positions of all sites in the system. Most importantly, $\{ \beta_{\vect{e}_n} \}_{n = 1, 2, \ldots, d}$ is a set of real factors determining the increase of the ``spatial wave function'' corresponding to $\gamma$ in each of the directions defined by the primitive vectors $\vect{e}_n$. Note that $\beta_{j-i} = (\beta_{i-j})^{-1}$ is satisfied owing to the translation invariance discussed above ($\beta_{j-i}$ being a shorthand notation for $\beta(\vect{r}_j - \vect{r}_i)$)~\footnote{Note that $\beta_{j-i}$ is defined with respect to the site $i$ associated with $L_i$, whereas $u_{j-i}$ and $v_{j-i}$ are defined with respect to the center of symmetry of $L_i$.}.

Equation~\eref{eq:generalSolutionZeroMode} states that $\gamma$---or, more precisely, its spatial ``wave function''---is completely delocalized when $\abs{\beta_{\vect{e}_n}} = 1$ for all $n$. In the less pathological case where $\abs{\beta_{\vect{e}_n}} \neq 1$ for some $n$, $\gamma$  grows exponentially in the direction $\sigma_n \vect{e}_n$ with $\sigma_n = \pm 1$ defined such that $\abs{\beta_{\sigma_n \vect{e}_n}} > 1$ (or, equivalently, such that $\abs{\beta_{-\sigma_n \vect{e}_n}} = \abs{\beta_{\sigma_n \vect{e}_n}}^{-1} < 1$). In that case, $\gamma$ can be normalized if and only if the system is finite in the direction of ``steepest'' exponential increase, i.e., as long as the edge of the system is perpendicular to the direction defined by the unit vector $\vect{e}_\perp \propto \sum_{n = 1}^d \log{(\abs{\beta_{\sigma_n \vect{e}_n}})} \sigma_n \vect{e}_n$, as shown in figure~\ref{fig:edgePhysics}. The corresponding Majorana zero-damping mode $\gamma$ is then exponentially localized along that edge, decaying in every direction $-\sigma_n \vect{e}_n$ away from the edge (into the bulk) on a characteristic length scale $\xi_n = 1 / \log{(\abs{\beta_{\sigma_n \vect{e}_n}})}$.

In general, the existence of Majorana zero-damping modes $\gamma$ of the form~\eref{eq:generalSolutionZeroMode} can be assessed from the knowledge of the Lindblad operators by looking for solutions of the equation
\begin{eqnarray} \label{eq:generalConditionZeroMode}
    0 & = & \sum_{j \in \mathcal{I}(i)} \left[ \mathrm{e}^{-\rmi \phi_i} u_{j-i} (\beta_{j-i} - \beta_{\overline{j}-i}) + v_{j-i} (\beta_{j-i} + \beta_{\overline{j}-i}) \right]
\end{eqnarray}
which can be expressed solely in terms of $\{ \beta_{\vect{e}_n} \}_{n = 1, 2, \ldots, d}$ using the relations
\begin{eqnarray*}
    \text{(i)} & \quad & \beta_{k-i} = \beta_{k-j} \beta_{j-i} \quad \text{(for any triple of indices $(i, j, k)$)}; \\
    \text{(ii)} & \quad & \beta_{i-i} \equiv \beta_0 = 1,
\end{eqnarray*}
where $u_{j-i}$ and $v_{j-i}$ are fixed coefficients defining the Lindblad operators (see equation~\eref{eq:translationInvariantLindbladOp}) and $(j, \overline{j})$ denotes pairs of sites located symmetrically around the center of symmetry $\mathcal{S}_i$ of the Lindblad operators. If there exists a set $\{ \beta_{\vect{e}_n} \}_{n = 1, 2, \ldots, d}$ of real factors satisfying equation~\eref{eq:generalConditionZeroMode} for some value of the phase $\phi_i$, the system supports at least a single Majorana zero-damping modes of the form~\eref{eq:generalSolutionZeroMode}. We remark that the phase $\phi_i$ has a direct physical meaning in 2D systems where time-reversal symmetry is broken (symmetry class D), where it is simply defined by the orientation of the edge. We refer to~\ref{app:explicit} for further details as well as for an explicit derivation of the above results.

\subsection{Robustness of Majorana zero-damping modes and role of topology}
\label{sec:robustnessMZMs}

We now investigate the robustness of Majorana zero-damping modes against weak quasi-local \emph{dissipative} perturbations in the general case where the steady state is not necessarily pure but is characterized by a finite bulk purity gap, so that its topological nature is well-defined. We will assume that the dissipative dynamics giving rise to such modes consists of $n$ independent dissipative processes (described by Lindblad operators $L_i$ ($i = 1, 2, \ldots, n$)) and will distinguish two types of perturbations: (i) weak quasi-local perturbations affecting each of these $n$ dissipative processes and (ii) perturbations that introduce additional weak quasi-local dissipative processes (or Lindblad operators). Distinguishing spurious from genuine Majorana zero-damping modes as in section~\ref{sec:dissipativeGap}, we will demonstrate that topology does not guarantee the robustness of genuine Majorana zero-damping modes against perturbations, although it is crucial to be able to isolate such modes spatially. Robustness may additionally be ensured by geometrical properties of the system imposed through quantum reservoir engineering.

Majorana modes are mathematically defined in the mode space $\mathcal{M} \cong \mathbb{R}^{2N}$ consisting of real linear combinations of local Majorana basis operators $c_j$ ($j = 1, 2, \ldots, 2N$), where $N$ is the number of lattice sites in the system. In order for a particular Majorana mode $\gamma = \gamma^\dagger$ to be an exact zero-damping mode of the dissipative dynamics, one must have $\{ L_i, \gamma \} = \{ L^\dagger_i, \gamma \} = 0$ for all $i$ (see section~\ref{sec:dissipativeGap}), which translates, in mode space, as two independent conditions $\{ L_i + L^\dagger_i, \gamma \} = 0$ and $\{ L_i - L^\dagger_i, \gamma \} = 0$ for each $i$. Every Lindblad operator therefore imposes \emph{two} independent constraints for Majorana zero-damping modes in $\mathbb{R}^{2N}$ as long as it is neither Hermitian nor antihermitian, which can safely be assumed, e.g., in the presence of disorder. Consequently, the dynamics generated by $n < N$ dissipative processes necessarily gives rise to $2(N - n)$ \emph{exact} Majorana zero-damping modes, independently of the form---quasi-local or not, with or without specific symmetries---of the corresponding Lindblad operators, as illustrated in figure~\ref{fig:edgePhysics}. We emphasize, however, that such modes which trivially do not take part in the dynamics need not be spatially localized and isolated from each other (as is required for them to exhibit non-Abelian exchange statistics; see section~\ref{sec:nonAbelianStatistics}). As embedded in the dissipative bulk-edge correspondence, it is the topological nature of the bulk that plays a crucial role in isolating Majorana zero-damping modes away from each other, allowing to ``fractionalize'' the fermionic degrees of freedom of the system. Robust isolated Majorana zero-damping modes may therefore arise from the interplay between topology and the ``incomplete'' nature of the dissipative dynamics (i.e., $n < N$), as we will demonstrate through explicit examples in section~\ref{sec:2D} below. Robustness in that case stems from the fact that the number $n$ of dissipative processes (or Lindblad operators) taking part in the dynamics is a well-controlled quantity in typical physical implementations. Intuitively, this can be understood from the fact that driven-dissipative processes considered here require a finite (typically large) amount of energy to occur and therefore do not arise unless they are deliberately introduced via quantum reservoir engineering (see section~\ref{sec:physicalRealization}). In light of this physical constraint, dissipative perturbations of the type (i) discussed above are the most relevant ones to consider in assessing the robustness of Majorana zero-damping modes. We will focus on the latter.

We remark that there is \emph{a priori} no guarantee to find Majorana zero-damping modes in the case where the system is driven by $n \geq N$ Lindblad operators---independently of the topological nature of the bulk---since, as discussed above, every generic dissipative process introduced in the dynamics leads to two additional constraints in mode space and further reduces the subspace available for Majorana zero-damping modes. Whereas spurious Majorana zero-damping modes can generally be removed by introducing additional dissipative processes---leaving no trace in the purity or damping spectra---genuine Majorana zero-damping modes, in contrast, must either survive or give rise to intrinsic Majorana zero-purity modes as dictated by the dissipative bulk-edge correspondence. Which of these two outcomes actually occurs cannot be determined from topological properties but rather depends on the dissipative boundary conditions, i.e., on the specific form of the edge dissipative dynamics resulting for the combined properties of bulk engineered dissipative dynamics and the physical edge delimiting it spatially. As we will demonstrate using explicit examples in sections~\ref{sec:1D},\ref{sec:2D} below, the dissipative dynamics can be constrained by e.g. geometric properties of the edge in such a way that the purity gap closes in a robust, controlled manner at that edge, thereby giving rise to an odd number of Majorana zero-damping modes in a phase whose bulk topology would naively imply the existence of an even number of such modes. Interesting phenomena with no Hamiltonian counterparts may therefore arise from Liouvillian dynamics as far as the edge physics is concerned, with intriguing effects such as the transformation of Abelian vortices into non-Abelian ones (see section~\ref{sec:2D}).

\section{Non-Abelian statistics of dissipative Majorana modes}
\label{sec:nonAbelianStatistics}

A key property of Majorana modes in the Hamiltonian context is their non-Abelian statistics under spatial exchange. An important question is therefore whether or not this property is also present in the case of dissipative Majorana modes (more precisely, of Majorana zero-damping modes). Here we provide a simple affirmative argument following reference~\cite{Diehl11}: Physically, this may be understood from the fact that Majorana modes in the context of topological insulators or superfluids are not dynamical, but are purely static degrees of freedom. The actual bulk dynamics is irrelevant; it can be either unitary or dissipative.

In order to verify non-Abelian exchange statistics, we first explain in more detail how, in a dissipative setting, the edge subspace of the density matrix---containing the Majorana modes---is isolated from the bulk subspace of the latter. We then consider the effect of adiabatic parameter changes of the Liouville operator or, more generally, of the operator generating the system dynamics.

\subsection{Dissipative isolation of the edge mode subspace}
\label{sec:dissipativeIsolation}

As argued in sections \ref{sec:purityGap}, \ref{sec:dissipativeGap} and \ref{sec:edgePhysics}, an important prerequisite for the presence of stable dissipative Majorana modes is the existence of a zero-mode subspace as well as its isolation from the bulk. Here we formulate the conditions for such a situation in a general framework that is valid beyond the quadratic setting introduced in section~\ref{sec:interactingLiouvillians}.  To this end, we introduce projectors $p$ and $q =1-p$ on the edge and bulk subspaces, respectively. A decoupled edge subspace then appears if the Lindblad operators $\ell_i$ are block diagonal in this projection, i.e., $\ell_{i,pq}=\ell_{i,qp} =0$, with an edge block identical to zero, i.e., $\ell_{i,pp}=0$. The dissipative evolution then reads
 \begin{eqnarray}\label{GenEq}
\hspace{-0.4cm} \partial_t \left(   \begin{array}{cc}
 \rho_{pp} & \rho_{pq} \\
 \rho_{qp} & \rho_{qq} 
 \end{array} 
 \right) =    \hspace{-0.05cm}
\left( \begin{array}{cc}
0  & - \frac{1}{2} \rho_{pq}  \sum_i  \ell^\dag_{i, qq} \ell_{i, qq} \\
- \frac{1}{2}  \sum_i  \ell^\dag_{i, qq} \ell_{i, qq} \rho_{qp} &    \mathcal L_{ qq}[\rho_{qq}]
 \end{array}
 \right).\nonumber\\
 \end{eqnarray}
Here, the bulk dissipative evolution $\mathcal L_{ qq} [\rho_{qq}]$ has a Lindblad form and $\rho_{pp}$ is a constant of motion, by construction. Crucially, the density matrix elements $\rho_{qp}$ and $\rho_{pq}$ coupling these sectors damp out according to $\rho_{qp} = e^{- \sum_i \ell_{i,qq}^\dag \ell_{i,qq} t} \rho_{qp} (t=0)$. In the presence of a dissipative gap, this process is exponentially fast. In steady state, the density matrix then takes a factorized form $\rho = \rho_{\text{edge}} \otimes \rho_{\text{bulk}}$, with $\rho_{\text{edge}} = \rho_{pp}$ and $\rho_{\text{bulk}} = \rho_{qq}$.
Clearly, the absence of dynamics in the decoherence-free subspace does not allow for a controlled initialization of its occupation. This property is in complete analogy with a Hamiltonian ground-state scenario, where the zero-energy edge subspace is decoupled from Hamiltonian dynamics. The preparation and detection ideas developed in \cite{Kraus12} for the ground states of fermionic atoms can be directly applied to the dissipative setting. 

Moving from a second- to a first-quantized representation (for a quadratic setting and using a Majorana basis as in section~\ref{sec:gaussianMasterEquations}), we obtain the evolution
\begin{eqnarray}\label{specevol}
\partial_t \left(\hspace{-0.1cm}\begin{array}{cc}
\Gamma_{pp}  & \Gamma_{pq}\\
\Gamma_{qp} & \Gamma_{pp} 
\end{array} \hspace{-0.1cm}\right)\hspace{-0.1cm}
&=&\hspace{-0.1cm} \left( \hspace{-0.1cm}\begin{array}{cc}
0 & - \Gamma_{pq} X_{qq}   \\
- X_{qq} \Gamma_{qp}  & - \{X_{qq} , \Gamma_{qq}\} +  Y_{qq} 
\end{array}\hspace{-0.1cm} \right),
\end{eqnarray}
showing explicitly that the fadeout of bulk-edge coherences is indeed controlled by the dissipative gap, i.e., by the slowest rate in $X_{qq}$.

\subsection{Adiabatic parameter changes and braiding}
\label{sec:braiding}

We consider a steady state with Majorana zero-damping modes whose corresponding eigenvectors $| \alpha \rangle$ span a non-local decoherence-free subspace described by the density matrix elements $(\rho_{\text{edge}} )_{\alpha\beta} = \langle \alpha | \rho | \beta \rangle  =\rho_{\alpha\beta}$. The decoherence-free subspace has the property $\dot{\rho}_{\alpha\beta}= 0$.
We now study the time evolution of the density matrix in a co-moving basis $|a(t) \rangle= U(t) | a(0) \rangle$ that follows the decoherence-free subspace of the edge modes, i.e., that preserves the property $\dot{\rho}_{\alpha\beta}= 0$. Without specifying the actual dynamics generating the physical evolution, the time evolution in that basis is given by
\begin{eqnarray}
\partial_t \rho &= \sum_{a,b} \left( |\dot {a} \rangle \rho_{ab} \langle b| +  | a \rangle \dot{\rho}_{ab} \langle b| +  | a \rangle \rho_{ab} \langle \dot{b} |  \right) \nonumber \\
&= \sum_{a,b,c} |c\rangle \langle c |\dot {a} \rangle \rho_{ab} \langle b|  +  \sum_{a,b}  | a \rangle \dot{\rho}_{ab} \langle b| + \sum_{a,b,c}  | a \rangle \rho_{ab} \langle \dot{b} | c \rangle \langle c| .
\end{eqnarray}
We define the vector potential $A_{ab} = \langle a(0) | U^{\dagger} \dot{U} | b(0) \rangle =\frac{ \mathrm{i} }{2}(\langle a | \dot{b} \rangle - \langle \dot{a} | b \rangle)  $, which is antisymmetric and Hermitian, by construction. Taking into account the normalization $\partial_{t} \langle b(t) |a(t) \rangle= 0$, we then obtain
\begin{equation}\label{eq:adiab}
\fl \qquad\qquad \partial_t \rho= - i [A , \rho] + \sum_{a,b} |a\rangle\dot{\rho}_{ab} \langle b | =-i [H+A,\rho] +  \sum_i (\ell_i \rho \ell_i^{\dagger} - \frac{1}{2} \{ \ell_i^{\dagger} \ell_i , \rho \}) ,
\end{equation}
where $\dot{\rho}_{ab} \equiv \langle a(t) | \partial_{t} \rho| b(t) \rangle$ is the time evolution in the instantaneous basis. The Heisenberg commutator clearly reflects the appearance of a gauge structure \cite{Berry84,Simon83,WilczekZee83,Carollo2003a,Pachos99} in the density matrix formalism. Crucially, this structure emerges \emph{independently} of whether the physical dynamics encoded in $\dot{\rho}_{ab}$---inserted in the last inequality---is unitary or dissipative.

The transformation applied in the zero-mode subspace of either a Hamiltonian or a Liouvillian starting from the initial condition $\rho_{\alpha\beta}(0)$ is given by $\rho_{\alpha\beta}(t) = (  V(t) \rho(0) V(t)^{\dagger})_{\alpha\beta}$, with time-ordered $V(t) = T\exp{(-\mathrm{i }\int^{t}_{0} d\tau A(\tau))  } $ where $A(t)_{\alpha\beta}$ is the projection of the vector potential onto the decoherence-free subspace. The adiabatic change of the parameters is the key feature for such a state transformations to be realized while preserving the zero-mode subspace. Here one crucially requires the ratio between the rate of parameter changes encoded in $A$ and the dissipative gap to be very small. Due to this separation of time scales---enabled by the non-evolving subspace---the decoherence-free subspace is never left. This phenomenon is sometimes referred to as the Quantum Zeno effect \cite{Beige00}.

In the first-quantized formulation, the eigenvalues and eigenvectors of $X$ now depend on time. The transformation into the instantaneous basis $|a(t)\rangle= U(t) |a_{0}\rangle$, where $|a_{0}\rangle$ is the initial reference basis, is now orthogonal. The vector potential is given by the Hermitian antisymmetric matrix $A_{ab} = \frac{\mathrm i}{2}  ( \langle a(t) | \dot{b}(t) \rangle - \langle \dot{a}(t) | b(t)  \rangle )$, and the correlation matrix evolves in this basis according to ($h$ is defined above equation \eref{eq:gammah})
\begin{equation}
\partial_t \Gamma  =  [ h +  A ,  \Gamma  ] -  \{ X ,  \Gamma  \} +   Y.
\end{equation}

Starting form this basic understanding, one can construct adiabatic local parameter changes of the Lindblad operators to perform elementary dissipative Majorana moves~\cite{Diehl11}. Such procedure can then be applied sequentially in order to exchange two particular modes, keeping them sufficiently far apart from each other during the process (in networks of 1D wires, such exchange can be made using a T-junction geometry~\cite{Alicea11}). The unitary braiding matrix describing the complete process is $B_{ij}=\exp{\left( \frac{\pi}{4} \gamma_{i} \gamma_{j} \right)}$ for two Majorana modes $i,j$. Since $\left[ B_{ij}, B_{jk} \right] \neq 0$ for $i \neq j$, non-Abelian statistics is obtained.

\section{Illustrative examples in 1D}
\label{sec:1D}

In this section, we apply our results to several dissipative models in one dimension. In particular, we investigate the interplay between criticality and purity at the onset of non-equilibrium topological phase transitions, exemplifying the results of section~\ref{sec:phenomenology}. We focus on 1D lattice systems of spinless fermions with translational symmetry evolving under a dissipative dynamics described by Lindblad operators $L_n = \sum_m v_{n-m} a^\dagger_m + u_{n-m} a_m$, where $a^\dagger_m$ and $a_m$ are creation and annihilation operators associated with a lattice site $m$ and $v_{n-m}$ and $u_{n-m}$ are translation-invariant functions. The system is then most conveniently described in momentum space with Fourier-transformed Lindblad operators $L_k = \sum_m \rme^{\rmi k m} L_m$ of the form
\begin{equation}
    L_k = v_k a^\dagger_k - u_k a_{-k}.
\end{equation}
Here we assume that
\begin{eqnarray} \label{eq:vkukInterest}
    \text{$v^*_k = v_{-k}$ and $u^*_k = u_{-k}$},
\end{eqnarray}
as will be satisfied in the examples examined below. The dissipative dynamics obviously occurs in decoupled momentum sectors corresponding to the fermionic modes $a^\dagger_k$ and $a_{-k}$. Defining a corresponding basis of Majorana operators $(c_{2k-1}, c_{2k}, c_{2(-k)-1}, c_{2(-k)})$, the matrices $X_k$ and $Y_k$ describing the dynamics in each of these sectors take the following form (see section~\ref{sec:gaussianMasterEquations}):
\begin{eqnarray} \label{eq:Xmatrix}
    X_k & = 2 \left( \begin{array}{cc}
        (\abs{u_k}^2 + \abs{v_k}^2) \mathbb{I}_2 & 2 \, \mathrm{Re}(u^*_k v_k) \sigma_z \\
        2 \, \mathrm{Re}(u^*_k v_k) \sigma_z & (\abs{u_k}^2 + \abs{v_k}^2 ) \mathbb{I}_2
    \end{array} \right), \\
    Y_k & = 4 \left( \begin{array}{cc}
        (\abs{u_k}^2 + \abs{v_k}^2) (-\rmi \sigma_y) & 2 \, \mathrm{Im}(u_k^* v_k) \sigma_z \\
        -2 \, \mathrm{Im}(u^*_k v_k) \sigma_z & (\abs{u_k}^2 + \abs{v_k}^2)(-\rmi \sigma_y)
    \end{array} \right), \label{eq:Ymatrix}
\end{eqnarray}
where $\sigma_{\mu = x, y, z}$ denotes the usual Pauli matrices. The system exhibits critical behavior if at least one of the eigenvalues $\lambda_k^{(1,2)} = \pm \abs{u^*_k \pm v_k}^2$ of $X_k$ vanishes for some $k$, i.e., if the dissipative gap closes. Note that in that case the matrix $Y_k$ identically vanishes, since its eigenvalues are given by $\pm (\abs{u^*_k + v_k})(\abs{u^*_k - v_k})$.

All matrices $X_k$ of the form~\eref{eq:Xmatrix} can be diagonalized through the same unitary transformation. This allows us to obtain the steady-state correlation matrix in a generic form (solving the steady-state equation $\{ X_k, \Gamma_k \} = Y_k$; see section~\ref{sec:gaussianMasterEquations}); namely,
\begin{equation}
    \Gamma_k = \frac{1}{\abs{v_k}^2 + \abs{u_k}^2} \left( \begin{array}{cc}
        (\abs{v_k}^2 - \abs{u_k}^2) \rmi \sigma_y & 2 \, \mathrm{Im}(u^*_k v_k) \sigma_z \\
        -2 \, \mathrm{Im}(u^*_k v_k) \sigma_z & (\abs{v_k}^2 - \abs{u_k}^2) \rmi \sigma_y
    \end{array} \right). \label{eq:Cmatrix}
\end{equation}
The symmetry class to which the steady state belongs can be readily determined from this expression. We recall that one can associate a fictitious free-fermion Hamiltonian $H_\Gamma = \rmi \sum_{i, j} \Gamma_{ij} c_i c_j$ with the steady-state correlation matrix $\Gamma$ (see section~\ref{sec:topClassification}). Here we write $H_\Gamma = 2 \sum_k \Psi^\dagger_k \mathcal{H}_k \Psi_k$ in a momentum-space Nambu representation with $\Psi^\dagger_k = (a^\dagger_k, a_{-k})$ and obtain
\begin{equation}
    \mathcal{H}_k \equiv \left( \begin{array}{cc}
        \xi_k & \Delta_k \\
        \Delta^*_k & -\xi_{-k}
    \end{array}\right)
    = \frac{1}{N^2_k} \left(\begin{array}{cc}
        \abs{v_k}^2 - \abs{u_k}^2 & 2 \rmi \, \mathrm{Im}(u^*_k v_k) \\
        -2 \rmi \, \mathrm{Im}(u^*_k v_k) & \abs{u_k}^2 - \abs{v_k}^2
    \end{array}\right),
\end{equation}
where $N^2_k = \abs{v_k}^2 + \abs{u_k}^2$. Equation~\eref{eq:vkukInterest} implies that $\xi_{-k} = \xi_k$ and $\Delta^*_{-k} = \Delta_k$, and one can easily verify that time-reversal and particle-hole symmetries are satisfied as $\mathcal{H}^*_{-k} = +\mathcal{H}_k$ and $\sigma_x \mathcal{H}^*_{-k} \sigma_x = - \mathcal{H}_k$, respectively (see equations~\eref{eq:PHS} and~\eref{eq:TRS}). Consequently, all (Gaussian) steady states resulting from a translation-invariant dissipative dynamics described by matrices $X_k$ and $Y_k$ of the form~\eref{eq:Xmatrix} and~\eref{eq:Ymatrix} have chiral symmetry and accordingly belong to the symmetry class BDI of Altland and Zirnbauer.

In what follows, we present three examples of dissipative systems belonging to the symmetry class BDI exhibiting a topological phase transition due to the change of some external parameter $\kappa$. We first consider an example where the dissipative dynamics gives rise to a pure steady state in the whole parameter range of $\kappa$ and the system exhibits critical behavior at the point where the topological phase transition occurs. We then discuss an example in which the topological phase transition is also driven by criticality but accompanied by the closure of the purity gap. Finally, we provide a third example in which the topological phase transition does not lead to any critical behavior. The three examples presented below therefore illustrate the three possibilities for non-equilibrium topological phase transitions identified in section~\ref{sec:phenomenology}.

We illustrate our results by plotting, as a function of $\kappa$, the dissipative gap $\Delta_\text{d}$ (given by the minimum eigenvalue of $X$) and the purity gap which we define here as $\Delta_{\text{p}} = \min_k \{ \Tr(\Gamma_k^2)/4 \}$. We also consider finite systems with open boundary conditions but translation-invariant Lindblad operators and investigate the behavior of the edge zero-damping modes in the vicinity of the topological phase transition.

\subsection{Example 1: topological phase transition of a pure state with criticality}

We consider translation-invariant Lindblad operators $L_n$ that act on three consecutive sites:
\begin{equation} \label{eq:Chern}
    L_n = \frac{1}{\sqrt{4 + \kappa^2}} \left[ \kappa a^\dagger_n + (a^\dagger_{n+1} + a^\dagger_{n-1}) + (a_{n+1} - a_{n-1}) \right],
\end{equation}
where $\kappa$ is a real parameter controlling the coupling to the central site $n$~\footnote{Note that the overall multiplicative factor is introduced for convenience and does not affect the physical properties of the system.}. In momentum space, these Lindblad operators take the form $L_k = v_k a^\dagger_k - u_k a_{-k}$ with $v_k = \kappa + 2 \cos k$ and $u_k = -2 \rmi \sin k$ (note that equation~\eref{eq:vkukInterest} is satisfied). The corresponding matrix $X_k$ is diagonal, namely, $X_k =  (8 + 2 \kappa^2 + 8 \kappa \cos k)/(4 + \kappa^2) \mathbb{I}$, and one can readily see that the dissipative gap closes at $\kappa = \pm 2$, giving rise to critical behavior around these points. The steady-state correlation matrix can be written explicitly as
\begin{eqnarray}
    \fl \qquad\qquad \Gamma_k & = g_k \left(\begin{array}{cc}
        (\kappa^2 + 4 \kappa \cos k + 4 \cos 2k) \rmi \sigma_y & 4(\sin k + \kappa \sin 2k) \sigma_z \\
        -4(\sin k + \kappa \sin 2k) \sigma_z & (\kappa^2 + 4 \kappa \cos k + 4 \cos 2k) \rmi \sigma_y
    \end{array}\right),
\end{eqnarray}
with $g_k = 1/(4 + \kappa^2 + 4 \kappa \cos k)$ and one can easily verify that $\Gamma^2_k = -\mathbb{I}$ for all $k$, meaning that the steady state is always pure.

\begin{figure}[t]
    \begin{center}
        \includegraphics[width=\columnwidth]{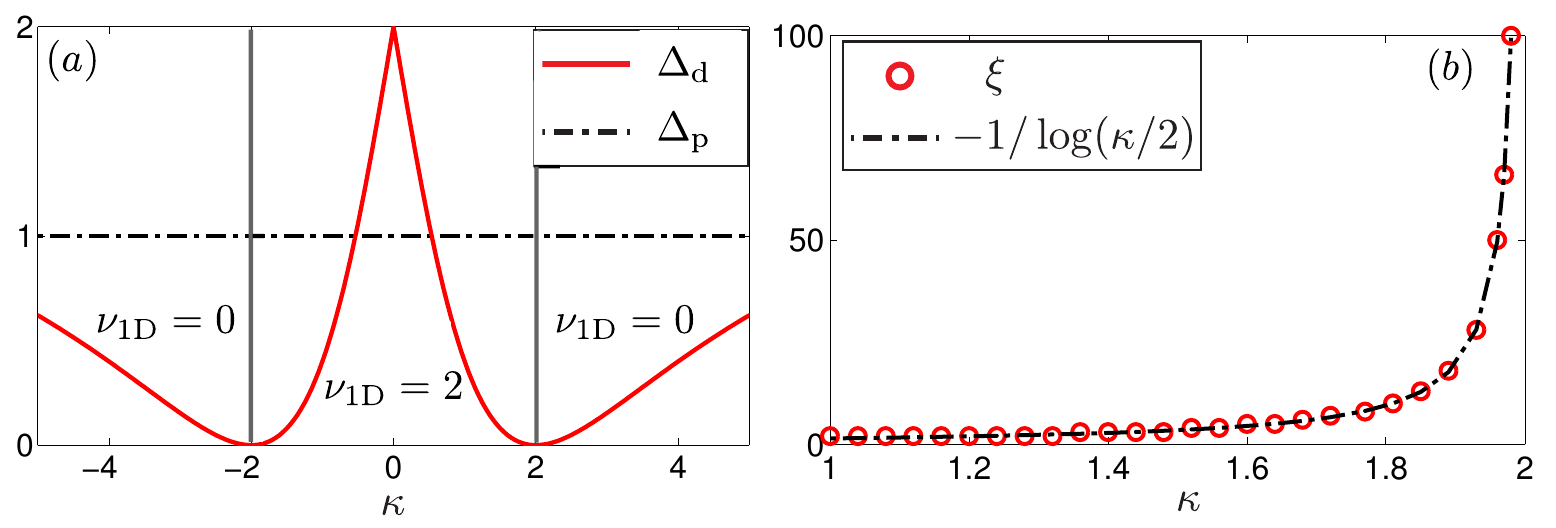}
        \caption{Topological phase transition driven by criticality. (a) The system driven by the translation-invariant Lindblad operators of equation~\eref{eq:Chern} exhibits critical behavior at $\kappa = \pm 2$, as indicated by the closure of the dissipative gap $\Delta_\text{d}$. At these points, the system undergoes a transition from a topologically trivial ($\nu_\text{1D} = 0$) to a topologically ordered ($\nu_\text{1D} = 2$) phase. Since the purity gap $\Delta_\text{p} = 1$, the steady state of the dissipative dynamics is always pure. (b) Divergence of the localization length associated with one of the two Majorana zero-damping modes found at the edges upon approaching the topological phase transition at $\kappa = 2$.}
        \label{fig:1DExample1}
    \end{center}
\end{figure}

Expressing the momentum-space correlation matrix in the form $\Gamma_k = \rmi (\vect{n}_k \cdot \boldsymbol{\sigma})$ with $\vect{n}_k \in \mathbb{R}^3$ similarly as in section~\ref{sec:topPropertiesBulk}, we obtain $\vect{n}_k = g_k (0, 4 \kappa (\sin k + \sin 2 k), \kappa^2 + 4 \kappa \cos k + 4 \cos 2k)$. As expected, this vector $\vect{n}_k$ is non-zero for all $k$ in the whole parameter range of $\kappa$ where the dissipative gap is finite, i.e., for $\abs{\kappa} \neq 2$. In that case one can ``spectrally flatten'' the steady-state correlation matrix $\Gamma_k$ by normalizing the vector $\vect{n}_k$ to unity in order to calculate the corresponding winding number topological invariant $\nu_\text{1D}$ (see equation~\eref{eq:windingNumber} and discussion above). Here we find $\nu_\text{1D} = 2$ for $\abs{\kappa} < 2$ and $\nu_\text{1D} = 0$ for $\abs{\kappa} > 2$ (for an explicit calculation, see our previous work~\cite{Bardyn12}). We therefore identify a topological phase transition at $\abs{\kappa} = 2$ where the dissipative gap closes and the system becomes critical. These results are illustrated in figure~\ref{fig:1DExample1}(a).

We now examine the case of a finite system (or ``chain'') of $N$ sites with open boundary conditions. We first notice that only $N-2$ three-site Lindblad operators of the form~\eref{eq:Chern} can be ``placed'' on such an open chain. Hence, according to our discussion of section~\ref{sec:robustnessMZMs}, four exact Majorana zero-damping modes (two at each edge, by symmetry) must be present independently of the topological properties of the system (in particular, for any value of the parameter $\kappa$). Obviously, this number which remains constant across the topological phase transition cannot be used to identify the latter. However, the topological phase transition can be revealed in two other ways: (i) One can remove spurious Majorana zero-damping modes with no topological origin by introducing additional arbitrary Lindblad operators acting at the edges; assuming that these extra Lindblad operators preserve chiral symmetry~\footnote{Note that time-reversal symmetry only must be ensured since particle-hole symmetry is automatically satisfied in our dissipative framework.}, the system must then exhibit a total of $\nu_\text{1D} = 2$ Majorana zero-damping and/or zero-purity modes in the parameter range $\abs{\kappa} < 2$, as dictated by the dissipative bulk-edge correspondence (see section~\ref{sec:bulkEdge}). (ii) One can examine the behavior of the localization length associated with the Majorana zero-damping modes as the parameter $\kappa$ is tuned across the topological phase transition; owing to criticality, the characteristic length scale associated with at least one of the two modes found at an edge must diverge at the phase transition point $\abs{\kappa} = 2$.

Here we follow strategy (ii): Since the Lindblad operators $L_n$ are translation-invariant and have a symmetric form around their central site $n$~\footnote{Namely, the creation and annihilations parts of $L_n$ (see equation~\eref{eq:Chern}) have $s$-wave and $p$-wave symmetry, respectively.}, we use the results of section~\ref{sec:generalFormMZMs} and assess the existence of Majorana zero-damping modes that decay into the bulk by solving equation~\eref{eq:generalConditionZeroMode}. Choosing $\phi_i = 0$, the latter equation takes the explicit form $0 = \kappa + 2 \beta_{\vect{e}_x}$ ($\vect{e}_x$ being a unit vector joining neighboring sites of the chain) and we readily obtain $\beta_{\vect{e}_x} = -\kappa/2$, showing that a least one of the two Majorana zero-damping modes at the edge must decay (exponentially) into the bulk when $\abs{\kappa} < 2$ (as required by $\abs{\beta_{\vect{e}_x}} < 1$) on a characteristic length scale $\xi = -1 / \log{(\abs{\beta_{\vect{e}_x}})} = -1 / \log{(\abs{\kappa}/2)}$. As expected, $\xi$ diverges at $\abs{\kappa} = 2$ and thus reveals the topological phase transition. Figure~\ref{fig:1DExample1}(b) illustrates numerical results corroborating this behavior.

\subsection{Examples 2 and 3: topological phase transition of a mixed state with and without criticality}

\begin{figure}[t]
    \begin{center}
        \includegraphics[width=\columnwidth]{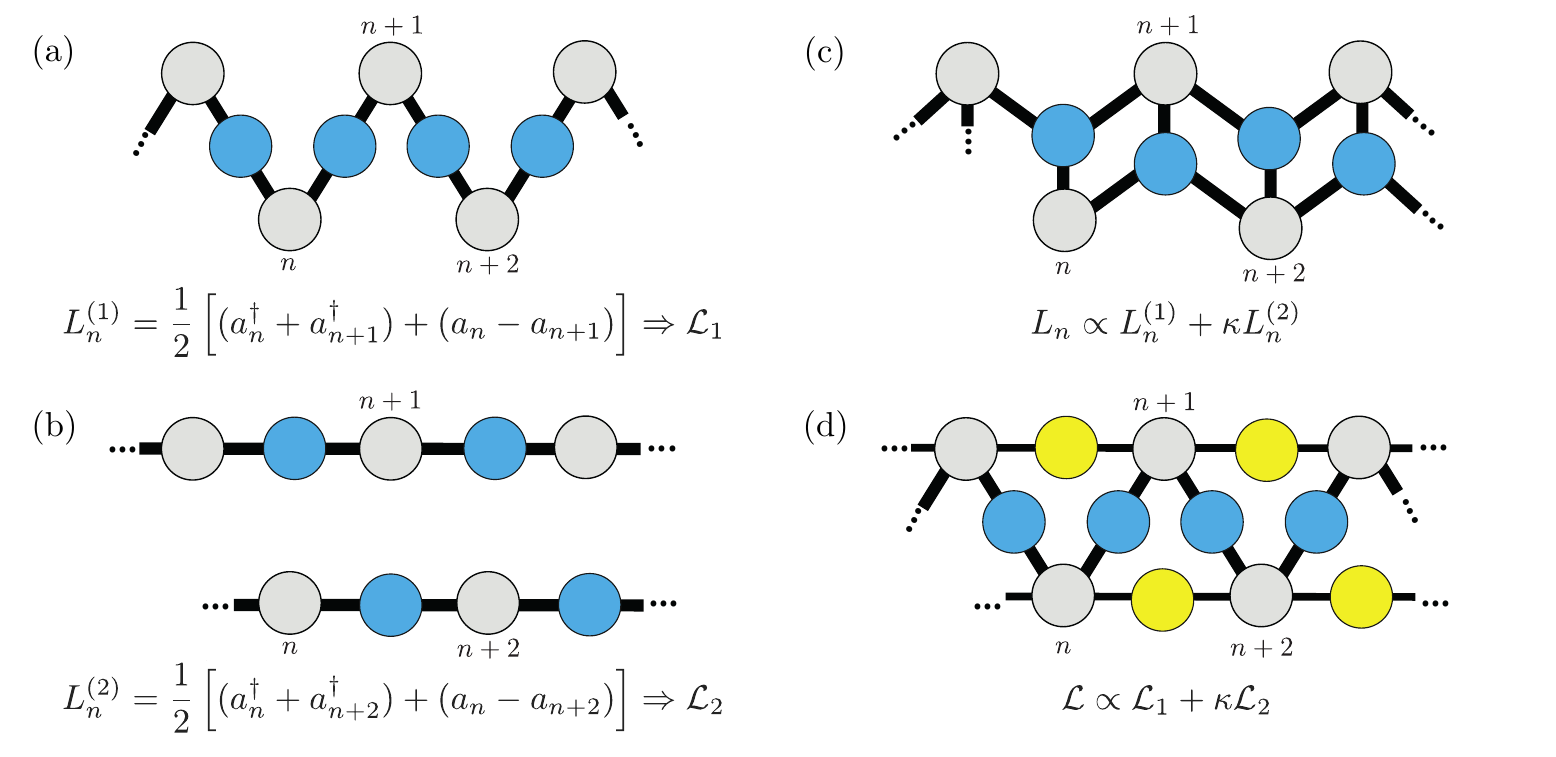}
        \caption{Generalizations of the dissipative Kitaev chain. By positioning the sites of the 1D chain (grey disks) in a zigzag geometry, three different types of dissipative processes can be engineered (see section~\ref{sec:physicalRealization}) from auxiliary sites (blue circles): (a) Lindblad operators $L^{(1)}_n$ driving the system into the ground state of the Kitaev chain. (b) Lindblad operators $L^{(2)}_n$ driving the system into two decoupled Kitaev chains. (c) By moving the auxiliary sites relative to the physical ones in the zigzag geometry, one can engineer a coherent superposition $L_n \propto L^{(1)}_n + \kappa L^{(2)}_n$ of the Lindblad operators $L^{(1)}_n$ and $L^{(2)}_n$. (d) Furthermore, by doubling the number of auxiliary sites (introducing the yellow ones), one can engineer a dissipative dynamics $\mathcal{L} \propto \mathcal{L}_1 + \kappa \mathcal{L}_2$ corresponding to two ``competing'' Liouvillians $\mathcal{L}_1$ and $\mathcal{L}_2$ associated with Lindblad operators $L^{(1)}_n$ and $L^{(2)}_n$, respectively.}
        \label{fig:zigzagGeometry}
    \end{center}
\end{figure}

We now consider a dissipative dynamics that involves two types of translation-invariant Lindblad operators:
\begin{eqnarray}
    L^{(1)}_n = \frac{1}{2} \left[ (a^\dagger_n + a^\dagger_{n+1}) + (a_n - a_{n+1}) \right], \\
    L^{(2)}_n = \frac{1}{2} \left[ (a^\dagger_n + a^\dagger_{n+2}) + (a_n - a_{n+2}) \right],
\end{eqnarray}
acting on neighboring and next-to-nearest neighboring sites, respectively. The Lindblad operators $L^{(1)}_n$ generate a dissipative dynamics that drives the system into a pure steady state corresponding to the ground state of a topologically non-trivial Kitaev chain~\cite{Kitaev01}, as demonstrated in our previous work~\cite{Diehl11}; two Majorana zero-damping modes are found in that case (one at each edge). The Lindblad operators $L^{(2)}_n$ describe two decoupled Kitaev chains, with four Majorana zero-damping modes (two at each edge). Below we discuss two scenarios: (i) We first assume that both dissipative processes $L^{(1)}_n$ and $L^{(2)}_n$ occur coherently, so that the relevant Lindblad operators are $L_n = (L^{(1)}_n + \kappa L^{(2)}_n)/(2 (\kappa^2 + \kappa + 1))$ with $\kappa \in \mathbb{R}$. (ii) We then consider the case where both dissipative processes ``compete'' against each other, so that the relevant Liouvillian to describe the dissipative dynamics is $\mathcal{L} = (\mathcal{L}_1 + \kappa \mathcal{L}_2)/(1+ \kappa)$, where $\mathcal{L}_\alpha$ denotes the Liouvillian corresponding to a single Lindblad operator $L^{(\alpha)}_n$ and $\kappa \geq 0$. In practice, $L^{(1)}_n$ and $L^{(2)}_n$ as well as their combinations (i) and (ii) can be realized in a ``zigzag'' geometry, e.g., as depicted in Fig.~\ref{fig:zigzagGeometry}.

\begin{figure}[t]
    \begin{center}
        \includegraphics[width=\columnwidth]{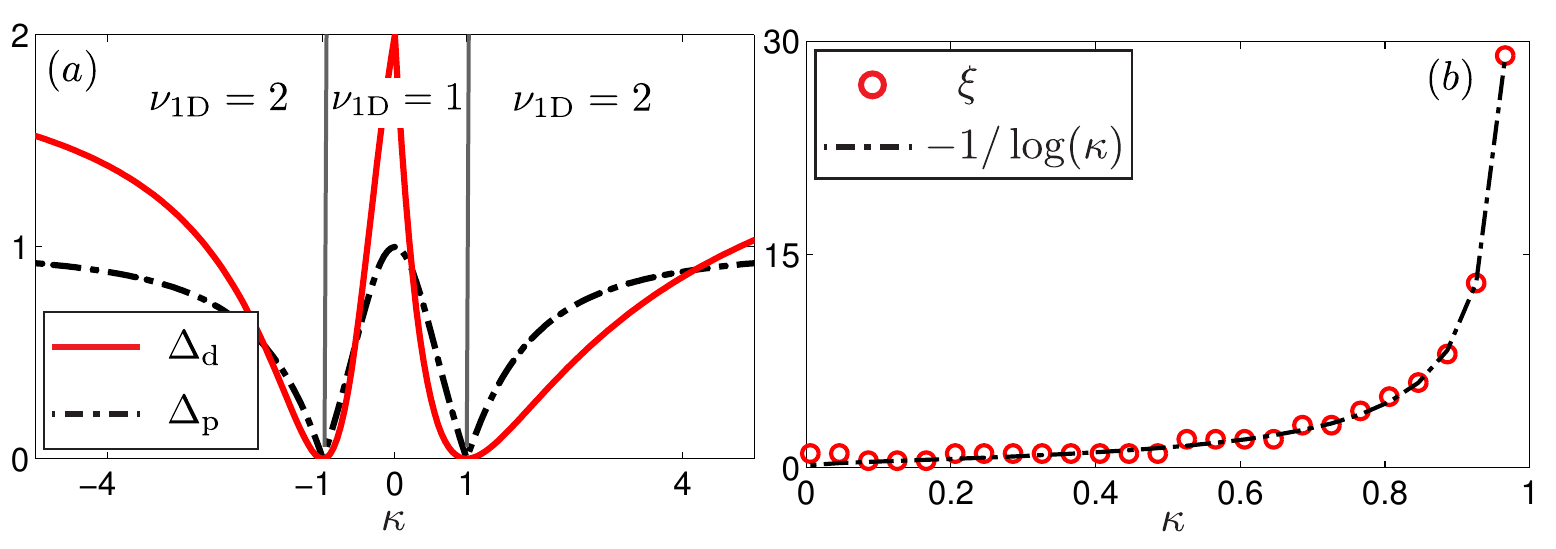}
        \caption{Topological phase transition of a mixed state with criticality. (a) The system driven by translation-invariant Lindblad operators of the form $L_n \propto L^{(1)}_n + \kappa L^{(2)}_n$ exhibits critical behavior at $\kappa = \pm 1$, as indicated by the closure of the dissipative gap $\Delta_\text{d}$. At these points, the system undergoes a topological phase transition from a state characterized by $\nu_\text{1D} = 1$ to a state with $\nu_\text{1D} = 2$. The steady state is mixed (unless $\kappa = 0$ or $\kappa \to \pm \infty$) and the phase transition is crucially accompanied by the closure of the purity gap $\Delta_\text{p}$. (b) Divergence of the localization length associated with the Majorana zero-damping mode found at each of the edges upon approaching the topological phase transition at $\kappa = 1$.}
        \label{fig:1DExample2}
    \end{center}
\end{figure}

In momentum space, the Lindblad operators take the generic form $L^{(\alpha)}_k = v^{(\alpha)}_k a^\dagger_k - u^{(\alpha)}_k a_{-k}$ ($\alpha = 1, 2$) with $v_k = \rme^{\rmi k \alpha/2} \cos(k \alpha/2)$ and $u_k = \rmi \rme^{\rmi k \alpha/2} \sin(k \alpha/2)$ (note that equation~\eref{eq:vkukInterest} is again satisfied). We first examine case (i), in which the eigenvalues of the matrix $X_k$ take the explicit form
\begin{equation}
    \lambda_1 = 2 \frac{(1 + \kappa)^2}{1+ \kappa + \kappa^2 }, \qquad \lambda_2 = 2 \frac{1 + 2 \kappa \cos k + \kappa^2 }{1 + \kappa + \kappa^2},
\end{equation}
showing that the dissipative gap closes at $\kappa = \pm 1$, and the steady-state correlation matrix reads
\begin{eqnarray}
    \Gamma_k = g_k \left( \begin{array}{cc}
        (\cos k + \kappa \cos 2k) \rmi \sigma_y & -(\sin k + \kappa \sin 2k) \sigma_z \\
        (\sin k + \kappa \sin 2k) \sigma_z & (\cos k + \kappa \cos 2k) \rmi \sigma_y
    \end{array} \right)
\end{eqnarray}
with $g_k = (1+ \kappa)/(1 + \kappa + \kappa^2 + \kappa \cos k)$ and eigenvalues $\pm (1 + \kappa) \sqrt{1 + \kappa^2 + 2 \kappa \cos k}/(1 + \kappa + \kappa^2 + \kappa \cos k)$, indicating that the steady state is pure for $\kappa = 0$ and $\kappa \to \pm \infty$ and mixed otherwise (as expected from the fact that $\kappa = 0$ ($\kappa \to \pm \infty$) corresponds to the case where $L^{(1)}_n$ (respectively $L^{(2)}_n$) alone generates the dynamics). The vector $\vect{n}_k$ defined so that $\Gamma_k = \rmi (\vect{n}_k \cdot \boldsymbol{\sigma})$ is non-zero for all $k$ provided that $\kappa \neq \pm 1$ and takes the explicit form $\vect{n}_k = g_k (0, -(\sin k + \kappa \sin 2k), -(\cos k + \kappa \cos 2k))$. The corresponding winding number topological invariant is $\nu_\text{1D} = 1$ for $\abs{\kappa} < 1$ and $\nu_\text{1D} = 2$ for $\abs{\kappa} > 1$, in accordance with the fact that $\nu_\text{1D} = 1$ for a single topologically non-trivial Kitaev chain. These results, which are depicted in figure~\ref{fig:1DExample2}(a), illustrate the possibility of a topological phase transition driven by criticality \emph{and} by the closure of the purity gap.

To conclude our investigation of case (i), we consider a finite chain of $N$ sites with open boundary conditions. Since only $N-2$ Lindblad operators of the form $L_n \propto L^{(1)}_n + \kappa L^{(2)}_n$ can be applied on such a chain, we again obtain four exact Majorana zero-damping modes that are either genuine or spurious. A numerical analysis below the topological phase transition point $\kappa = 1$ reveals that two of these modes (one at each edge) decay exponentially into the bulk on a characteristic length scale $\xi = -1/\log(\kappa)$ which diverges at $\kappa = 1$, as expected (see figure~\ref{fig:1DExample2}(b)).

\begin{figure}[t]
    \begin{center}
        \includegraphics[width=\columnwidth]{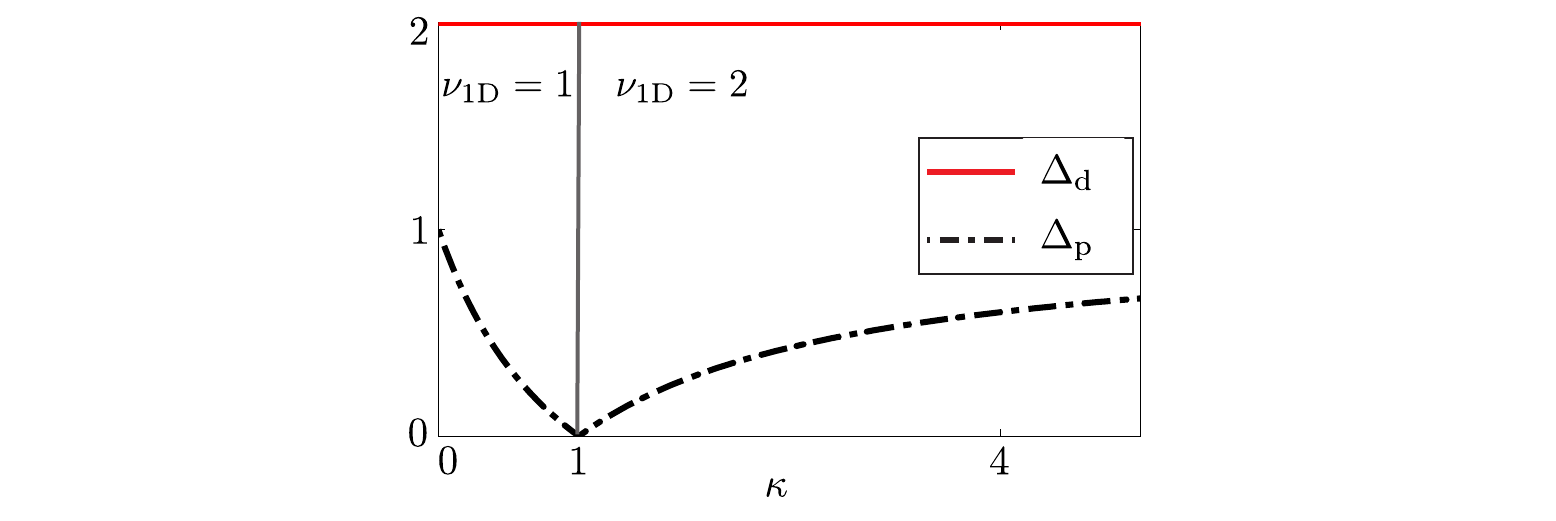}
        \caption{Topological phase transition of a mixed state without criticality. The system evolving under a dissipative dynamics $\mathcal{L} \propto \mathcal{L}_1 + \kappa \mathcal{L}_2$ generated by two ``competing'' Liouvillians $\mathcal{L}_1$ and $\mathcal{L}_2$ corresponding to the Lindblad operators $L^{(1)}_n$ and $L^{(2)}_n$, respectively, exhibits a topological phase transition at $\kappa = 1$ from a state with winding number $\nu_\text{1D} = 1$ to a state with winding number $\nu_\text{1D} = 2$ that results from the closure of the purity gap $\Delta_\text{p}$ only. The system never becomes critical, since the dissipative gap $\Delta_\text{d}$ remains constant in the whole parameter range of $\kappa$.}
        \label{fig:1DExample3}
    \end{center}
\end{figure}

We now turn to the investigation of case (ii) in which the two dissipative processes $L^{(1)}_n$ and $L^{(2)}_n$ compete against each other. In that scenario, one can readily verify that $X_k = 2 \mathbb{I}$, such that the damping spectrum is gapped (and ``flat'') for all $k$. The steady-state correlation matrix is then given by
\begin{equation}
    \Gamma_k = g \left( \begin{array}{cc}
        (\cos k + \kappa \cos 2k) \rmi \sigma_y & -(\sin k + \kappa \sin 2k) \sigma_z \\
        (\sin k + \kappa \sin 2k) \sigma_z & (\cos k + \kappa \cos 2k) \rmi \sigma_y
    \end{array} \right),
\end{equation}
with $g = 1/(1 + \kappa)$ and eigenvalues $\pm \sqrt{1 + \kappa^2 + 2 \kappa \cos k}/(1 + \kappa)$, implying that the steady state is pure if and only if $\kappa = 0$ or $\kappa \to \infty$. The corresponding vector $\vect{n}_k$ is non-zero (for all $k$) except at $\kappa = 1$ in which case $\Gamma_{k = \pm \pi} = 0$. For $\kappa \neq 1$, one finds $\vect{n}_k = g (0, -(\sin k + \kappa \sin 2k), -(\cos k + \kappa \cos 2k))$ and the corresponding winding number topological invariant is $\nu_\text{1D} = 1$ for $\kappa < 1$ and $\nu_\text{1D} = 2$ for $\kappa > 1$. These results, which are presented in figure~\ref{fig:1DExample3}, demonstrate that case (ii) provides an example of a topological phase transition that is not driven by criticality but rather by the closure of the purity gap only.

Examining case (ii) on a finite chain with open boundary conditions, we find four Majorana zero-damping modes (two at each edge) in the whole parameter range of $\kappa$. In accordance with the dissipative bulk-edge correspondence (see section~\ref{sec:bulkEdge}), all of these modes must be genuine for $\kappa > 1$ where $\nu_\text{1D} = 2$, while two of them (one at each edge) must be spurious for $\kappa < 1$ where $\nu_\text{1D} = 1$. Crucially, the spatial wave function of these four modes does not exhibit any critical behavior~\footnote{In fact, it does not change at all in this particular example.} upon approaching the topological phase transition point $\kappa = 1$.

\section{Illustrative example in 2D}
\label{sec:2D}

In the previous section, we have exemplified the phenomenology of Majorana zero-damping modes at physical edges in the context of 1D systems. Here we focus instead on the physics that arises when topological defects are introduced in the bulk of the system, away from physical edges. We consider topological defects in the form of \emph{dissipative} vortices in infinite 2D systems in order to illustrate such physics in the simplest possible scenario.

\subsection{Dissipative vortices}
\label{sec:dissipativeVortices}

\emph{Dissipative} vortices are most conveniently defined starting from an infinite 2D lattice system evolving under a dissipative dynamics generated by translation-invariant Lindblad operators $L_i = \sum_j u_{j-i} \, a_j + v_{j-i} \, a^\dagger_j$ (using notations as in section~\ref{sec:generalFormMZMs} above; see equation~\eref{eq:translationInvariantLindbladOp}, in particular). We assume that the corresponding Liouvillian exhibits a finite dissipative gap and that its steady state is characterized by a finite purity gap, so that the topological nature of the system is well-defined (i.e., the steady state belongs to a specific topological class and is characterized by quantized topological invariants); in particular, the dissipative bulk-edge correspondence introduced in section~\ref{sec:edgePhysics} is satisfied~\footnote{One can assume, without loss of generality, that the Lindblad operators form a complete set of anticommuting operators, so that the system is driven into a pure steady state independently of the initial conditions (see discussion of section~\ref{sec:steadyStateSymmetries}).}. We then introduce a \emph{dissipative vortex} at a position defined by the vector $\vect{R}_0$~\footnote{Note that this position need not coincide with a lattice site.} by modifying the translation-invariant coefficients $u_{j-i} \equiv u_{ij}$ defining the annihilation part of the Lindblad operators in the following way:
\begin{eqnarray} \label{eq:dissipativeVortex}
    u_{ij} \to u_{ij} f(r_j) \rme^{-\rmi \ell \varphi_j},
\end{eqnarray}
where $(r_j, \varphi_j)$ are polar coordinates defining the position of each site $j$ with respect to $\vect{R}_0$, $f(r)$ is a real and positive function that vanishes as $r \to 0$ and reaches a constant value as $r \gg \delta$ ($\delta$ being a characteristic length scale associated with the dissipative vortex, defining the \emph{vortex core}), and $\rme^{-\rmi \ell \varphi_j}$ describes the \emph{vortex phase} winding $\ell$ times around the origin $\vect{R}_0$ ($\ell$ defining the so-called \emph{vorticity}). While vortices exhibit similar properties in Hamiltonian systems, the specific form of a dissipative vortex defined above is motivated by its natural realization in typical implementation schemes with cold atoms based on optical vortex imprinting (we refer to our previous work~\cite{Bardyn12} for further details).

We remark that the qualitative features of a dissipative vortex do not depend on the explicit form of the function $f(r)$ describing its core. Of crucial importance is the fact that $f(r)$ vanishes as $r \to 0$, so that the vortex core can be identified with a region of space in which the system is topologically equivalent to the vacuum, with all sites essentially occupied~\footnote{This can be understood from the fact that the Lindblad operators acting in the vortex core have an annihilation part that essentially vanishes, namely, $L_i = C^\dagger_i + A_i$ with $A_i \approx 0$.}. If the topological nature of the bulk is non-trivial, a small circular edge (or domain wall) of radius $\sim \delta$ must then appear around the vortex core, separating this vacuum-like region from the surrounding bulk which is potentially topologically non-trivial. Remembering the dissipative bulk-edge correspondence discussed in section~\ref{sec:edgePhysics}, one then naively expects Majorana zero-damping modes and/or intrinsic Majorana zero-purity modes to appear at this domain wall. We argue below that the possibility of having such modes, however, additionally and crucially depends on the phase winding $\ell$ of the dissipative vortex, as expected by analogy to the Hamiltonian setting.

The crucial role of the vortex phase winding $\ell$ in prohibiting or allowing the existence of Majorana zero modes in the vortex core can be understood in analogy to the Hamiltonian setting. In the Hamiltonian context, circular domain walls typically trap low-energy modes with quantized angular momentum $m$ and a corresponding energy $E \sim m$~\cite{Read00}. When a flux of the $U(1)$ gauge field associated with the fermions corresponding to $n$ quanta of angular momentum is inserted in the region enclosed by the domain wall, the angular momentum of these low-energy modes shifts to values $m + n$ and the energy of the latter shifts accordingly, thereby prohibiting or allowing for zero-energy modes. We argue that the same argument applies here in the case of a dissipative vortex, namely: the phase winding $\ell$ defines a flux $\pi \ell$ threading the vortex core and shifts the eigenvalues corresponding to low-\emph{damping} or low-\emph{purity} modes bound to the vortex core (in the damping or purity spectrum, respectively). This can be understood from the fact that the Hamiltonian phenomenology discussed above automatically applies to the parent Hamiltonian associated with the dissipative dynamics, whose zero modes either correspond to Majorana zero-damping modes or to intrinsic Majorana zero-purity modes (see discussion of section~\ref{sec:dissipativeGap}). In analogy with the Hamiltonian setting (see, e.g.,~\cite{Read00}), we then expect Majorana zero-damping modes and/or intrinsic Majorana zero-purity modes to appear in dissipative vortices with \emph{odd} vorticity $\ell$ only, as we corroborate below in an explicit model.

\begin{figure}[t]
    \begin{center}
        \includegraphics[width=\columnwidth]{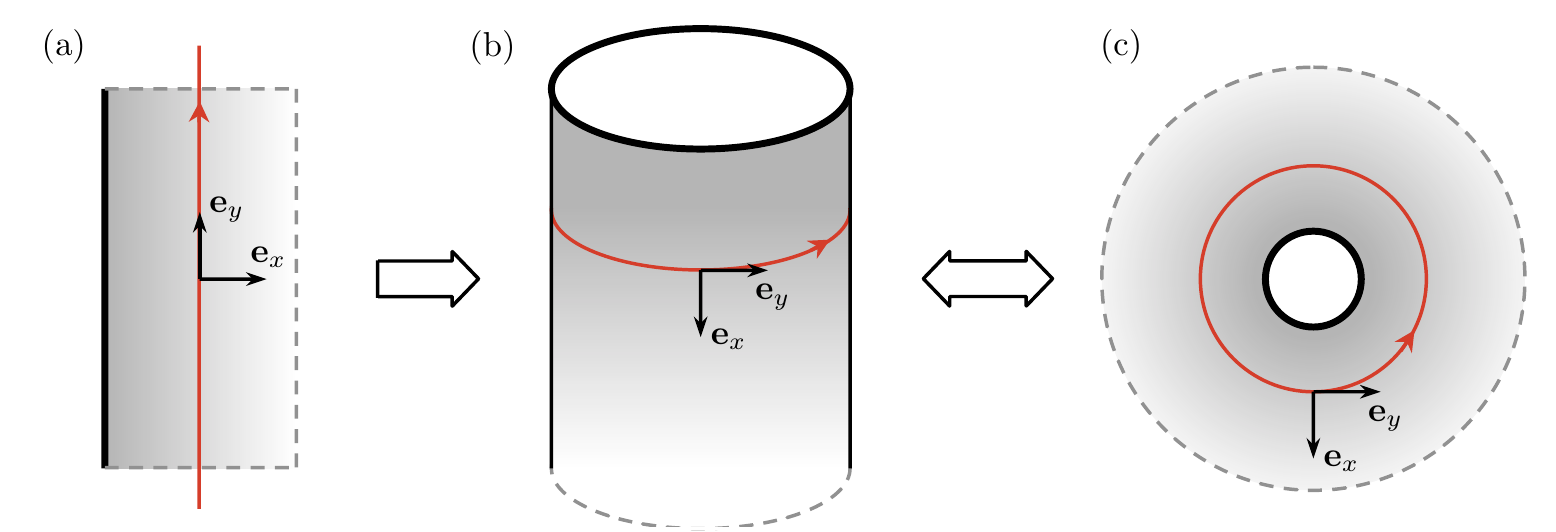}
        \caption{Left: Semi-infinite plane ``wrapped up'' into a semi-infinite cylinder in the $\vect{e}_y$ direction (shown in red). Right: Topological equivalence between the semi-infinite cylinder geometry and an infinite planar geometry with a single dissipative vortex ``puncturing'' the plane.}
        \label{fig:cylinderVortexMapping}
    \end{center}
\end{figure}

We remark that the physics associated with a dissipative vortex remains qualitatively the same upon insertion or removal of $2\pi$ fluxes in the vortex core since such modifications can be seen as gauge transformations~\cite{Read00}. The physical properties of a dissipative vortex are therefore determined by the vorticity $\ell$ modulo $2$, and one can restrict oneself to $\ell = 0$ or $\ell = 1$, without loss of generality. Since the case of a dissipative vortex with $\ell = 0$ is physically equivalent to that of a small circular physical edge, we will focus exclusively on dissipative vortices with $\ell = 1$. Remarkably, the physics arising in a single dissipative vortex with a $\pi$ flux ($\ell = 1$) in an infinite system on the plane can be shown to be equivalent that occurring at the edge of a semi-infinite system with a cylinder geometry and no flux (see figure~\ref{fig:cylinderVortexMapping} and our previous work~\cite{Bardyn12} for further details). This equivalence---which will be useful in the sections below---can intuitively be understood from the fact that the $\pi$ flux of the $U(1)$ gauge field introduced by a vortex with $\ell = 1$ naturally arises in cylinder geometry due to the (extrinsic) curvature of the latter.

\subsection{Explicit 2D model and bulk properties}
\label{sec:2DModel}

We now consider, in the framework defined above, an explicit model defined on a square lattice with primitive vectors $\vect{e}_x$ and $\vect{e}_y$, with translation-invariant Lindblad operators $L_i = \sum_j u_{j-i} \, a_j + v_{j-i} \, a^\dagger_j$ whose only non-zero coefficients are
\begin{eqnarray} \label{eq:crossLindbladOpCoeff}
    v_0 = \beta; \qquad & v_{\pm \vect{e}_x} = 1; \qquad & v_{\pm \vect{e}_y} = 1; \nonumber \\
    & u_{\pm \vect{e}_x} = \pm 1; & u_{\pm \vect{e}_y} = \pm \rmi,
\end{eqnarray}
where $\beta$ is a real parameter which can be used to tune the system across phase transitions. Every site $i$ therefore has an associated Lindblad operator $L_i$ of the form
\begin{eqnarray} \label{eq:crossLindbladOp}
    L_i = C^\dagger_i + A_i = \beta \, a^\dagger_i \; & + & \; ( a^\dagger_{i + \vect{e}_x} + \hphantom{\rmi} a^\dagger_{i + \vect{e}_y} + \hphantom{\rmi} a^\dagger_{i - \vect{e}_x} + \hphantom{\rmi} a^\dagger_{i - \vect{e}_y} ) \nonumber \\
    & + & \; ( a_{i + \vect{e}_x} + \rmi a_{i + \vect{e}_y} - \hphantom{\rmi} a_{i - \vect{e}_x} - \rmi a_{i - \vect{e}_y} ),
\end{eqnarray}
where the creation part $C^\dagger_i$ is isotropic or $s$-wave symmetric with respect to the central site $i$ (as naturally arises in typical implementation schemes; see section~\ref{sec:physicalRealization}) and the annihilation part $A_i$ has a $p$-wave symmetry, i.e., $A_i = (\nabla_x + \rmi \nabla_y) a_i$ with $\nabla_\lambda a_i \equiv a_{i + \vect{e}_\lambda} - a_{i - \vect{e}_\lambda}$. In momentum space, these translation-invariant Lindblad operators take the form $L_\bk = u_\bk a_\bk + v_\bk a^\dagger_{-\bk}$, where
\begin{eqnarray} \label{eq:ukvk}
    u_\bk & = 2 \mathrm{i} ( \sin{(k_x)} + \mathrm{i} \sin{(k_y)} ); \nonumber \\
    v_\bk & = \beta + 2 ( \cos{(k_x)} + \cos{(k_y)} ).
\end{eqnarray}
Since the above functions satisfy $u_\bk v_\bk = -u_{-\bk} v_{-\bk}$, the Lindblad operators form a complete set of anticommuting operators and the system is driven into a pure Gaussian steady state corresponding to the ground state of the parent Hamiltonian $H_\mathrm{parent} = \sum_\bk L^\dagger_\bk L_\bk$ (see equation~\eref{eq:oddConstraintFromPurity} and discussion thereof). The damping spectrum, which coincides with the spectrum of $H_\mathrm{parent}$ in that case, is then given by the norm
\begin{eqnarray}
    \mathcal{N}_\bk = \sqrt{\{ L^\dagger_\bk, L_\bk \}} = \sqrt{\abs{u_\bk}^2 + \abs{v_\bk}^2}.
\end{eqnarray}
Remembering the explicit form of the functions $u_\bk$ and $v_\bk$ defined in equation~\eref{eq:ukvk} above, one can easily verify that the corresponding dissipative gap closes at the parameter values $\beta = 0$, $\pm 4$. We thus expect the existence of four distinct---possibly topological---phases depending on the value of $\beta$.

We remark that the Lindblad operators are not invariant under time-reversal symmetry due to the $p$-wave symmetry embedded in their annihilation part. The system thus belongs to the symmetry class D of Altland and Zirnbauer (see section~\ref{sec:topClassification}) and is characterized by the Chern number topological invariant defined in equation~\eref{eq:chernNumber}. One can easily verify that, in accordance with the discussion of section~\ref{sec:chernNumber}, the Chern number vanishes in the whole $\beta$ parameter range where the system exhibits a finite dissipative gap (we refer to our previous work~\cite{Bardyn12} for an explicit proof). As dictated by the dissipative bulk-edge correspondence introduced in section~\ref{sec:edgePhysics}, we therefore naively do not expect to find Majorana zero-damping modes in the system if edges are introduced. Despite this conclusion, let us now try to construct such modes explicitly in the translation-invariant setting of section~\ref{sec:generalFormMZMs} anyway.

\subsection{Construction of Majorana zero-damping modes}
\label{sec:constructionMZMs}

The possibility of having Majorana zero-damping modes at a physical edge when the dissipative dynamics is generated by translation-invariant Lindblad operators as in section~\ref{sec:generalFormMZMs} above can be assessed from the explicit form of the Lindblad operators (see equation~\eref{eq:crossLindbladOpCoeff}) by looking for a solution of equation~\eref{eq:generalConditionZeroMode}. Choosing $\phi_i = 0$ (which corresponds to choosing a particular orientation of the edge; see section~\ref{sec:generalFormMZMs}), the real and imaginary parts of equation~\eref{eq:generalConditionZeroMode} here take the explicit form
\begin{eqnarray*}
    0 & = & \beta + 2 \beta_{\vect{e}_x} + \beta_{\vect{e}_y} + (\beta_{\vect{e}_y})^{-1}, \\
    0 & = & \beta_{\vect{e}_y} - (\beta_{\vect{e}_y})^{-1}.
\end{eqnarray*}
One thus finds $\beta_{\vect{e}_y} = \pm 1$ and $\beta_{\vect{e}_x} = -(\beta/2 \pm 1)$, which implies the existence of a solution with $\abs{\beta_{\vect{e}_x}} < 1$ if and only if $0 < \abs{\beta} < 4$. Consequently, there exists at least one Majorana zero-damping mode $\gamma$ of the form defined by equation~\eref{eq:generalSolutionZeroMode} when $0 < \abs{\beta} < 4$ if the system possesses an edge in the $-\vect{e}_x$ direction~\footnote{Note that a different orientation of the edge would be obtained for $\phi_i \neq 0$.}, in which case $\gamma$ is uniformly spread along the edge (since $\abs{\beta_{\vect{e}_y}} = 1$) and exponentially localized in the $+\vect{e}_x$ direction away from the edge on a characteristic length scale $\xi = -1 / \log{(\abs{\beta_{\vect{e}_x}})} = -1 / \log{(\abs{\beta/2 \pm 1})}$. The semi-infinite planar geometry of the system in that case is depicted in figure~\ref{fig:cylinderVortexMapping}(a). Remarkably, the special points $\beta = 0$, $\pm 4$ where the dissipative gap closes coincide with those at which $\abs{\beta_{\vect{e}_x}} = 1$, i.e., at which the localization length associated with $\gamma$ diverges. This suggests the possible existence of topological phase transitions occurring at the gap-closing points $\abs{\beta} = 0$, $\pm 4$ which are not identified by the (vanishing) Chern number topological invariant.

\subsection{Topological origin}
\label{sec:topOrigin}

We have demonstrated above that the system can support Majorana zero-damping modes at an edge (in the parameter range $0 < \abs{\beta} < 4$) despite its vanishing Chern number. We argue below that this apparent contradiction stems from the fact that we have assumed translational symmetry, and establish the topological origin of Majorana zero-damping modes in the above translation-invariant setting. It is  clear that translational symmetry is not robust against disorder. However, assuming such a symmetry will allow us to demonstrate, in the simple model above, a key mechanism with no Hamiltonian counterpart through which spatially isolated Majorana zero-damping modes can be obtained in a topological phase characterized by an even integer topological invariant, in contradiction with the Hamiltonian bulk-edge correspondence.

We first remark that the construction of section~\ref{sec:constructionMZMs} above remains unchanged if we assume that the system is finite in the $\vect{e}_y$ direction, ``wrapped up'' into a cylinder (as depicted in figure~\ref{fig:cylinderVortexMapping}(b)). Since such a semi-infinite cylinder geometry is topologically equivalent to that of an infinite plane ``punctured'' by a single dissipative vortex (figure~\ref{fig:cylinderVortexMapping}(c)), Majorana zero-damping modes of the form found in section~\ref{fig:cylinderVortexMapping} must therefore similarly appear in the core of a dissipative vortex on the plane~\footnote{Note that the translation symmetry along the circumference of the cylinder translates as a \emph{rotational} symmetry around the vortex core.}. In these two equivalent geometries, the topological origin of Majorana zero-damping modes can be identified by moving to momentum space in the $\vect{e}_y$ direction associated with translational (or rotational, in the vortex case) symmetry (see figures~\ref{fig:cylinderVortexMapping}(b) and (c)). The system then reduces to a ``stack'' of semi-infinite 1D wires (in the $\vect{e}_x$ direction) associated with different momenta $k_y$. In particular, the Fourier-transformed counterpart of the Lindblad operators defined in equation~\eref{eq:crossLindbladOp} take the form
\begin{eqnarray} \label{eq:1Dwire}
	L_i = \kappa \, a^\dagger_i \; + \; ( a^\dagger_{i + \vect{e}_x} + a^\dagger_{i - \vect{e}_x} ) \; + \; ( a_{i + \vect{e}_x} - a_{i - \vect{e}_x} )
\end{eqnarray}
in the momentum sectors corresponding to $k_y = 0$ or $\pi$, where $i$ indexes the lattice sites in the remaining spatial direction $\vect{e}_x$ and $\kappa = \beta + 2$ for $k_y = 0$ and $\beta - 2$ for $k_y = \pi$, respectively. In these two sectors, the system thus reduces to the 1D wire with chiral symmetry investigated in section~\ref{sec:1D} above (up to an overall dissipation rate $\sqrt{4 + \kappa^2}$ which does not affect any of the topological properties of the system). Owing to chiral symmetry, such a system belongs to the symmetry class BDI of Altland and Zirnbauer and is characterized by the winding number topological invariant $\nu_\text{1D}$ defined by equation~\eref{eq:windingNumber}. As discussed in section~\ref{sec:1D} and depicted in figure~\ref{fig:1DExample1} thereof, one finds $\nu_\text{1D} = 2$ for $\abs{\kappa} < 2$ and $\nu_\text{1D} = 0$ for $\abs{\kappa} > 2$. Discontinuities in the winding number at $\kappa = \pm 2$ (or, equivalently, at $\beta = 0$, $\pm 4$) clearly establish the occurrence of non-equilibrium \emph{topological} phase transitions at the dissipation gap-closing points identified in section~\ref{sec:constructionMZMs}.

In summary, we have identified a non-trivial topological invariant $\nu_\text{1D} = 2$ in the parameter range $0 < \abs{\beta} < 4$~\footnote{Note that $\beta > 0$ ($\beta < 0$) corresponds to the 1D wire in the momentum sector $k_y = 0$ ($k_y = \pi$).} by taking advantage of translation invariance to reduce the original 2D problem in semi-infinite cylinder geometry or, equivalently, in planar geometry with a single $\ell = 1$ dissipative vortex, to a 1D wire defined along the axis of the cylinder or in the radial direction away from the vortex core ($\vect{e}_x$ direction shown in figures~\ref{fig:cylinderVortexMapping}(b) and (c)). This provides a topological explanation as to why we were able to demonstrate the existence of at least one Majorana zero-damping mode for $0 < \abs{\beta} < 4$ in section~\ref{sec:constructionMZMs}. In fact, invoking the dissipative bulk-edge correspondence of section~\ref{sec:bulkEdge}, we can now argue that a total of $\nu_\text{1D} = 2$ Majorana zero-damping modes and/or Majorana zero-purity modes must be present at the edge of the semi-infinite system in cylinder geometry or, equivalently, in the core of the $\ell = 1$ dissipative vortex on the (infinite) plane, as long as the symmetries of the system are preserved~\footnote{More precisely, translational symmetry along the circumference of the cylinder or, equivalently, rotational symmetry around the vortex core must be preserved, as well as the chiral symmetry of the dimensionally reduced problem (1D wire). In the presence of disorder, one expects Majorana zero-damping modes to become ``softer'', acquiring a small but finite damping rate.}.

The precise number $(m_\text{damping})_\textit{edge} \leq 2$ (see equation~\eref{eq:dissipativeBulkEdgeCorrespondence} and discussion thereof) of genuine Majorana zero-damping modes that is found at the edge of the cylinder or in the vortex core depends on the specific dissipative boundary conditions that appear there. Crucially, physically distinct boundary conditions naturally emerge in these two systems, although their geometries are topologically equivalent. In the case of the dissipative vortex, in particular, one can demonstrate explicitly that a \emph{single} Majorana zero-damping mode is found in the vortex core (alongside with a single Majorana zero-purity mode such that the dissipative bulk-edge correspondence is satisfied; see equation~\eref{eq:dissipativeBulkEdgeCorrespondence}). The key mechanism through which the dissipative vortex isolates exactly one Majorana zero-damping mode despite the fact that the bulk is characterized by an even (integer) topological invariant ($\nu_\text{1D}$) originates from the geometry of the vortex core alone and is therefore generic. We refer to our previous work~\cite{Bardyn12} for an explicit proof of this statement.

\begin{figure}[t]
    \begin{center}
        \includegraphics[width=\columnwidth]{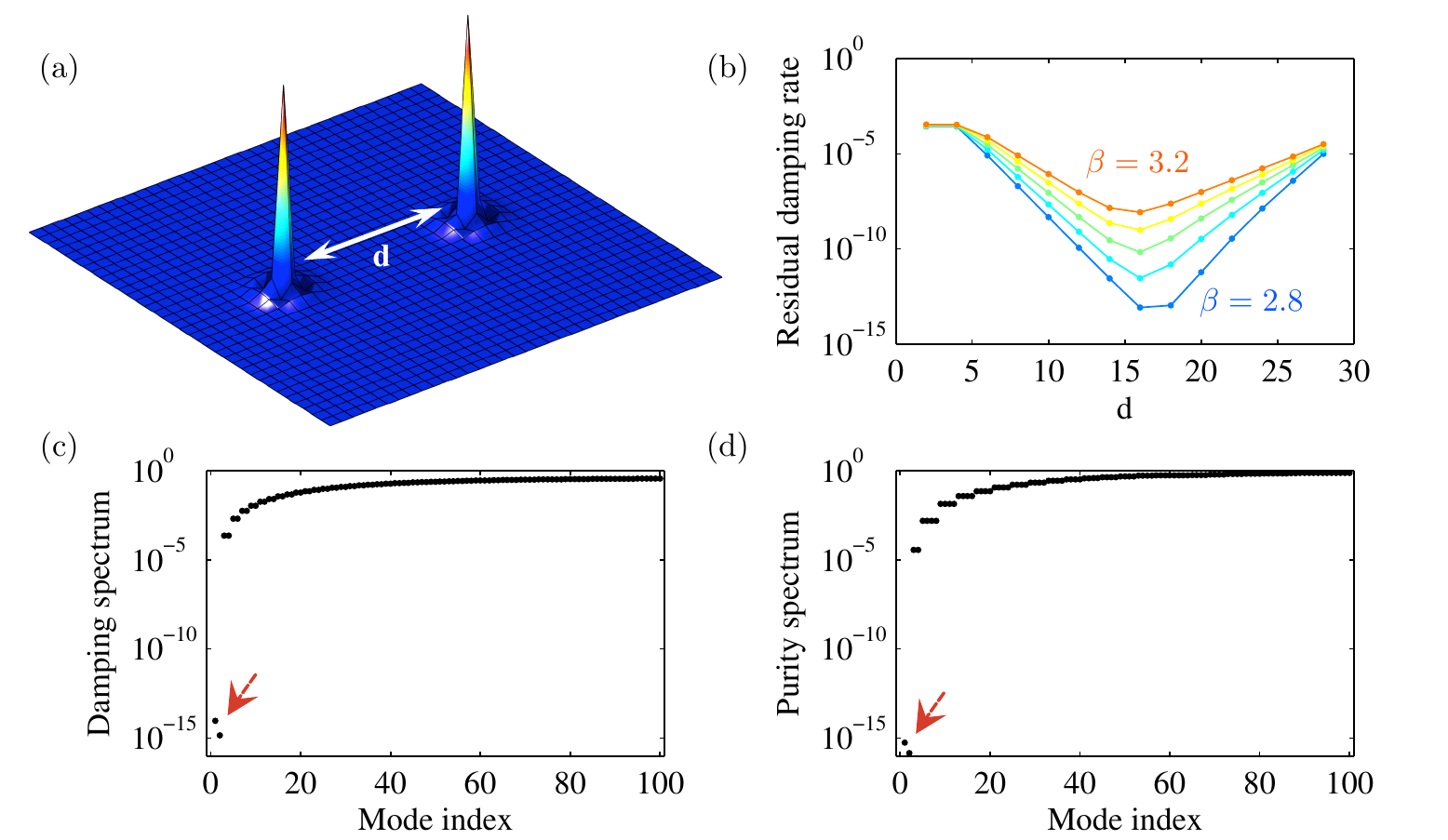}
        \caption{Numerical results for a configuration of two $\ell = 1$ dissipative vortices on a square lattice of $35 \times 35$ sites with unit spacing. (a) Generic form of the Majorana zero-damping modes localized in each of the vortex cores (here for $\beta = 2$ and a vortex-vortex separation $d = 16$), and (b) exponential behavior of their residual damping rate as a function of the vortex-vortex separation $d$ for $\beta = 2.8$ to $3.2$ by steps of $0.1$ (from bottom to top). Note that, due to the finite size of the system, vortices ``interact'' more strongly (via the edge) as they come closer to the edge, giving rise to the symmetric behavior that can be seen around $d \approx 16$. (c) and (d): Low-lying part of the damping (respectively purity) spectrum for $\beta = 2$ featuring a gap with two quasi-zero ($\sim 10^{-15}$) eigenvalues (indicated by a red arrow) corresponding to Majorana zero-damping (zero-purity) modes trapped in each of the vortex cores. All results were obtained for vanishingly small vortex cores, i.e., with $f(r) = 1$ everywhere except at $r = 0$ where it vanishes (see equation~\eref{eq:dissipativeVortex}).}
        \label{fig:numericsTwoVortices}
    \end{center}
\end{figure}

\subsection{Physical properties of dissipative vortices carrying single Majorana zero-damping modes}
\label{sec:physPropDissipativeVortices}

In the Hamiltonian context, $\pi$ flux vortices carrying single Majorana zero-energy modes have been demonstrated to exhibit non-Abelian exchange statistics~\cite{Ivanov01}. The underlying proof exclusively relies on the algebraic properties of Majorana modes and, crucially, does not depend on the dynamics giving rise to them (see section~\ref{sec:braiding}). As a result, the same conclusions apply in the dissipative setting of interest in this work. In particular, $\pi$ flux (i.e., $\ell = 1$) dissipative vortices carrying single Majorana zero-damping modes must exhibit non-Abelian exchange statistics (when moved around through adiabatic parameter changes as discussed in section~\ref{sec:braiding}). In light of the mechanism illustrated in the above explicit model, we thus conclude that vortices can have, in our dissipative setting, \emph{non-Abelian} exchange statistics in a phase characterized by an \emph{even} integer topological invariant, in stark contrast to what can occur in the Hamiltonian context.

In addition to their non-Abelian statistics, vortices carrying single Majorana zero modes behave, in our dissipative setting, in a fully analogous way as in the Hamiltonian context. In particular, two such vortices ``interact'' as the distance $d$ between the latter becomes small, acquiring an exponentially small residual damping rate $\sim \rme^{-d/\xi}$, where $\xi$ is the localization length associated with the Majorana zero-damping mode that they carry. We have verified such phenomenology through numerical simulations. Figure~\ref{fig:numericsTwoVortices} summarizes the results obtained in the case of a finite system on the plane with two ``interacting'' $\ell = 1$ dissipative vortices.

\section{Conclusions}
\label{sec:conclusions}

In this paper we have provided a discussion of non-equilibrium superfluid topological states generated by means of engineered dissipation, highlighting analogies and differences to the more conventional Hamiltonian setting. Similarities mainly stem from the fact that topological properties are properties of the state of the system---encoded in the static correlations---with some care to be taken regarding the purity of the state. In particular, identifying the correlation matrix as a tool for the symmetry-based topological classification of arbitrary Gaussian states analogous to a first-quantized quadratic Hamiltonian, it is clear that the classification of topological states according to the symmetry classes of Altland and Zirnbauer extends to a general non-equilibrium steady-state situation, highlighting the universality of this classification. Differences, on the other hand, mostly stem from the fact that the spectral or dynamical properties of the system and the properties of the steady state are not as tightly related as in the Hamiltonian ground-state scenario. The interplay between the well-known characteristics of topological states and the new elements brought in by the dissipative nature of the dynamics can give rise to effects with no Hamiltonian counterpart. Examples are topological phase transitions which do not require the dissipative spectral gap to close, or the trapping of unpaired Majorana modes in a bulk with vanishing Chern number.

We have presented in this work a detailed discussion of dissipatively induced topological superfluid states in 1D and 2D systems which is a first step towards a complete picture of non-equilibrium topological order. In this respect, various directions remain to be explored in future work:

From the viewpoint of dissipation engineering, we may ask ourselves whether other symmetry classes than the ones identified in this work can be achieved in realistic setups. So far we have concentrated on topological superfluids of spinless fermions. Adding spin degrees of freedom opens up the perspective of reaching topological insulating phases with potentially more robust edge modes. It will also be interesting to investigate whether going beyond quasi-local Lindblad operators is possible in realistic cold atom experiments, allowing to reach phases characterized by a non-vanishing Chern number.

From a many-body perspective, an interesting direction will be to explore the nature of topological phases and phase transitions in more detail. This concerns, in particular, the interplay of dissipative and Hamiltonian dynamics, which we did not examine in this work. While this intentional omission is not a severe limitation in the context of cold atomic gases where the Hamiltonian dynamics can be suppressed, including the effects of such dynamics would give a more complete picture of the robustness of topological phases under general combined unitary and dissipative dynamics. From the viewpoint of critical phenomena, topological phase transitions accompanied by a closure of the dissipative gap offer intriguing examples of criticality in fermionic systems. Going beyond a mean-field approach by including the effects of Hamiltonian and dissipative interactions will be necessary to reach a detailed understanding of this issue.

From a quantum information point of view, it will be important to investigate in more detail the analogies and differences between dissipatively induced non-local decoherence-free subspaces and more conventional Hamiltonian edge mode subspaces. To answer these questions, a more systematic classification of harmful dissipative perturbations as well as a more detailed understanding of the dynamics of excitations on top of the dissipatively stabilized state will be necessary. The propagation of excitations or defects on top of a state which is entirely induced by dissipation can be expected to be vastly different from a Hamiltonian ground state. Also, manipulation and readout tools tailored to the dissipative setting will have to be developed.

\emph{Acknowledgements} -- The authors thank A. Akhmerov, M. Fleischhauer, S. Habraken, V. Gurarie, V. Lahtinen and M. Troyer for helpful discussions. This work was supported by NCCR Quantum Science and Technology (NCCR QSIT), research instrument of the Swiss National Science Foundation (SNSF), an ERC Advanced Investigator Grant (A. I.), the Austrian Science Fund (FWF) through SFB FOQUS F4016-N16 and the START grant Y 581-N16 (S. D.), the European Commission (AQUTE, NAMEQUAM), the Institut f\"ur Quanteninformation GmbH and the DARPA OLE program.

\appendix

\titlecontents{section}
  [7.5em]{}
  {\contentslabel{7.5em}}
  {\hspace*{-6em}}
  {\titlerule*[0.6pc]{.}\contentspage}
\titlecontents{subsection}
  [7.5em]{}
  {\contentslabel{7.5em}}
  {\hspace*{-6em}}
  {\titlerule*[0.6pc]{.}\contentspage}

\section{Dissipative mean-field theory}
\label{app:meanField}

We show in more detail how the microscopically number conserving implementation of the Lindblad operators, leading to an interacting (quartic in the fermion operators) Liouvillian, are related to the quadratic setting analyzed in view of its topological properties. To this end, we first present a general relation of the number conserving operators to their fixed phase counterpart on the level of the dark state wave functions. We then show within a properly devised mean-field theory how these two settings are connected dynamically, with main result stating that the quadratic dynamics emerges naturally in the long time limit of the microscopically number conserving quartic dissipative evolution.

\subsection{General relation of fixed number and fixed phase Lindblad operators}

We show that for a given real space quadratic master equation with pure dark state, we can directly construct a number conserving quartic version and vice versa, and specify the respective fixed phase or fixed number exact dark state wave functions. In particular, in the thermodynamic limit, both descriptions become equivalent as made explicit below.

We study a general Liouville operator generated by Lindblad quantum jump operators $J_i$ which exhibits a  pure dark state: 
\begin{eqnarray}
\mathcal L [\rho ] &=& \kappa \sum_{i} \big(J_i \rho J_i^\dag  - \frac{1}{2} \{  J_i^\dag J_i ,\rho \}\big),\quad
J_i |D\rangle = 0 \,\, \forall i , \,\,\rho_D = |D\rangle \langle D |.
\end{eqnarray}
We will show the following: For each fixed phase setting, specified by Lindblad operators and dark state
\begin{eqnarray}
J_i &=& L_i = A_i +   \alpha C^\dag_i ,\quad |D\rangle =  |\text{BCS},\theta\rangle \propto \exp(\alpha G^\dag  )  |\text{vac}\rangle,
\end{eqnarray}
there is a fixed number setting with \footnote{Working with the functions $|\text{BCS},N\rangle$ for finite $N$ may require an infrared momentum cutoff $q_L \sim 1/L$, so that $N = n L^d$ ($n$ the particle filling, $L$ the number of sites in each direction, $d$ the spatial dimension), which is consistent with the thermodynamic limit $N \to \infty,  L\to \infty , n= N/ L^d  \to \mathrm{const}$.}
\begin{eqnarray}
J_i &=& \ell_i = C_i^\dag A_i, \quad
|D\rangle = |\text{BCS},N\rangle \propto G^{\dag\,N}|\text{vac}\rangle.
\end{eqnarray}
The reverse statement holds as well. Here, as in the main text the creation (annihilation) parts $C_i^\dag (A_i)$ are linear in the spinless fermion operators and obey the following requirements: \\
(i) \emph{Translation invariance} -- $C_i^\dag = \sum_j v_{i-j} a_j^\dag$ and $A_i = \sum_j u_{i-j} a_j$ with translation invariant complex quasi-local position space functions $v_{i-j},u_{i-j}$. The Fourier transforms for the creation and annihilation part are local in momentum space, 
\begin{eqnarray}
C^\dag_\mathbf{k} = \sum_i e^{-\mathrm i \mathbf{k} \mathbf{x} _i} C^\dag_i = v_\mathbf{k}  a^\dag_\mathbf{k} , \quad A_\mathbf{k}  = \sum_i e^{\mathrm i \mathbf{k} \mathbf{x} _i} A_i = u_\mathbf{k}  a_\mathbf{k} .
\end{eqnarray}
The Fourier transformed fixed phase and fixed number Lindblad operators then read, respectively
\begin{eqnarray}\label{eq:Lklk}
L_\mathbf{k}  = A_\mathbf{k} + \alpha C^\dag_\mathbf{k}   , \quad   \ell_\mathbf{k}  = \sum_\mathbf{q}  C^\dag_{\mathbf{q} -\mathbf{k} } A_\mathbf{q} .
\end{eqnarray}
(ii) \emph{Antisymmerty} -- The functions have the property $u_{-\mathbf{k} } = \pm u_\mathbf{k}  , v_{-\mathbf{k} } =\mp v_\mathbf{k} $, such that the related ``wave function'' $\varphi$ is antisymmetric,
\begin{eqnarray}
\varphi_\mathbf{k}  = v_\mathbf{k} /u_\mathbf{k}  =  - \varphi_{-\mathbf{k} }.
\end{eqnarray}
This implies that the set $L_\mathbf{k} $ form a full Dirac algebra up to a normalization,
\begin{eqnarray}
 \{ L_\mathbf{k} ,L_{\mathbf{k} '} \} = \{ L^\dag_\mathbf{k} ,L^\dag_{\mathbf{k}'} \} = 0 , \quad \{ L_\mathbf{k} ,L^\dag_{\mathbf{k}'} \} = (|u_\mathbf{k} |^2 + |\alpha v_\mathbf{k} |^2)\delta(\mathbf{k}  - \mathbf{k}')  
\end{eqnarray}
for all $\mathbf{k} ,\mathbf{k}'$. We may introduce normalized operators via 
\begin{eqnarray} \label{eq:Lkbar}
    \bar L_\mathbf{k}  = \bar u_\mathbf{k}  a_\mathbf{k}  + \alpha \bar v_\mathbf{k}  a^\dag_{-\mathbf{k} }, \quad \bar u_{\bk} = u_{\bk}/\sqrt{N_{\bk}}, \quad \bar v_{\bk} = v_{\bk}/\sqrt{N_{\bk}}, \nonumber \\
    N_\mathbf{k}  = |u_\mathbf{k} |^2 + |\alpha v_\mathbf{k} |^2 ,
\end{eqnarray}
such that $|\bar u_{\bk}|^2 + |\alpha \bar v_{\bk}|^2 = 1, \{ \bar L_\mathbf{k} ,\bar L^\dag_{\mathbf{k}'} \} =\delta(\mathbf{k}  - \mathbf{k}') $.

The generator of the state is bilinear and reads in terms of the momentum space wave function
\begin{eqnarray}
 G^\dag &=& \sum_\mathbf{k}  \varphi_\mathbf{k}  a_{-\mathbf{k} }^\dag a_\mathbf{k} ^\dag.
\end{eqnarray}

The fixed number operators satisfy $[\ell_i,\hat N] = 0$ and thus conserve total particle number $\hat N = \sum_i a_i^\dag a_i$. The corresponding dark states $ |\text{BCS},N\rangle$ contain $2N$ fermions. Instead, in the fixed phase setting, $[L_i,\hat N]\neq 0$. $\alpha = r e^{\mathrm i \theta}$ is an arbitrary complex number. Within the mean field described below, the modulus $r$ can be related to the average particle number. $\theta$ has an interpretation in terms of a fixed, superfluid phase. This is made explicit from the expansion of the coherent state wave function
\begin{eqnarray}\label{bcsth}
    |\text{BCS},\theta\rangle = \mathcal N \exp(\alpha G^\dag  )  |\text{vac}\rangle = \mathcal N\prod_{\mathbf{k} } \nolimits'(1+\alpha\varphi_{\mathbf{k} }a_{-\mathbf{k} }^{\dag}a_{\mathbf{k} }^{\dag})|\text{vac}\rangle, \nonumber \\
    \mathcal N = \prod_\mathbf{k}\nolimits' (1 + |\alpha \varphi_\mathbf{k} |^2 )^{-1/2},
\end{eqnarray}
where $\mathcal N$ is a global normalization. $\prod'_{\mathbf{k} }$ is restricted to $\mathbf{k}$ such that $k_\lambda>0$ in all primitive lattice directions $\lambda$.

With these preparations, we can prove the above statement. We work in momentum space and start with the number conserving setting. The Lindblad operators $\ell_\mathbf{k}  = \sum_\mathbf{q}  C^\dag_{\mathbf{q} -\mathbf{k} } A_\mathbf{q} $ are normal ordered, such that $\ell_\mathbf{k} |\text{vac}\rangle =0$. The dark state property $\ell_i |\text{BCS},N\rangle = \ell_\mathbf{k}  |\text{BCS},N\rangle =0$ for all $i$ or $\mathbf{k} $ is therefore equivalent to following relation for all $\mathbf{k} $,
\begin{eqnarray}
[ \ell_\mathbf{k} ,G^\dag ] &=& \sum_\mathbf{q}  v_{\mathbf{q} -\mathbf{k} } u_\mathbf{q}  \varphi_\mathbf{q}  a^\dag_{\mathbf{q} -\mathbf{k} }a^\dag_{-\mathbf{q} } = - \sum_\mathbf{q}  v_{\mathbf{q} } u_{\mathbf{q} - \mathbf{k} } \varphi_{\mathbf{q} -\mathbf{k} } a^\dag_{\mathbf{q} - \mathbf{k} }a^\dag_{- \mathbf{q} } \stackrel{!}{=} 0.
\end{eqnarray}
This is true if and only if  $\frac{v_\mathbf{q}   u_{\mathbf{q} -\mathbf{k} }}{u_\mathbf{q}  v_{\mathbf{q} - \mathbf{k} }} = \frac{\varphi_\mathbf{q} }{\varphi_{\mathbf{q} -\mathbf{k} }}$ holds for all $\mathbf{k} $, i.e., for the wave function $\varphi_\mathbf{q}  = v_\mathbf{q} /u_\mathbf{q} $ up to a momentum independent complex number. The latter can be absorbed into the constant $\alpha$. 

It is easy to show the uniqueness of the dark state in the fixed phase ensemble. The coherent state wave function~\eref{bcsth} can be written as $|\text{BCS},\theta\rangle \propto \prod_{\mathbf{k} }\nolimits' L_{\mathbf{k} }L_{-\mathbf{k} }|\text{vac}\rangle$, so that it is indeed the unique wave function annihilated by the full set of Dirac operators $L_\mathbf{k} $ up to normalization. We are not aware of an analogous argument for the fixed number case (for each given particle number $2N$), since the bilinears $\ell_\mathbf{k} $ do not obey a simple algebra. Numerical simulations in the context of spinful Lindblad operators \cite{Diehl10b,Yi} however suggest a unique steady dark state.

The two wave functions $|\text{BCS},N\rangle ,|\text{BCS},\theta\rangle$ are equivalent in the thermodynamic limit $N\to \infty$: We consider the relative number fluctuations of the fixed phase state,
\begin{eqnarray}
\Delta N^2 = \frac{\langle \hat N^2 \rangle - \langle \hat N \rangle^2}{\langle \hat N \rangle^2} = \frac{\sum_\bk n_\bk(1- n_\bk)}{(\sum_\bk n_\bk)^2} = \frac{\sum_\bk |  \bar  u_\bk \bar  v_\bk|^2}{(\sum_\bk |\bar v_\bk|^2)^2} \sim \frac{1}{N},
\end{eqnarray}
where $\hat N = \sum_\bk \hat n_\bk , \hat n_\bk = c_\bk^\dag c_\bk$ is the total particle density, the average is taken in $|\text{BCS},\theta\rangle$, and $n_\bk = \langle \hat n_\bk\rangle = |\bar v_\bk|^2$. Here we have used $\hat n_\bk^2= \hat n_\bk$, implying $\langle\hat n_\bk^2\rangle= \langle\hat n_\bk\rangle$, as well as $\langle \hat n_\mathbf{q} \hat n_\bk\rangle = \langle \hat n_\mathbf{q} \rangle\langle \hat n_\bk\rangle$ for $\bk\neq \mathbf{q}$ due to the product nature of the BCS state. Since $0\leq n_\bk \leq 1$ and $n_\bk=0$ only for a non-extensive set of modes, the scaling of the numerator is $\sim N$, and that of the denominator $\sim N^2$ in the thermodynamic limit. It is the mean-field (product) nature of the state which is responsible for this scaling. 

The fixed phase wave function is an exact solution of the number conserving equation (if the particle number is not fixed to a specific $N$), since $|\text{BCS},\theta\rangle \propto \exp(\alpha G^\dag  )  |\text{vac}\rangle  = \sum_N (\alpha G^\dag )^N/N! |\text{vac}\rangle $. The above discussion shows that for fixed $N\to\infty$, the fixed phase wave function approximates the fixed number one in a controlled way. We will rely on this fact, which is solely related to large $N$, in the mean-field theory to be discussed next.

\subsection{Mean-field theory}

Here we describe a mean-field theory for quartic, number conserving fermionic master equations. The mean-field description, represented by an effective quadratic master equation,  is valid in the thermodynamic plus long-time limit, where the system is already close to the steady state. It is instructive to compare this mean-field theory to BCS theory for Hamiltonian ground states. The role of the ground state is played by the dark state. The long time limit is analogous to a low energy limit in condensed matter systems, where the description of interacting fermions in terms of quadratic effective BCS Hamiltonians becomes appropriate. We emphasize that unlike the BCS problem of locally interacting fermions, our mean-field theory is controlled by the fact that the exact dark state for the interacting quartic dissipative problem with fixed particle number is known. The effective linear Lindblad operators take the form of annihilation operators for the dark state, in complete analogy to the Bogoliubov operators for the BCS problem. The momentum dependent rate premultiplying the Lindblad form describes the damping in the vicinity of the dark state, replacing the excitation eigenenergies in the BCS Hamiltonian. The mean-field theory allows us to calculate the damping rates and number equation from the microscopic quartic description. 

In our mean-field approach, as in standard BCS theory we give up exact particle number conservation and work in the fixed phase ensemble, exploiting the discussion above. Our mean-field ansatz is then defined by a factorization of the density matrix in momentum space, $\rho= \prod_\mathbf{k}'\rho_\mathbf{k}$, where $\rho_\mathbf{k}$ describes the mode pair $\{\mathbf{k},-\mathbf{k}\}$. This is motivated by the form of the coherent state steady density matrix $\rho_D = |\text{BCS},\theta\rangle \langle \text{BCS},\theta |$ sharing this property. We implement this ansatz,  focusing e.g. on the recycling term,
\begin{eqnarray}
    \fl \mathcal L [\rho] \ni \sum_{\mathbf{k}} \ell_\mathbf{k} \rho \ell_\mathbf{k}^\dag & = &\quad  \sum_{\mathbf{q}_1, ... , \mathbf{q}_4} v_{\mathbf{q}_1} u_{\mathbf{q}_2} u^*_{\mathbf{q}_3}v^*_{\mathbf{q}_4} a_{\mathbf{q}_1}^\dag a_{\mathbf{q}_2} \rho  a_{\mathbf{q}_3}^\dag a_{\mathbf{q}_4}\delta (\mathbf{q}_1 + \mathbf{q}_3 - \mathbf{q}_2 - \mathbf{q}_4)  \\
& = & \quad \sum_{\mathbf{q}_2,  \mathbf{q}_3 \neq \mathbf{p}} v_{\mathbf{p}} u_{\mathbf{q}_2} u^*_{\mathbf{q}_3}v^*_{\mathbf{p}} a_{\mathbf{p}}^\dag a_{\mathbf{q}_2} \rho  a_{\mathbf{q}_3}^\dag a_{\mathbf{p}}\delta ( \mathbf{q}_3 - \mathbf{q}_2 ) \nonumber \\
&  & +  \sum_{\mathbf{q}_1, \mathbf{q}_4\neq \mathbf{p}} v_{\mathbf{q}_1} u_{\mathbf{p}} u^*_{\mathbf{p}}v^*_{\mathbf{q}_4} a_{\mathbf{q}_1}^\dag a_{\mathbf{p}} \rho  a_{\mathbf{p}}^\dag a_{\mathbf{q}_4}\delta (\mathbf{q}_1  - \mathbf{q}_4) \nonumber \\
& & + \sum_{\mathbf{q}_2 , \mathbf{q}_4\neq \mathbf{p}} v_{\mathbf{p}} u_{\mathbf{q}_2} u^*_{-\mathbf{p}}v^*_{\mathbf{q}_4} a_{\mathbf{p}}^\dag a_{\mathbf{q}_2} \rho  a_{-\mathbf{p}}^\dag a_{\mathbf{q}_4}\delta (\mathbf{q}_2 +\mathbf{q}_4) \nonumber \\
&  & + \sum_{\mathbf{q}_1,\mathbf{q}_3\neq \mathbf{p}} v_{\mathbf{q}_1} u_{\mathbf{p}} u^*_{\mathbf{q}_3}v^*_{-\mathbf{p}} a_{\mathbf{q}_1}^\dag a_{\mathbf{p}} \rho  a_{\mathbf{q}_3}^\dag a_{-\mathbf{p}}\delta (\mathbf{q}_1 + \mathbf{q}_3 )   +  \{\mathbf{p}\to -\mathbf{p}\} + \text{h. o. t.}\nonumber
\end{eqnarray}
For the last equality, we have selected the contributions which are of second order in the operators for some selected momentum mode $\mathbf p$ and neglect the higher order terms  in the following, as appropriate in the thermodynamic limit. Now we perform the partial traces using $[\rho_\mathbf{k},\rho_\mathbf{q}] = [\rho_\mathbf{k},a_\mathbf{q}] =  [\rho_\mathbf{k},a^\dag_\mathbf{q}] = 0$ for $\mathbf{q}\neq \mathbf{k}$, since $\rho_\mathbf{k}$ contains an even number of fermions. We furthermore reorder the fermion operators using that the sums do not contain the modes $\pm \mathbf{p}$; i.e., we only have to use the anticommutation property $\{a_\mathbf{q},a_\mathbf{p}^\dag\} =0$ for $\mathbf{q}\neq \mathbf{p}$. That is, the procedure will only yield signs depending on how many reorderings are necessary:
\begin{eqnarray}
\hspace{-2.5cm} \mathrm{Tr}_{\neq \mathbf{p}} \mathcal L [\rho] \hspace{-0.5cm} & \ni & \mathrm{Tr}_{\neq \mathbf{p}}\big[ v_\mathbf{p} v_\mathbf{p}^* a_\mathbf{p}^\dag \rho_\mathbf{p} a_\mathbf{p} \sum_{\mathbf{q}\neq \mathbf{p} } u_\mathbf{q} u_\mathbf{q}^* a_\mathbf{q} \prod_{\mathbf{k}\neq \mathbf{p}} \rho_\mathbf{k} a^\dag_\mathbf{q}  + u_\mathbf{p} u_\mathbf{p}^* a_\mathbf{p} \rho_\mathbf{p} a_\mathbf{p}^\dag \sum_{\mathbf{q}\neq \mathbf{p} } v_\mathbf{q} v_\mathbf{q}^* a^\dag_\mathbf{q} \prod_{\mathbf{k}\neq \mathbf{p}} \rho_\mathbf{k} a_\mathbf{q} \nonumber \\
&  &\qquad -  v_\mathbf{p} u_{-\mathbf{p}}^* a_\mathbf{p}^\dag \rho_\mathbf{p} a_{-\mathbf{p}}^\dag \sum_{\mathbf{q}\neq \mathbf{p} } u_\mathbf{q} v_{-\mathbf{q}}^* a_\mathbf{q} \prod_{\mathbf{k}\neq \mathbf{p}} \rho_\mathbf{k} a_{-\mathbf{q}} \nonumber \\
&&\qquad  - u_\mathbf{p} v_{-\mathbf{p}}^* a_\mathbf{p} \rho_\mathbf{p} a_{-\mathbf{p}} \sum_{\mathbf{q}\neq \mathbf{p} } u^*_\mathbf{q} v_{-\mathbf{q}} a^\dag_{-\mathbf{q}} \prod_{\mathbf{k}\neq \mathbf{p}} \rho_\mathbf{k} a^\dag_\mathbf{q} +  \{\mathbf{p}\to -\mathbf{p}\}\big] \nonumber \\
& = & \qquad\,\, v_\mathbf{p} v_\mathbf{p}^* a_\mathbf{p}^\dag \rho_\mathbf{p} a_\mathbf{p} \sum_{\mathbf{q}\neq \mathbf{p} } u_\mathbf{q} u_\mathbf{q}^* \langle a^\dag_\mathbf{q}  a_\mathbf{q}\rangle 
  + u_\mathbf{p} u_\mathbf{p}^* a_\mathbf{p} \rho_\mathbf{p} a_\mathbf{p}^\dag \sum_{\mathbf{q}\neq \mathbf{p} } v_\mathbf{q} v_\mathbf{q}^* \langle a_\mathbf{q}  a^\dag_\mathbf{q}\rangle \nonumber\\
& & - v_\mathbf{p} u_{-\mathbf{p}}^* a_\mathbf{p}^\dag \rho_\mathbf{p} a_{-\mathbf{p}}^\dag \sum_{\mathbf{q}\neq \mathbf{p} } u_\mathbf{q} v_{-\mathbf{q}}^* \langle a_{-\mathbf{q}}  a_\mathbf{q}\rangle 
  - u_\mathbf{p} v_{-\mathbf{p}}^* a_\mathbf{p} \rho_\mathbf{p} a_{-\mathbf{p}} \sum_{\mathbf{q}\neq \mathbf{p} } u^*_\mathbf{q} v_{-\mathbf{q}} \langle a^\dag_\mathbf{q}  a_{-\mathbf{q}}^\dag\rangle \nonumber\\
  &&  +  \{\mathbf{p}\to -\mathbf{p}\},
\end{eqnarray}
where in the last equality we have chosen the normalization $\mathrm{Tr}_\mathbf{k}\rho_\mathbf{k} = 1$ and the cyclic invariance of the trace. In the thermodynamic limit, $\sum_{\mathbf{q}\neq \mathbf{p} }(.) = \sum_{\mathbf{q}}(.)$. Close to the steady state we can evaluate the correlation functions using the stationary form equation (\ref{bcsth}) for the state, which take a factorized form, i.e., $\langle a^\dag_\mathbf{q}  a_\mathbf{q}\rangle = |\bar v_\mathbf{q}|^2,  \langle a_\mathbf{q}  a^\dag_\mathbf{q}\rangle = |\bar u_\mathbf{q}|^2, \langle a_{-\mathbf{q}}  a_\mathbf{q}\rangle = - \bar u_\mathbf{q}^* \bar v_\mathbf{q}, \quad \langle a^\dag_\mathbf{q}  a_{-\mathbf{q}}^\dag\rangle = - \bar v_\mathbf{q}^* \bar u_\mathbf{q}$. We can thus factorize a real positive number $\kappa_0$ from the last equation, such that  mean-field Liouvillian operator for the mode pair $\{\mathbf{k},-\mathbf{k}\}$  takes the form
\begin{eqnarray}
\mathcal L_\mathbf{k} [\rho_\mathbf{k}] &=& \mathrm{Tr}_{\neq \mathbf{p}} \mathcal L [\rho] = \kappa_0 \sum_{\sigma = \pm} (L_{\sigma \mathbf{k} } \rho_\mathbf{k}  L^\dag_{\sigma \mathbf{k} }  - \frac{1}{2} \{L^\dag_{\sigma \mathbf{k} }  L_{\sigma \mathbf{k} } , \rho_\mathbf{k} \})\\\nonumber
&=& \sum_{\sigma = \pm} \kappa_\mathbf{k} \big( \bar L_{\sigma \mathbf{k} } \rho_\mathbf{k}  \bar L^\dag_{\sigma \mathbf{k} }  - \frac{1}{2} \{\bar L^\dag_{\sigma \mathbf{k} }  \bar L_{\sigma \mathbf{k} } , \rho_\mathbf{k} \}\big), \\\nonumber
\kappa_\mathbf{k}  &=& \kappa_ 0  N_\mathbf{k} , \quad \kappa_0 = \sum_\mathbf{q} |\bar v_\mathbf{q} \bar u_\mathbf{q} |^2 =  \sum_\mathbf{q}\frac{ | v_\mathbf{q}  u_\mathbf{q} |^2}{N^2_\mathbf{q}},
\end{eqnarray}
with $L_\mathbf{k} , \bar L_\mathbf{k} $ specified in Eqs. (\ref{eq:Lklk}), (\ref{eq:Lkbar}).

In addition to the effective dissipative rate, we can also calculate the amplitude of the relative coefficient $\alpha$ in dependence of the particle filling $n = \sum_\mathbf{q} \langle a^\dag_\mathbf{q}  a_\mathbf{q}\rangle$ from mean-field theory. While for the number conserving equation the particle number operator is a constant of motion due to $[\ell_i, \hat N ] =0$ (i.e., the expectation values of any power of the total particle number are conserved, $\partial_t \langle \hat N^m\rangle=0$), no such property is found for the mean-field dynamics. However, for any initial average particle filling $n= \langle \hat N\rangle/L^d$ there is an $\alpha = re^{\mathrm i \theta}$ such that the first moment, i.e., the average particle density, is conserved, and $\partial_t \langle \hat N\rangle =0$ holds. In other words, the value of $\alpha$ fixes the average particle number, which is intuitive since $\alpha$ measures the relative strength of creating fermions vs. annihilating fermions. The number equation of state for which $ n(t) = \text{const}.$ is given by
\begin{eqnarray}
n &=& \sum_\mathbf{q}|\bar v_\mathbf{q}|^2
= \sum_\mathbf{q}  \frac{|v_\mathbf{q}|^2}{|v_\mathbf{q} |^2 + |\alpha v_\mathbf{q} |^2} .
\end{eqnarray}
While, therefore, the modulus $|\alpha| = r$ is fixed by an additional physical condition from the initial state, no such condition exists which fixes the phase $\theta$. This is physically meaningful, since the original number conserving operators are invariant under global $U(1)$ phase rotations. 

The interpretation of $\alpha = r e^{\mathrm i \theta}$ therefore is the following: (i) the macroscopic phase is fixed spontaneously by the dynamics, i.e., it indicates spontaneous symmetry breaking of $U(1)$ phase rotations in the dissipative setting. (ii) The amplitude relates to the average particle number in the sample, which can be chosen such as to match the average particle number of the number conserving setting $n = N/L^d$ in the initial state.

\subsection{Explicit derivation of the general form of Majorana zero-damping modes for translation-invariant dissipative processes}
\label{app:explicit}

In this section, we provide an explicit derivation of equation~\eref{eq:generalConditionZeroMode} in the framework introduced in section~\ref{sec:generalFormMZMs} (using the same notations).

An arbitrary Majorana operator $\gamma = \gamma^\dagger$ can be expressed, in terms of the fermionic creation and annihilation operators $a^\dagger_i$ and $a_i$ corresponding to distinct lattice sites $i$ of the system, in the form
\begin{eqnarray} \label{appeq:generalFormZeroMode}
    \gamma = \sum_j (\alpha_j a_j + h.c.),
\end{eqnarray}
where the sum runs over all sites $j$ belonging to the system and $\alpha_j \in \mathbb{C}$ with $\sum_j \abs{\alpha_j}^2 = 1$ (so that $\{ \gamma, \gamma^\dagger \} = 2$ as required for a Majorana operator). The operator $\gamma$ then corresponds to a Majorana zero-damping mode of the dissipative dynamics described by the translation-invariant Lindblad operators $L_i = \sum_{j \in \mathcal{I}(i)} u_{j-i} \, a_j + v_{j-i} \, a^\dagger_j$ (equation~\eref{eq:translationInvariantLindbladOp}) if and only if the following conditions are satisfied (see section~\ref{sec:dissipativeGap}):
\begin{eqnarray} \label{appeq:conditionZeroMode}
    0 = \{ L_i, \gamma \} = \{ L_i, \gamma_i \} \qquad \text{(for all $i$)},
\end{eqnarray}
where $\gamma_i$ denotes the restriction of the Majorana operator $\gamma$ to the sites $j \in \mathcal{I}(i)$ where $L_i$ acts non-trivially. These anticommutation conditions obviously constrain the possible form of $\gamma_i$ in a way that solely depends on the form of $L_i$. Since the form Lindblad operator $L_i$ does not depend on $i$ due to translation invariance, $\gamma_i$ must be of the form
\begin{eqnarray} \label{appeq:majoranaMode}
    \gamma_i = \sum_{j \in \mathcal{I}(i)} (\beta_{j-i} \alpha_i a_j + h.c.),
\end{eqnarray}
where $\beta_{j-i}$ are complex factors ($\beta_{j-i}$ being a shorthand notation for $\beta(\vect{r}_j - \vect{r}_i)$, as defined in section~\ref{sec:generalFormMZMs}) which, for consistency with equation~\eref{appeq:generalFormZeroMode}, must satisfy the following set of equations:
\begin{eqnarray}
    \beta_{j-i} \alpha_i = \alpha_j; \qquad & \beta^*_{j-i} \alpha^*_i = \alpha^*_j; \\
    \beta_{i-j} \alpha_j = \alpha_i; \qquad & \beta^*_{i-j} \alpha^*_j = \alpha^*_i,
\end{eqnarray}
leading to a single ``consistency relation''
\begin{eqnarray} \label{appeq:consistencyRelation}
    \beta_{k-i} = \beta_{k-j} \beta_{j-i},
\end{eqnarray}
which is valid for any triple of indices $(i, j, k)$ associated with lattice sites belonging to the system, with $\beta_{i-i} \equiv \beta_0 = 1$. Since $\beta_{j-i} = (\beta_{i-j})^{-1}$ and $\beta^*_{j-i} = (\beta^*_{i-j})^{-1}$, the factors $\beta_{j-i}$ must be real.

Equation~\eref{appeq:consistencyRelation} implies that any Majorana zero-damping mode $\gamma$ can be fully constructed from the only knowledge of $d$ real factors $\beta_{\vect{e}_n}$, where $\vect{e}_n$ ($n = 1, 2, \ldots, d$) are primitive vectors associated with the $d$-dimensional Bravais lattice on which the system is defined. Such a construction can be made starting from any lattice site $i$, choosing a phase $\phi_i \in [0, 2\pi)$ for the coefficient $\alpha_i$ of $\gamma$ (see equation~\eref{appeq:generalFormZeroMode}) corresponding to that particular site. This allows us to express $\gamma$ in the generic form presented in the main text (see section~\ref{sec:generalFormMZMs}):
\begin{eqnarray} \label{appeq:generalSolutionZeroMode}
    \gamma = \mathcal{N} \rme^{\rmi \phi_i / 2} \sum_{\{ m_1, m_2, \atop \phantom{\{} \ldots, m_d \}} (\beta_{\vect{e}_1})^{n_1} (\beta_{\vect{e}_2})^{n_2} \ldots (\beta_{\vect{e}_d})^{n_d} a(\vect{r}_i + \sum_{n = 1}^d m_n \vect{e}_n) + h.c. \; ,
\end{eqnarray}
where $\mathcal{N} > 0$ is a normalization factor and $\{ m_1, m_2, \ldots, m_d \}$ a set of integers defined such that the vectors $\vect{r}_i + \sum_{n = 1}^d m_n \vect{e}_n$ span the positions of all sites in the system~\footnote{Note that $\gamma = \gamma^\dagger$ is defined up to a sign, which allows us to introduce the phase $\phi_i$ as $\rme^{\rmi \phi_i / 2}$.}.

The explicit form of the elementary ``building blocks'' $\beta_{\vect{e}_n}$ defining the form of $\gamma$ can be found by expressing equation~\eref{appeq:conditionZeroMode} explicitly using equation~\eref{appeq:majoranaMode}, namely,
\begin{eqnarray} \label{appeq:conditionZeroMode2}
    0 & = & \sum_{j \in \mathcal{I}(i)} \sum_{k \in \mathcal{I}(i)} \left\{  u_{j-i} a_j + v_{j-i} a^\dagger_j, \beta_{k-i} \alpha_i a_i + \beta_{k-i} \alpha^*_i a^\dagger_i \right\} \nonumber \\
    & = & \sum_{j \in \mathcal{I}(i)} \left( \alpha^*_i u_{j-i} \beta_{j-i} + \alpha_i v_{j-i} \beta_{j-i} \right) \nonumber \\
    & = & \frac{1}{2} \sum_{j \in \mathcal{I}(i)} \left[ \alpha^*_i (u_{j-i} \beta_{j-i} + u_{\overline{j}-i} \beta_{\overline{j}-i}) + \alpha_i (v_{j-i} \beta_{j-i} + v_{\overline{j}-i} \beta_{\overline{j}-i}) \right],
\end{eqnarray}
where in the last step we have symmetrized the expression using the fact that $\mathcal{I}(i)$ contains pairs of sites $(j, \overline{j})$ located symmetrically around the center of symmetry $\mathcal{S}_i$ of $L_i$ (see section~\ref{sec:generalFormMZMs}). The fact that the Lindblad operators are symmetric around $\mathcal{S}_i$ and that the dissipative dynamics leads to a pure steady state as assumed in the main text implies that $u_{j-i} = -u_{\overline{j}-i}$ and $v_{j-i} = v_{\overline{j}-i}$ (see also equation~\eref{eq:oddConstraintFromPurity} and discussion thereof). Equation~\eref{appeq:conditionZeroMode2} therefore reduces to
\begin{eqnarray} \label{appeq:generalConditionZeroMode}
    0 & = & \sum_{j \in \mathcal{I}(i)} \left[ \mathrm{e}^{-\rmi \phi_i} u_{j-i} (\beta_{j-i} - \beta_{\overline{j}-i}) + v_{j-i} (\beta_{j-i} + \beta_{\overline{j}-i}) \right],
\end{eqnarray}
which is the result presented in the main text (see equation~\eref{eq:generalConditionZeroMode}). We note that this equation can be further simplified when the center of symmetry $\mathcal{S}_i$ of $L_i$ coincides with a lattice site. In that case, $\beta_{\overline{j}-i} = \beta_{i-j} = (\beta_{j-i})^{-1}$ and we find
\begin{eqnarray} \label{appeq:generalConditionZeroMode2}
    0 = v_0 + \frac{1}{2} \sum_{j \in \mathcal{I}(i) \atop j \neq i} [ \mathrm{e}^{-\rmi \phi_i} u_{j-i} (\beta_{j-i} - (\beta_{j-i})^{-1}) + v_{j-i} (\beta_{j-i} + (\beta_{j-i})^{-1}) ].
\end{eqnarray}
We remark that equation~\eref{appeq:generalConditionZeroMode} (as well as equation~\eref{appeq:generalConditionZeroMode2}) can be expressed in terms of the factors $\beta_{\vect{e}_n}$ only using equation~\eref{appeq:consistencyRelation}. If there exists a set $\{ \beta_{\vect{e}_n} \}_{n = 1, 2, \ldots, d}$ of real factors $\beta_{\vect{e}_n}$ satisfying equation~\eref{appeq:generalConditionZeroMode} for some particular phase $\phi_i \in [0, 2 \pi)$, the system can support \emph{at least one} Majorana zero-damping mode of the form~\eref{eq:generalSolutionZeroMode}. Otherwise there can be no such mode. In order to assess the possibility of having Majorana zero-damping modes, one therefore has to solve a polynomial equation in $d$ real variables ($\{ \beta_{\vect{e}_n} \}_{n = 1, 2, \ldots, d}$) with coefficients that may be real or complex depending on the coefficients $\mathrm{e}^{-\rmi \phi_i} u_{j-i}$, since $v_{j-i}$ can always be chosen as real (see section~\ref{sec:physicalConstraints}). The existence of solutions thus depends on the phase $\phi_i$ that is chosen as an ``initial condition'' for the construction of $\gamma$ starting from site $i$ according to equation~\eref{appeq:generalSolutionZeroMode}.

We finally discuss the physical meaning of the phase $\phi_i$ in the two types of dissipative systems that are most relevant to this work (see section~\ref{sec:physicalConstraints}): (i) 1D dissipative systems with TRS (symmetry class BDI) and (ii) 2D dissipative systems without TRS (symmetry class D)~\footnote{Note that PHS is automatically satisfied in our dissipative framework (see section~\ref{sec:topPropertiesBulk}).}. In the 1D case, TRS constrains the coefficients $u_{j-i}$ to be real \emph{up to a global phase} and the phase $\phi_i$ crucially allows to compensate for such a phase so that all coefficients $\mathrm{e}^{-\rmi \phi_i} u_{j-i}$ are real. Equation~\eref{appeq:generalConditionZeroMode} then reduces to a univariate polynomial equation with \emph{real} coefficients which, as opposed to a univariate polynomial equation with complex coefficients (as would generically be obtained in the absence of TRS), can have robust solutions (i.e., solutions that survive if the coefficients $u_{j-i}$ and $v_{j-i}$ are slightly modified). In the 2D case, $v_{j-i}$ typically exhibits an isotropic (or $s$-wave symmetric) form due to physical constraints (see section~\ref{sec:physicalConstraints}) and TRS can only be broken if $u_{j-i} \equiv u(\vect{r}_j - \vect{r}_i)$ transforms under rotations as an eigenfunction of angular momentum with odd integer eigenvalue $\ell \neq 0$, namely, $u(\vect{r}) = u(r, \varphi) \sim f(r) \mathrm{e}^{-\rmi \ell \varphi}$ with polar coordinates $(r, \varphi)$, $f(r)$ being an arbitrary positive function of $r$. We then find $\mathrm{e}^{-\rmi \phi_i} u_{j-i} = \mathrm{e}^{-\rmi \phi_i} u(r, \varphi) = u(r, \varphi + \phi_i / \ell)$, showing that modifying the phase $\phi_i$ simply corresponds to rotating the reference frame or, equivalently, to modifying the orientation of the edge of the system. If there exists a solution of equation~\eref{eq:generalConditionZeroMode} for some $\phi_i$, there must exist a solution for any $\phi_i \in [0, 2\pi)$ or, equivalently, for any orientation of the edge. TRS breaking thus crucially allows for the existence of Majorana zero-damping modes at edges that are curved arbitrarily, as expected physically.

\section*{References}
\bibliography{iopart-num}

\end{document}